\documentclass[aps,10pt,twocolumn,tightenlines,nofootinbib,floatfix]{revtex4-1}
\usepackage[utf8]{inputenc}
\usepackage{amsmath}
\usepackage{appendix}
\usepackage{natbib}
\usepackage{graphicx}
\usepackage{geometry}
\usepackage{multirow}
\usepackage{ragged2e}
\usepackage{amsfonts,amssymb}
\usepackage[table]{xcolor}
\usepackage{array}
\newcolumntype{C}[1]{>{\centering\let\newline\\\arraybackslash\hspace{0pt}}m{#1}}
\usepackage[backref,breaklinks,colorlinks]{hyperref}
\hypersetup{linkcolor=blue,citecolor=blue,urlcolor=blue}

\definecolor{pink1}{RGB}{226, 24, 166}

\definecolor{hookgreen}{rgb}{0.0,0.44,0.0}

\newcommand{\sideheader}[1]{\noindent --- \textit{\textbf{#1}} ---}

\newcommand{\taon}{$\tau$-lepton }
\newcommand{\taons}{$\tau$-leptons }

\newcommand{\grandacro}{GRAND200k}
\newcommand{\kmthreenet}{KM3NeT}
\newcommand{\bnsevent}{GW170817}
\newcommand{\bnseventgrb}{GRB170817A}
\newcommand{\icevent}{IC-170922A}
\newcommand{\iceventtde}{IC-191001A}
\newcommand{\txsblazar}{TXS0506+056}

\newcommand{\lsim}{\mathrel{\hbox{\rlap{\lower.75ex \hbox{$\sim$}} \kern-.3em \raise.4ex \hbox{$<$}}}}
\newcommand{\gsim}{\mathrel{\hbox{\rlap{\lower.75ex \hbox{$\sim$}} \kern-.3em \raise.4ex \hbox{$>$}}}}

\begin{document}

\title{POEMMA's target of opportunity sensitivity \\ to cosmic neutrino transient sources}

\author{Tonia M. Venters}
\affiliation{Astrophysics Science Division, NASA Goddard Space Flight Center, Greenbelt, MD 20771, USA}
\author{Mary Hall Reno}
\affiliation{Department of Physics and Astronomy, University of Iowa, Iowa City, IA 52242, USA}
\author{John F. Krizmanic}
\affiliation{CRESST/NASA Goddard Space Flight Center, Greenbelt, MD 20771, USA \\
University of Maryland, Baltimore County, Baltimore, MD 21250, USA}
\author{Luis A. Anchordoqui}
\affiliation{
Department of Physics, Graduate Center, City University of New York (CUNY), NY 10016, USA\\
Department of Physics and Astronomy,  Lehman College (CUNY), NY 10468, USA\\
Department of Astrophysics, American Museum of Natural History, NY 10024, USA}
\author{Claire Gu\'epin}
\affiliation{Joint Space-Science Institute, University of Maryland, College Park, MD 20742, USA}
\author{Angela V. Olinto}
\affiliation{Department of Astronomy \& Astrophysics, KICP, EFI, The University of Chicago, Chicago, IL 60637, USA}

\date{\today}

\begin{abstract}
\noindent We investigate the capability of the Probe Of Extreme Multi-Messenger Astrophysics (POEMMA) in performing Target-of-Opportunity (ToO) neutrino observations. POEMMA is a proposed space-based probe-class mission for ultrahigh-energy cosmic ray and very-high-energy neutrino detection using two spacecraft, each equipped with a large Schmidt telescope to detect optical and near-ultraviolet signals generated by extensive air showers (EASs). POEMMA will be sensitive to Cherenkov radiation from upward-moving EASs initiated by tau neutrinos interacting in the Earth. POEMMA will be able to quickly re-point ($90^\circ$ in $500$~s) each of the two spacecraft to the direction of an astrophysical source, which in combination with its orbital speed will provide it with unparalleled capability to follow up transient alerts. We calculate POEMMA's transient sensitivity for two observational configurations for the satellites (ToO-stereo and ToO-dual for smaller and larger satellite separations, respectively) and investigate the impact of variations arising due to POEMMA's orbital characteristics on its sensitivity to tau neutrinos in various regions of the sky. We explore separate scenarios for long ($\sim 10^{5-6}$~s) and short ($\sim 10^3$~s) duration events, accounting for intrusion from the Sun and the Moon in the long-duration scenario. We compare the sensitivity and sky coverage of POEMMA for ToO observations with those for existing experiments (\textit{e.g.}, IceCube, ANTARES, and the Pierre Auger Observatory) and other proposed future experiments (\textit{e.g.}, \grandacro). For long bursts, we find that POEMMA will provide a factor of $\gtrsim 7$ improvement in average neutrino sensitivity above $300$~PeV with respect to existing experiments, reaching the level of model predictions for neutrino fluences at these energies and above from several types of long-duration astrophysical transients (\textit{e.g.}, binary neutron star mergers and tidal disruption events). For short bursts, POEMMA will improve the sensitivity over existing experiments by at least an order of magnitude for $E_\nu \gsim 100$~PeV in the ``best-case'' scenario. POEMMA's orbital characteristics and rapid re-pointing capability will provide it access to the full celestial sky, including regions that will not be accessible to ground-based neutrino experiments. Finally, we discuss the prospects for POEMMA to detect neutrinos from candidate astrophysical neutrino sources in the nearby universe. Our results demonstrate that with its improved neutrino sensitivity at ultra-high energies and unique full-sky coverage, POEMMA will be an essential, complementary component in a rapidly expanding multi-messenger network.

\end{abstract}
\maketitle

\section{Introduction}

Astrophysical transients are now a staple of multi-wavelength observations of electromagnetic signals by ground-based and space-based telescopes. In the last few years, multi-messenger astronomy has blossomed with coincident observations of photons and gravitational waves or high-energy neutrinos. In 2017, LIGO reported the groundbreaking observation of gravitational waves from a binary neutron star (BNS) merger~\cite{TheLIGOScientific:2017qsa} coincident with a number of electromagnetic signals~\cite{GBM:2017lvd}. In 2018, the correlation of a neutrino event in IceCube with multi-wavelength observations of a flaring blazar~\cite{IceCube:2018dnn} heralded the beginning of multi-messenger programs using high-energy neutrinos. The next decade could pave the way for simultaneous observations of three astronomical messengers --- photons, neutrinos, and gravitational waves --- from the same astrophysical transients.

Here we derive the unique contributions to the multi-messenger studies of transient phenomena of a space-based mission designed to observe neutrinos above $10$~PeV. Below PeV energies, ground-based neutrino detectors~\cite{Aab:2015kma,Aab:2016ras,Abbasi:2019fmh,Anker:2019mnx,Aartsen:2018vtx,Ahnen:2018ocv,Allison:2015eky,Aab:2019gra} have the benefit of nearly full-sky coverage, but above such a critical energy, large areas of the sky become inaccessible to a given ground-based observatory because the Earth attenuates higher-energy neutrinos. Space-based neutrino detectors, while typically restricted in field-of-view (FoV), can be re-pointed to respond to astrophysical source alerts throughout the entire sky. For long transients, space-based instruments have the advantage of full-sky coverage, given the orbital motion and the precession of the orbit. For shorter transients, the capability to quickly reorient the instruments provides access to all sources that produce signals in the dark sky.

Astrophysical neutrino transient sources come from a wide range of phenomena~\cite{Guepin:2017dfi,Meszaros:2017fcs,Ackermann:2019ows}. Gamma-ray burst (GRB) emission is a textbook example~\cite{Waxman:1997ti,Murase:2007yt,Kimura:2017kan}. In tidal disruption events (TDEs), supermassive black holes (SMBHs) pull in stellar material that interacts with thermal and non-thermal photons to produce neutrinos~\cite{Dai:2016gtz,Lunardini:2016xwi}. Blazar flares, dominant sources of extragalactic gamma rays, may be important neutrino sources~\cite{Rodrigues:2017fmu,IceCube:2018dnn}. Neutrino fluence predictions from binary black hole (BBH)~\cite{Kotera:2016dmp} and BNS~\cite{Fang:2017tla} mergers may tie sources of gravitational waves and electromagnetic signals to neutrino signals. Neutrinos, not gamma rays, may be the primary signal of cosmic-ray acceleration in binary white dwarf (BWD) mergers~\cite{Xiao:2016man}. The spin-down of newly-born pulsars ultimately produces cosmic rays that may interact with the hadronic environment to produce neutrinos~\cite{Fang:2014qva}.

Neutrino and antineutrino production in these transient astrophysical sources is dominated by pion production for a large range of energies. For $E_\nu \gsim$ 10$^6$ GeV, the neutrino- and antineutrino-nucleon cross sections are effectively equal~\cite{Gandhi:1998ri}, so we do not distinguish between neutrinos and antineutrinos. To a first approximation, charged pion decay gives two muon neutrinos for each electron neutrino~\cite{Lipari:2007su}. The nearly maximal mixing of muon neutrinos and tau neutrinos in the Pontecorvo-Maki-Nakagawa-Sakata matrix of neutrino flavor mixing~\cite{Tanabashi:2018oca} results in approximately equal electron neutrino, muon neutrino, and tau neutrino fluxes at the Earth~\cite{Learned:1994wg}. Tau neutrinos that interact in the Earth produce \taons that can decay in the atmosphere producing upward-moving extensive air showers (EASs). They provide a unique signal for satellite-based or balloon-borne instruments~\cite{Domokos:1997ve,Fargion:2000iz,Fargion:2003ms,Fargion:2003kn,PalomaresRuiz:2005xw,Neronov:2016zou,Olinto:2017xbi,Olinto:2018avl,Allison:2018cxu,Gorham:2019guw,Krizmanic:2013pea}, and Earth-based instruments like the Pierre Auger Observatory~\cite{Bertou:2001vm,Feng:2001ue,Lachaud:2002sx,Aramo:2004pr,Abreu:2012zz} or other surface arrays~\cite{Hou:2002bh,Asaoka:2012em,Otte:2018uxj,Alvarez-Muniz:2018bhp,Neronov:2019htv}. 

At high elevation angles, the large path lengths through the Earth result in significant attenuation in the neutrino flux at high energies; however, Earth-skimming neutrinos that emerge with relatively small elevation angles can produce EAS signals. Tau neutrinos have the added feature that their attenuation through the Earth can be somewhat mitigated by regeneration, since the secondary \taon could decay and produce a third-generation tau neutrino, albeit at a lower energy~\cite{Halzen:1998be,Becattini:2000fj,Dutta:2000jv,Beacom:2001xn,Alvarez-Muniz:2017mpk}. 

The Probe Of Extreme Multi-Messenger Astrophysics (POEMMA)~\cite{Olinto:2017xbi} is a space-based mission described in the NASA Astrophysics Probe study report~\cite{POEMMAConcept}. POEMMA is optimized for measurements of EASs both from ultra-high--energy cosmic rays (UHECRs) using the stereo air fluorescence technique with the satellites in a quasi-nadir viewing configuration (POEMMA-stereo mode) and from upward-going tau neutrinos via Cherenkov signals in the optical band ($300 - 900$~nm) with the satellites pointed closer to the Earth limb (POEMMA-limb mode). The POEMMA instruments can quickly re-point towards the direction of a transient source and track it through the neutrino detection region, enabling follow-up of Target-of-Opportunity (ToO) alerts in neutrinos and/or other astrophysical messengers. POEMMA operates during astronomical night in order to measure the near-ultraviolet air fluorescence and optical Cherenkov EAS signals.

The POEMMA satellite-based instruments are planned to orbit in tandem with a separation of the order of $300$~km at an altitude of $h=525$~km, and with an orbital period of $T_s=95$~min. The orbital plane is oriented at an angle of $\xi_i=28.5^{\circ}$ relative to the Earth's polar axis, and the precession period is $T_p=54.3$~days. The spacecraft avionics will allow POEMMA to quickly slew its pointing by as much as $90^{\circ}$ in $500$~s. With these design features, POEMMA will have access to the entire dark sky within the timescale of one orbit. In the case of transients lasting longer than a day, the spacecraft propulsion systems will allow for adjusting the separation between the two satellites to bring a source within overlapping instrument light pools, lowering the energy threshold for detecting neutrinos. As such, POEMMA ToO observations will be conducted in one of two satellite configurations, depending on the duration of the transient event: the ToO-dual configuration with large satellite separation for short-duration events, and the ToO-stereo configuration with small satellite separation for long-duration events.

The focal plane of each POEMMA telescope contains an edge sector that is optimized for optical Cherenkov detection, with an FoV of $\sim 30^{\circ} \times 9^{\circ}$ for neutrino observations. In POEMMA-limb mode, the POEMMA instruments will be tilted to cover a viewing area extending from $7^{\circ}$ below the horizon to $2^{\circ}$ above it, equivalent to covering \taon trajectories emerging from the Earth with elevation angles $\beta_{\rm tr}\lsim 20^\circ$~\cite{Guepin:2018yuf,PhysRevD.100.063010} while measuring the background Cherenkov signal from potential above-the-limb UHECRs. To follow a ToO flaring neutrino source, the POEMMA telescopes can quickly
slew to larger angles below the horizon, keeping the source within the $\sim 30^{\circ} \times 9^{\circ}$ neutrino FoV, even after accounting for the few degree smearing due to the Cherenkov emission angle. 

In this paper, we calculate the neutrino sensitivity of POEMMA for both long and short transient events, and evaluate the prospects for detecting neutrinos from several candidate transient astrophysical source classes. Section~\ref{sec:2} presents the calculation of POEMMA's effective area, exposure and sensitivity to neutrino fluences. In Sec.~\ref{sec:3}, we describe our calculation of the expected numbers of events from flaring neutrino sources and discuss POEMMA's sky coverage in terms of detecting neutrinos according to two astrophysical models for two distinct ToO scenarios of multi-messenger follow-up observations and neutrino-only observations. Section~\ref{sec:3} also provides the maximum luminosity distances for detecting a single neutrino event for several astrophysical neutrino models and descriptions of the most promising source classes for ToO observations with POEMMA based on the occurrence of transient events, modeled as a Poisson process. We conclude in Sec.~\ref{sec:4}. Additional details for the effective area evaluation are included in Appendix~\ref{app:a}, and a discussion of considerations in setting the photo-electron (PE) threshold in the ToO-stereo and ToO-dual cases appears in Appendix~\ref{app:aprime}. Appendix~\ref{app:adoubleprime} provides detailed discussions of POEMMA's angular resolution and backgrounds for ToO observations. Appendix~\ref{app:b} discusses the relationship between isotropic equivalent source characteristics and the fluence observed at a source luminosity distance. Appendix~\ref{app:c} provides descriptions of additional proposed astrophysical neutrino source classes.

\section{POEMMA's Effective Area, Exposure, and Sensitivity}
\label{sec:2}

The effective area evaluation begins with the geometrical configuration of an instrument at $h=525~{\rm km}$ above the Earth. For measurements of the diffuse flux, more than $300~{\rm km^2 \, sr}$ of geometric aperture is accessible to POEMMA~\cite{PhysRevD.100.063010}. For point sources, the evaluation of the effective area depends on the elevation angle $\beta_{\rm tr}$ (with respect to the surface of the Earth) of the \taon trajectory and the elevation angle of the line of sight to the detectors from the point on the Earth at which the \taon emerges (the length of the line of sight is given by $v$ and makes an elevation angle $\beta_{v}$ with the spot on the ground). The decay length of the \taon along the line of sight is $s$. Details of the geometry are given in Ref.~\cite{PhysRevD.100.063010} and described here in Appendix~\ref{app:a}.

The ToO sensitivity at a given time depends on the area $A_{\rm Ch}$ subtended on the ground by the Cherenkov cone. For an EAS produced along the \taon trajectory emerging at angle $\beta_{\rm tr}$ and initiated by the \taon decay at altitude $a$, with a path length before decay $s(\beta_{\rm tr},a)$, we approximate
\begin{equation}
A_{\rm Ch} (s) \simeq \pi (v-s)^2 \times \left(\theta_{\rm Ch}^{\rm eff}\right)^2\ ,
\label{eq:ACh}
\end{equation}
where we take $\beta_{v}(t)\simeq \beta_{\rm tr}(t)$ and $\theta_{\rm Ch}^{\rm eff}$ is the effective Cherenkov angle that takes into account the altitude dependence and a broadening due to an increase in instrument acceptance for more intense Cherenkov signals from high-energy EASs (see Appendix~\ref{app:a}). For the purposes of calculating $\theta_{\rm Ch}^{\rm eff}$, we take the EAS energy, $E_{\rm shr}\simeq 0.5 E_\tau$, which provides a good estimate for the \taon decay channels \cite{PhysRevD.100.063010}. The effective area for $\nu_\tau$ detection is 
\begin{equation}
A(\beta_{\rm tr}(t),E_{\nu}) \simeq \int dP_{\rm obs}(E_{\nu},\beta_{\rm tr},s) A_{\rm Ch}(s) \,,
\label{eq:effA}
\end{equation}
where
the differential probability to observe the \taon EAS is
\begin{eqnarray}
\nonumber
dP_{\rm obs}(E_{\nu},\beta_{\rm tr},s)&=& ds\, P_{\rm exit}(E_{\nu},\beta_{\rm tr})\, p_{\rm dec}(s)\\ 
&\times & P_{\rm det}(E_{\nu},\beta_{\rm tr}, s)\ ,
\end{eqnarray}
where $P_{\rm exit}$ is the exit probability, $p_{\rm dec}$ is the decay distribution, and $P_{\rm det}$ is the detection probability.

The exit probability $P_{\rm exit}(E_{\nu},\beta_{\rm tr})$ depends on the tau neutrino cross section in Earth, the \taon energy distribution from the interaction, and \taon energy loss and decay as it transits through the Earth. Throughout this paper we evaluate the neutrino-nucleon cross section using the nCTEQ15 parton distribution functions~\cite{Kovarik:2015cma} and adopt the Abramowicz-Levin-Levy-Maor (ALLM) parameterization of the proton structure function~\cite{Abramowicz:1991xz,Abramowicz:1997ms} for photonuclear energy loss, as discussed in more detail in Ref.~\cite{PhysRevD.100.063010}. The \taon exit probabilities are shown in Fig.~\ref{fig:pexit} of Appendix~\ref{app:a}. For nadir angles down to $\sim 18^{\circ}$ below the horizon as viewed from POEMMA's altitude ($h = 525$~km), the emergent \taon trajectory elevation angles are $\beta_{\rm tr}\leq 35^\circ$.  For $\beta_{\rm tr}=35^{\circ}$, neutrino attenuation in the Earth gives the probability for a tau neutrino to produce an exiting \taon to be less than $10^{-5}$ for the energies of interest. Thus, our evaluation of Eq.~(\ref{eq:effA}) for $\beta_{\rm tr}\leq 35^{\circ}$ is a good approximation to the full angular range due to the minuscule \taon exit probability for larger angles.

The differential decay distribution is
\begin{equation}
p_{\rm dec}(s)\, ds = B_{\rm shr}\exp(-s/\gamma c \overline{\tau}_\tau)\frac{ds}{\gamma c \overline{\tau}_\tau}\ ,
\end{equation}
where $\overline{\tau}_\tau = \left(290.3 \pm 0.5\right) \times 10^{-15}$~s is the mean lifetime of the \taon and the \taon branching fraction to showers is $B_{\rm shr}=0.826$ (defined by excluding the muon channel with branching fraction $\sim 17.4$\%, based on the conservative assumption that muonic EASs yield Cherenkov signals below POEMMA's detection threshold; \textit{c.f.}, \citenum{Stanev:1989bc}).

Finally, the detection probability is approximated by 
\begin{equation}\label{eq:probdet}
P_{\rm det} \simeq 
H\left[N_{\rm PE} - N_{\rm PE}^{\rm min}\right]
\ ,
\end{equation}
in terms of the Heaviside function, $H(x)$:
\[ H\left(x\right) = \left\{  \begin{array}{ll}
                         0 & \mbox{if $x < 0$} \\
                         1 & \mbox{if $x \geqslant 0$}  \end{array}  \right. .
\]
The number of PEs, $N_{\rm PE}$, is determined from a model of the photon density from the \taon induced air showers as a function of shower energy (where $E_{\rm shr}=0.5 E_\tau$), decay altitude, and $\beta_{\rm tr}$,  multiplied by the collecting area of each detector and the quantum efficiency for photo-detection. The $N_{\rm PE}$ calculation depends on the Cherenkov signal intensity delivered to the POEMMA instruments, accounting for the effects of atmospheric attenuation. In this study, we use the same model for the atmospheric attenuation as in Ref.~\cite{PhysRevD.100.063010}. We use an 
\begin{figure}[htb]
\centering
	\includegraphics[width=0.95\columnwidth]{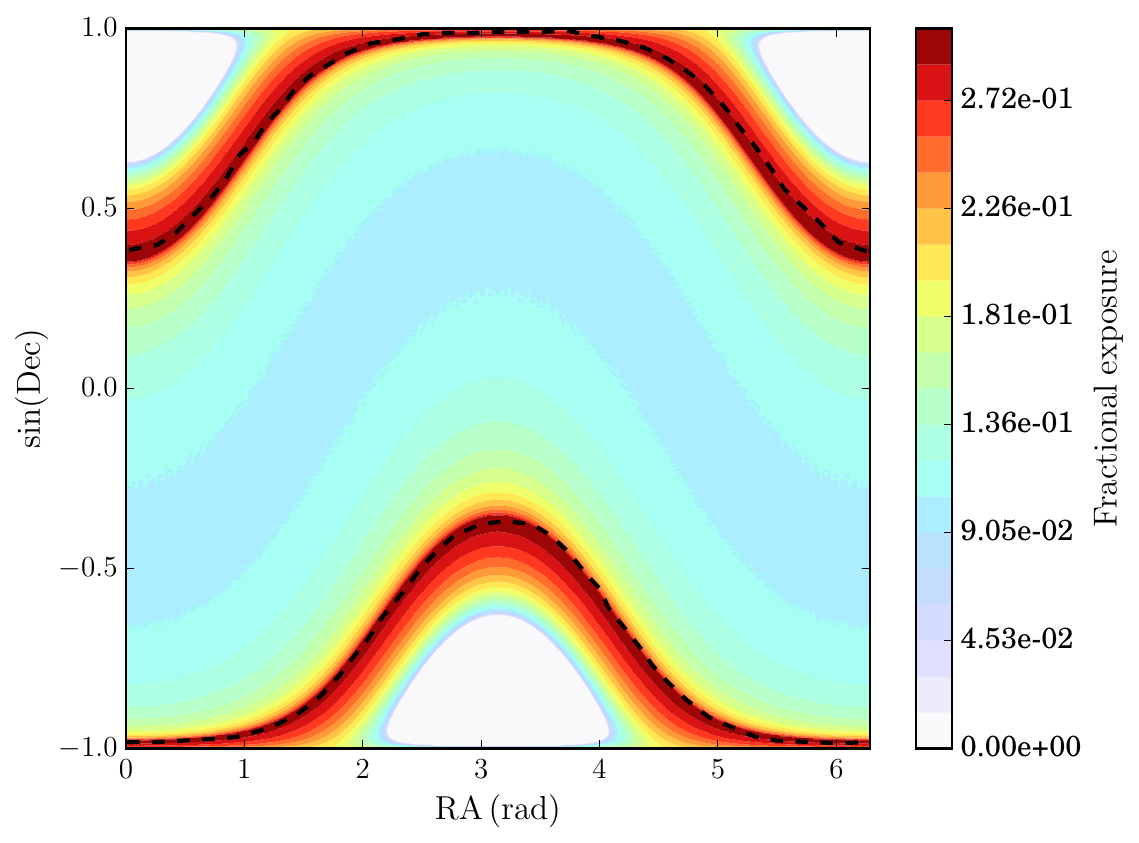}
	\includegraphics[width=0.99\columnwidth]{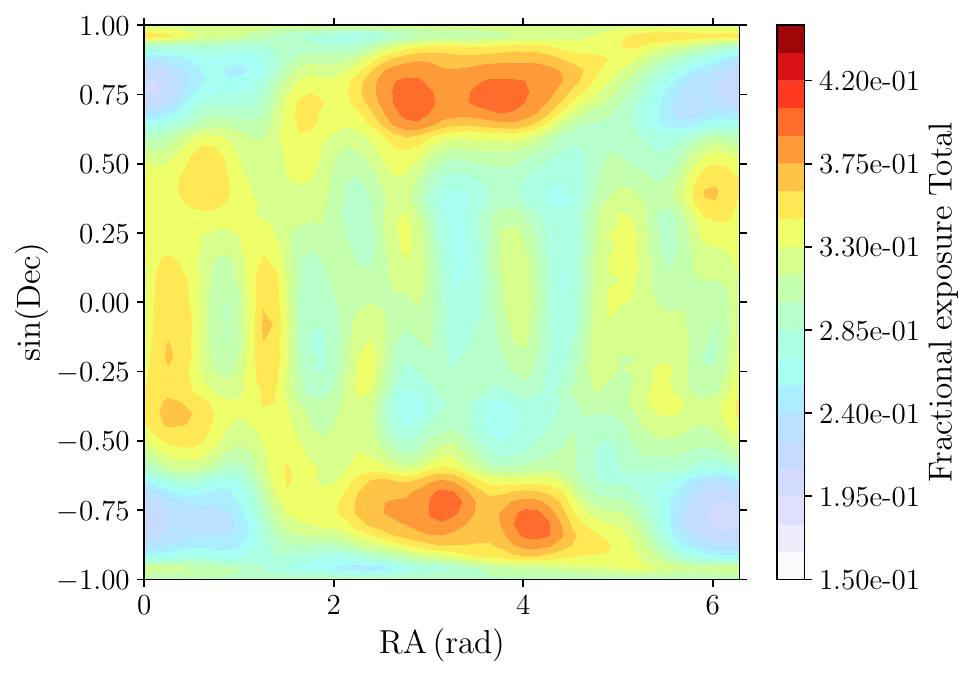}
	\caption{\textit{Upper}: Fractional exposure over one period for a given sky location at a particular time of the year plotted as a function of right ascension and sine of the declination. Viewing angles extend to $18.3^\circ$ below the Earth limb~\cite{Guepin:2018yuf}, and the effects of the Sun and the Moon have been neglected. \textit{Lower}: Range in values for $f_t$, the multiplicative factor that accounts for intrusion from the Sun and the Moon in Eq.~\ref{eq:sensitivity}. Here $f_t$ is plotted as the ratio of the fractional exposure accounting for Sun and Moon effects divided by the fractional exposure excluding Sun and Moon effects. Fractional exposures are calculated as averages over $7$ precession periods of POEMMA's orbital plane ($7\times 54.3\, {\rm days} \simeq 380\, {\rm days}$).}
		\label{fig:SunMoon}	
\end{figure}
optical collection area of $2.5$ m$^2$ and a quantum efficiency of $0.2$. Figures~\ref{fig:Cang} and \ref{fig:photd} in Appendix~\ref{app:a} show the effective Cherenkov angle and photon density as a function of elevation angle and altitude of \taon decay for $\beta_{\rm tr}\leq 40^\circ$. 

The PE threshold, $N_{\rm PE}^{\rm min}$, depends on the observing mode for the POEMMA satellites. It is set by requiring the false positive rate from the average night-sky air glow background (based on Refs. \cite{2003A&A...407.1157H, 2006JGRA..11112307C}) to be a fraction of an event per year ($\lesssim 0.03$~events per year for the entire POEMMA Cherenkov Camera or $\lesssim 0.0002$~events per year within a circle of radius $\sim$ the effective Cherenkov angle), based on the characteristics of Cherenkov signals and POEMMA's response to these signals. For long bursts, we assume the satellites are in the ToO-stereo configuration (within $\sim$ 25 km of each other and viewing the same light pool) with $N_{\rm PE}^{\rm min}=10$ threshold for the calculations. For the short bursts, we assume the satellites are in the ToO-dual configuration (assumed to be separated by $300$~km and not viewing the same light pool) with a higher PE threshold of $N_{\rm PE}^{\rm min}=20$ in each detector. However, the effective area in this mode is double the effective area in ToO-stereo mode for a fixed value of $N_{\rm PE}^{\rm min}$. A more detailed discussion of the ToO-dual and ToO-stereo configurations and their corresponding PE thresholds can be found in Appendix \ref{app:aprime}. A discussion of the PE threshold in POEMMA-limb mode can also be found in Ref.~\cite{PhysRevD.100.063010}.

In addition to the night-sky air glow, potential sources of background for POEMMA during ToO observations include the diffuse cosmic neutrino flux and reflected Cherenkov signals from UHECR showers when viewing away from the Earth's limb.\footnote{In the PeV energy range, the atmospheric neutrino spectrum falls as $E^{-\gamma}$ with $\gamma \sim 3$. At 1 PeV, the atmospheric muon neutrino flux is more than an order of magnitude below the diffuse neutrino flux, and the atmospheric tau neutrino flux is lower by an additional factor of $\sim 10$ \cite{Bhattacharya:2016jce}.} In the case of the diffuse cosmic neutrino flux, we expect contamination to be minuscule ($\lesssim 2.0 \times 10^{-4}$ events per long ToO observation) due to the level of the diffuse flux as compared with POEMMA's diffuse sensitivity~\cite{PhysRevD.100.063010} and the small solid angle defined by POEMMA's angular resolution and the Cherenkov angle. For reflected Cherenkov signals from UHECR EASs, we expect the time spreads for these signals to be much longer than expected for upward-going EASs from tau-neutrinos, making the background UHECR events easily distinguishable from the signal tau-neutrino events. Based on these considerations, we expect the background rate for POEMMA during ToO observations to be minuscule (combined total from air glow and diffuse comsic neutrinos $\lesssim 2.1 \times 10^{-4}$ events per long ToO observation), even allowing for a trials factor of $100$ observations (corresponding to $\lesssim 0.02$ events during long ToO observations over the course of the mission). For these reasons, we do not account for backgrounds in our calculations.

Direct Cherenkov signals from nearly horizontal UHECR EASs when POEMMA is viewing near the Earth's limb (above-the-limb UHECRs) are another potential source of background during ToO observations. However, we exclude these events from our estimates of the background rate as such estimates require a detailed study deserving of an independent publication. Preliminary studies of such events have provided geometrical constraints for their visibility by POEMMA that could lead to constraints on the ToO detection region. Future measurements by balloon-borne Cherenkov detectors such as EUSO-SPB2 will also help determine this background. More detailed discussions of potential backgrounds for POEMMA during ToO observations are provided in Appendix~\ref{app:adoubleprime}.

In calculating the detection probability, a more detailed Monte Carlo simulation was used in Ref.~\cite{PhysRevD.100.063010} to account for $\beta_v\neq\beta_{\rm tr}$ and to impose the requirement that \taon decay within an observation window that depends on the emergence angle and altitude of decay in order to produce detectable air showers. The simplification in Eq.~(\ref{eq:probdet}) is a very good approximation to the more detailed evaluation of the detection probability for the diffuse flux \cite{PhysRevD.100.063010}, so we use it here for the ToO sensitivity.

To determine the sensitivity for a burst, we calculate the time averaged effective area:
\begin{equation}
\langle A(E_{\nu},\theta,\phi)\rangle_{T_0} = \frac{1}{T_0}\int _{t_0}^{t_0+T_0}\, dt A(\beta_{\rm tr}(t),E_{\nu},\theta,\phi)
\ ,
\label{eq:avga}
\end{equation}
where $\theta$ and $\phi$ are the co-latitude and longitude of the source celestial position (\textit{i.e.}, $\phi$ is the right ascension in the equatorial celestial coordinate system and $\theta = \pi/2 - \delta$, where $\delta$ is the declination). For long-duration events during which the source emits neutrinos for a much longer time than the orbital period of POEMMA ($T_s=95$~min $=5.7 \times 10^3$~s), we use the orbit-averaged value, so $t_0=0$ and $T_0=T_s$. For short bursts, we find the average effective area for $T_0=T_{\rm burst}$. We use $T_{\rm burst}=10^3$~s as a representative short burst time in the results shown below. 

For sources that dip just below the horizon as the POEMMA satellites orbit, the effective area is optimal. Some sources, for a specific satellite orbit at a given instant of the orbital precession period, are not observable. The upper panel of Fig.~\ref{fig:SunMoon} shows the fractional exposure integrated over one orbit as a function of position in the sky in equatorial coordinates at a given instant of the orbital precession period, where the impacts of the Sun and the Moon on the observation time have been neglected.

In calculating the sensitivity, we account for the reduction in exposure due to intrusion by the Sun and/or the Moon by multiplying the time-averaged effective area $\langle A\rangle$ by a factor, $f_t$. To a first approximation, over long periods, the Sun eliminates half of the observing time. The bright Moon further reduces the observing time, again dependent on source location by a factor of $0.63 - 0.87$. 
\begin{figure}[htb]
\centering
	\includegraphics[width=1.0\columnwidth, trim = 2.5mm 2.5mm 2mm 1mm, clip]{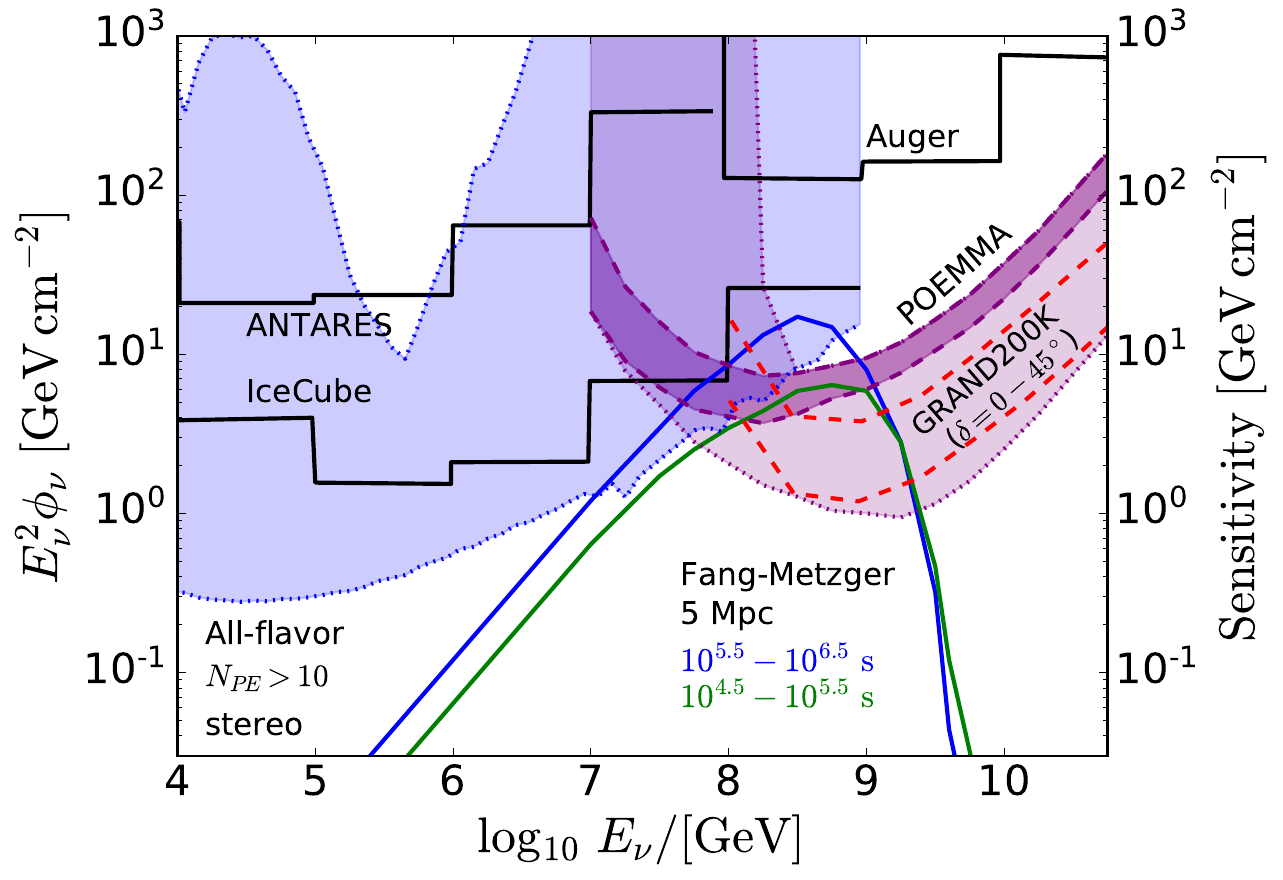}
	\caption{The POEMMA all-flavor $90$\% unified confidence level sensitivity per decade in energy for long-burst observations in ToO-stereo mode ($N_{PE}>10$) (purple bands), compared with sensitivities to \bnsevent\ from IceCube, Auger and ANTARES (scaled to three flavors) for $14$~days after its trigger time (solid black histograms)~\cite{ANTARES:2017bia}. The projected declination-averaged ($0^{\circ}-45^{\circ}$) sensitivity for \grandacro\ is denoted by the red dashed lines \cite{Alvarez-Muniz:2018bhp}. The blue shaded region shows the range of sensitivities based on IceCube's effective area as a function of energy and zenith angle. Bounds set over an e-fold energy interval~\cite{Anchordoqui:2002vb} are a factor of $2.3$ less restrictive. For comparison, the modeled all-flavor fluence from a BNS merger to a millisecond magnetar from Ref.~\cite{Fang:2017tla} is also plotted, assuming a source distance of $D=5$~Mpc. The effects of the Sun and Moon in reducing the effective area are incorporated using a factor of $f_t=0.3$. }
	\label{fig:sensitivity-sunmoon}
\end{figure}
The lower panel of Fig.~\ref{fig:SunMoon} demonstrates the combined effects of the Sun and the Moon in reducing the exposure for various points in the sky by plotting $f_t$ as the ratio of the fractional exposure accounting for the Sun and the Moon divided by the fractional exposure neglecting the Sun and Moon. The range of values is between $0.2 \alt f_t  \alt 0.4$.

For the neutrino sensitivity for long-duration events, we assume POEMMA is in the ToO-stereo configuration ($N_{\rm PE}^{\rm min}=10$), and we use the approximate relation 
\begin{equation}\label{eq:sensitivity}
{\rm Sensitivity} = \frac{2.44}{\ln(10)}
\times\frac{N_\nu E_{\nu}}{f_t \langle A(E_{\nu})\rangle_{T_0}} \ ,
\end{equation}
where $T_0=T_s$, the factor $N_\nu=3$ converts the tau-neutrino sensitivity to the all-flavor sensitivity, we have included the factor of $f_t$ that depends on sky location as discussed above, and we have taken the $90$\% unified confidence level~\cite{Feldman:1997qc} over a decade of energy $(2.44/\ln(10))$. In Fig.~\ref{fig:sensitivity-sunmoon}, we plot POEMMA's sensitivity to long bursts (purple shaded bands). For simplicity, we neglect the dependence on sky location for $f_t$ in calculating the sensitivity band plotted in Fig.~\ref{fig:sensitivity-sunmoon} and take $f_t = 0.3$ instead. The dark purple band in Fig.~\ref{fig:sensitivity-sunmoon} shows the range in POEMMA's sensitivity for most locations in the sky during a given orbit. 
\begin{table}[htb]
\caption{Minimum and maximum all-flavor sensitivities in units of [GeV/cm$^2$] for long bursts, taking the 90\% unified confidence level and location-dependent $f_t$ from $380$-day averages from Fig.~\ref{fig:SunMoon} and assuming the ToO-stereo configuration ($N_{\rm PE}^{\rm min}=10$) for POEMMA.}
\begin{center}
\begin{tabular}{|c|c|c|}
\hline
\hline
~~~~~~~~$E_\nu$\ [GeV]~~~~~~~~ & ~~~~~~~~min~~~~~~~~ & ~~~~~~~~max~~~~~~~ \\
\hline 
$10^7$ &$ 34.9$ &  $ 3.49\times 10^3$\\
\hline 
$10^8$ &$ 2.04$ &  $ 9.52$\\
\hline
$10^9$ & $ 1.99 $ & $ 11.7$\\
\hline
$10^{10}$ & $ 8.85 $ &  47.0\\
\hline
\hline
\end{tabular}
\end{center}
\label{table:longburst}
\end{table}
For example, for a given instant of the orbital precession period, over one orbit, the locations where this range in sensitivity applies is the region between the dashed curves in upper panel of Fig.~\ref{fig:SunMoon}. The extended lighter purple band shows the full range of the time-averaged sensitivity as a function of the tau neutrino energy. 

\begin{figure*}[htb]
\centering
	\includegraphics[width=1.0\columnwidth]{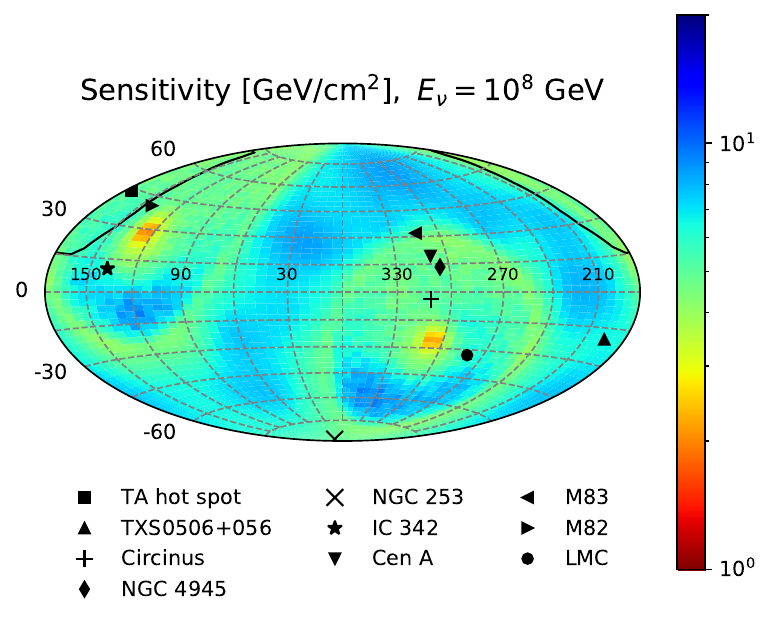}\includegraphics[width=1.0\columnwidth]{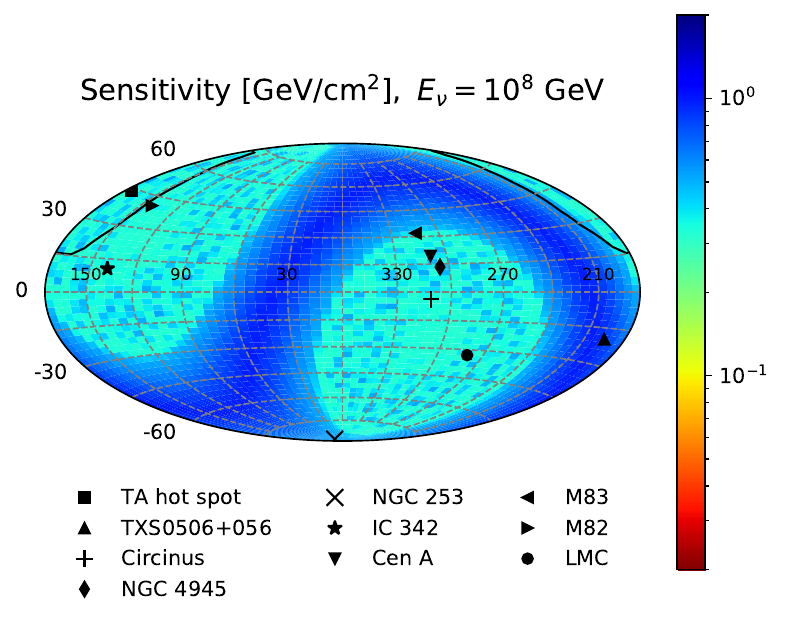}
	\includegraphics[width=1.0\columnwidth]{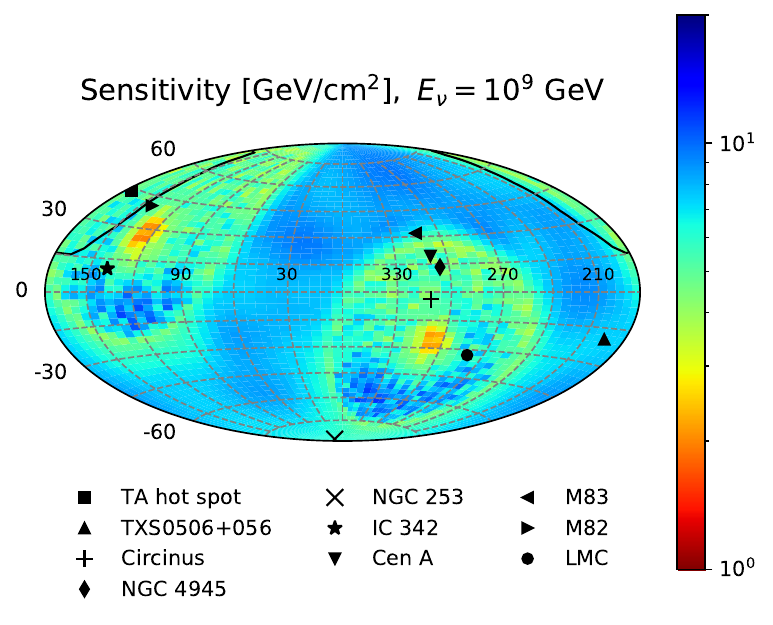}\includegraphics[width=1.0\columnwidth]{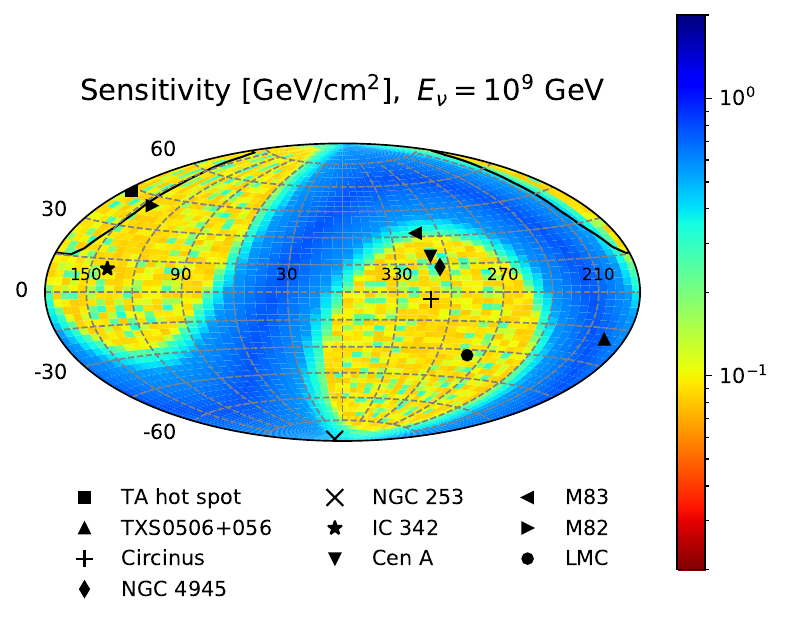}
	\caption{\textit{Left column}: Sky plots of the all-flavor 90\% unified confidence level sensitivity, for $E_{\nu}=10^8$ GeV (top) and $10^9$ GeV (bottom), for long bursts with a factor of $f_t$ that depends on sky location as plotted in Fig.~\ref{fig:SunMoon} for the time-averaged effective area, in galactic coordinates in a Hammer projection. \textit{Right column}: Sky plots of the all-flavor 90\% unified confidence level maximum sensitivity over a single POEMMA orbit during a $380$-day period for short ($10^3$~s) bursts, assuming optimal viewing conditions for the burst, for $E_{\nu}=10^8$ GeV (top) and $10^9$ GeV (bottom). Figures show the Hammer projection in galactic coordinates, with the sensitivity in units GeV/cm$^2$. Selected sources are shown, including: {\it (i)}~the Telescope Array's ``hot spot'' with a spherical cap of radius $28.43^\circ$ \cite{Abbasi:2014lda,Lundquist:2017fjo}, {\it (ii)}~nearby starburst galaxies featuring a possible correlation with UHECRs~\cite{Aab:2018chp,Abbasi:2018tqo,Anchordoqui:2018qom}, {\it (iii)}~the closest radiogalaxy Centaurus A (Cen A), {\it (iv)}~TXS~0506+056, the blazar observed by IceCube~\cite{IceCube:2018dnn,IceCube:2018cha}, and {\it (v)}~the Large Magellanic Cloud (LMC).}
	\label{fig:nusensitivityskyplot}
\end{figure*}

For comparison, we include in Fig.~\ref{fig:sensitivity-sunmoon} upper limits from IceCube, Auger and ANTARES (solid black histograms) scaled by a factor of three for the all-flavor comparison. 
These limits are based on a $14$-day window following the trigger on \bnsevent~\cite{ANTARES:2017bia}. The blue shaded region shows the range of IceCube's all-flavor sensitivity to bursts, based on their all-sky point-source effective area values tabulated as a function of energy and zenith angle for 2012 with 86 strings\footnote{Available at https://icecube.wisc.edu/science/data/PS-3years \cite[see also,][]{Aartsen:2016oji}.}. A background of zero events is assumed for IceCube, reasonable to within $20$\% even for long bursts~\cite{Meagher:2019edi}. For the purposes of rounding out the sample of experiments capable of detecting cosmic neutrinos through the widely discussed neutrino detection techniques, we also include a projected declination-averaged ($0^{\circ} < |\delta| < 45^{\circ}$) sensitivity band for \grandacro, denoted by the red dashed curves~\cite{Alvarez-Muniz:2018bhp}. A follow-on experiment to ANTARES that is currently being deployed in the Mediterranean Sea is \kmthreenet~\cite{Adrian-Martinez:2016fdl}. Based on the projected effective area for its ARCA site, we expect similar sensitivities for \kmthreenet\ as with IceCube, neglecting background; however, improvements in the angular resolution of \kmthreenet\ compared to IceCube ($0.2^{\circ}$ versus $1^{\circ}$ for track-like events; ~\cite{Adrian-Martinez:2016fdl}) will allow for improvements in the backgrounds at energies below $\sim 100$~TeV, particularly for observations lasting $\sim 10^{6}$~s or longer.

We also include in Fig.~\ref{fig:sensitivity-sunmoon} an example of a modeled all-flavor fluence from a long-duration transient event, the BNS merger model of Fang and Metzger~\cite{Fang:2017tla} scaled to a source distance of $5$~Mpc. While IceCube's best sensitivity in Fig.~\ref{fig:sensitivity-sunmoon} dips below the level of POEMMA's best sensitivity for energies below $\sim 10^8$ GeV, sensitivity depends on location in the sky as well as energy. Even considering optimal source locations, depending on the neutrino spectrum of the source, POEMMA may be able to detect bursts that IceCube will not.

In the left column of Fig.~\ref{fig:nusensitivityskyplot}, we provide sky plots of the all-flavor sensitivity for long bursts, including the location-dependent factor $f_t$ plotted in Fig.~\ref{fig:SunMoon}, as a function of sky position in galactic celestial coordinates for two fixed incident tau neutrino energies, $10^8$ GeV and $10^9$ GeV. For reference, we include several selected nearby sources and/or relevant sky regions (\textit{i.e.}, the Telescope Array ``hot spot''~[\citenum{Abbasi:2014lda,Lundquist:2017fjo}]) in the sky plots of Fig.~\ref{fig:nusensitivityskyplot}. In Table~\ref{table:longburst}, we list the minimum and maximum all-flavor sensitivities, assuming equal fluxes for the three neutrino flavors, for $E_{\nu}=10^7,\ 10^8,\ 10^9$, and $10^{10}$~GeV.

\begin{figure}[htb]
\centering
	\includegraphics[width=1.0\columnwidth, trim = 2mm 2mm 2mm 1mm, clip]{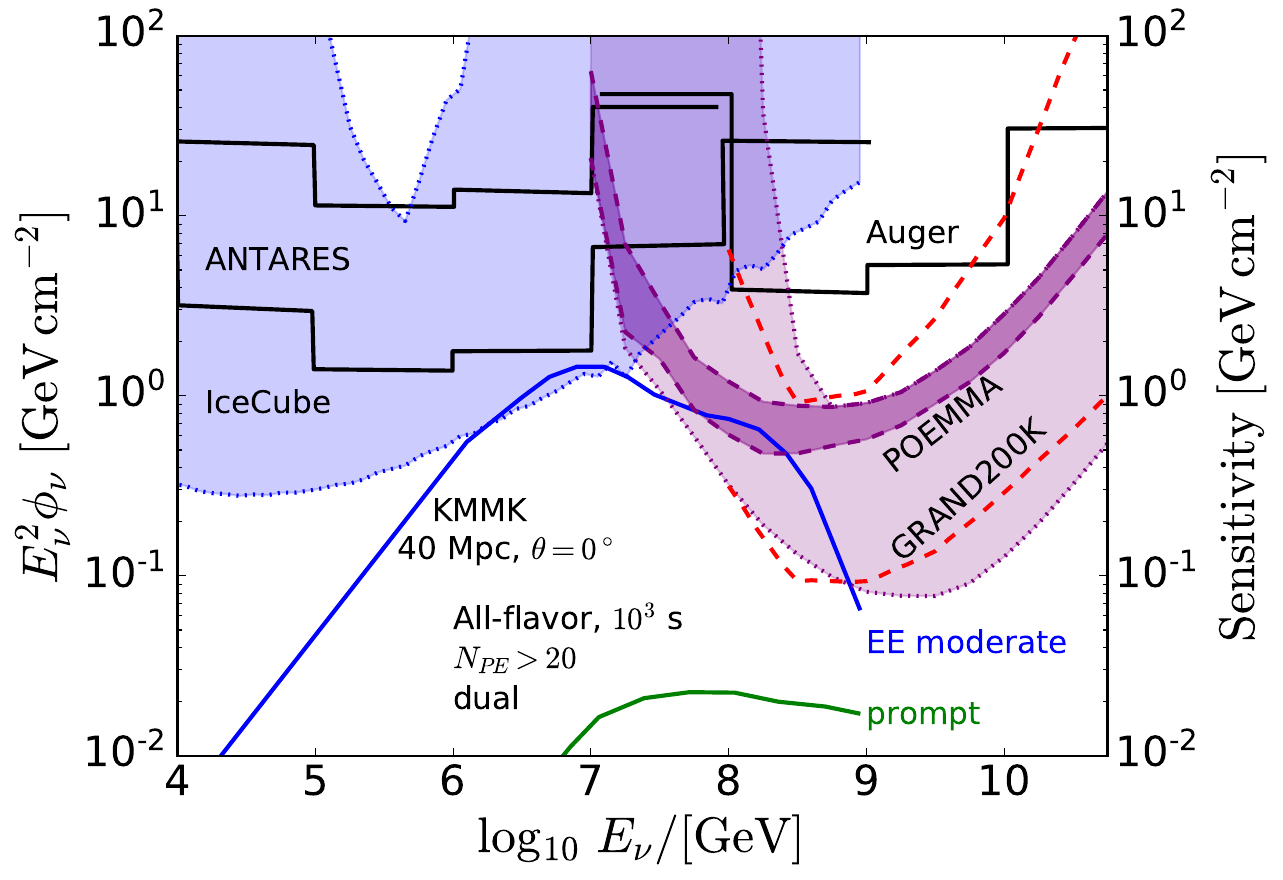}
	\caption{The POEMMA all-flavor $90$\% unified confidence level sensitivity per decade in energy for short-burst observations in ToO-dual mode ($N_{PE}>20$). The purple band shows the range of sensitivities accessible to POEMMA for a $10^3~{\rm s}$ burst in the ``best-case'' scenario (see text). The dark purple band corresponds to source locations in a large portion of the sky. The IceCube, Auger and ANTARES sensitivities to \bnsevent, scaled to three flavors, for $\pm 500$ s around the binary neutron star merger are shown with solid histograms~\cite{ANTARES:2017bia}. The red dashed curves indicate the projected instantaneous sensitivities of \grandacro\ at zenith angles $\theta=90^\circ$ and $94^\circ$~\cite{Alvarez-Muniz:2018bhp,martineau-huynh}. The blue shaded region shows the range of sensitivities that depend on location from  IceCube's effective area. Also plotted are examples of the all-flavor fluence for a short neutrino burst during two phases (extended and prompt) for an sGRB, as predicted by KMMK~\cite{Kimura:2017kan} for on-axis viewing ($\Theta = 0^{\circ}$) and scaled to $40$~Mpc.}
	\label{fig:sensitivity-burst}
\end{figure}

For the neutrino sensitivity for short bursts, several aspects of the calculations differ from those for the long bursts. The timing and location of the burst determines the extent to which POEMMA will be able to make observations. As such, we limit our considerations for short bursts to a ``best-case'' scenario in which POEMMA has started observations just as the source moves below the limb of the Earth, and the Sun and the Moon do not impede observations. In such a scenario, the sensitivity to short bursts, being in the optimal location for a given time, will be better than the sensitivity for long bursts. This optimal sensitivity is calculated by finding the time averaged effective area, now with $T_0=10^3$~s. For short-burst timescales ($T_{\rm burst} \sim 10^3$~s), we assume that the POEMMA satellites will be in the ToO-dual configuration ($N_{\rm PE}^{\rm min}=20$). We vary the satellite positions relative to sources and the Earth over a period of $380$~days in order to obtain a range of optimal POEMMA sensitivities.

In Fig.~\ref{fig:sensitivity-burst}, we plot the range of  POEMMA all-flavor sensitivities in the described ``best-case'' scenario for short bursts. For comparison, we include histograms for the IceCube, Auger and ANTARES sensitivities (scaled to three flavors) based on a $\pm 500$~s time window around the binary neutron star merger \bnsevent~\cite{ANTARES:2017bia}. 
\begin{table}[htb]
\caption{Minimum and maximum ``best-case'' all-flavor sensitivities in units of [GeV/cm$^2$] for bursts of $10^3~{\rm s}$, taking the 90\% unified confidence level and assuming observations during astronomical night ($f_t=1$) and the ToO-dual configuration ($N_{\rm PE}^{\rm min}=20$) for POEMMA.}
\begin{center}
\begin{tabular}{|c|c|c|}
\hline
\hline
~~~~~~~~$E_\nu$ \ [GeV]~~~~~~~~ & ~~~~~~~~min~~~~~~~ & ~~~~~~~~max~~~~~~~~ \\
\hline 
$10^7$ &$20.9$ &  $1.59\times 10^{6}$\\
\hline 
$10^8$ &$3.20\times 10^{-1}$ &  $9.90\times 10^{-1}$\\
\hline
$10^9$ & $8.15\times 10^{-2}$ & $7.64\times 10^{-1}$\\
\hline
$10^{10}$ & $1.28\times 10^{-1}$ & 2.41 \\
\hline
\hline
\end{tabular}
\end{center}
\label{table:shortburst}
\end{table}
We also include the projected instantaneous sensitivities of \grandacro\ for zenith angles $\theta=90^{\circ}$ and $94^{\circ}$~\cite{Alvarez-Muniz:2018bhp,martineau-huynh} to indicate the possible range in their sensitivity to short bursts. For reference, we also plot examples of the modeled all-flavor fluence for a short neutrino burst during two phases (extended and prompt) for a short gamma-ray burst (sGRB), as predicted by KMMK~\cite{Kimura:2017kan} for on-axis viewing ($\Theta = 0^{\circ}$). The modeled fluences in Fig.~\ref{fig:sensitivity-burst} are scaled to $40$~Mpc. 
In the right column of Fig.~\ref{fig:nusensitivityskyplot}, we provide sky plots of the ``best-case'' all-flavor sensitivity as a function of sky position in galactic celestial coordinates for $E_{\nu}=10^8$~GeV and $10^9$~GeV. In Table~\ref{table:shortburst}, we list the ``best case'' minimum and maximum sensitivities based on sky location. 

Figs.~\ref{fig:sensitivity-sunmoon} and \ref{fig:sensitivity-burst} show that the time-averaged sensitivity for long bursts and the ``best-case'' sensitivity for short bursts improve upon the Auger limits by more than an order of magnitude for most locations in the sky and by up to two orders of magnitude in the most favorable locations. A key feature of these satellite-based instruments is that they can track the source of tau neutrinos for a wider range of Earth-emergence angles ($\beta_{\rm tr}<35^{\circ}$) than capable with a ground-based observatory, such as Auger, that mostly detects neutrinos via Earth-skimming events ($\beta_{\rm tr}<6^\circ$)~\cite{Abreu:2012zz}.

\begin{figure}[htb]
\centering
	\includegraphics[width=0.95\columnwidth]{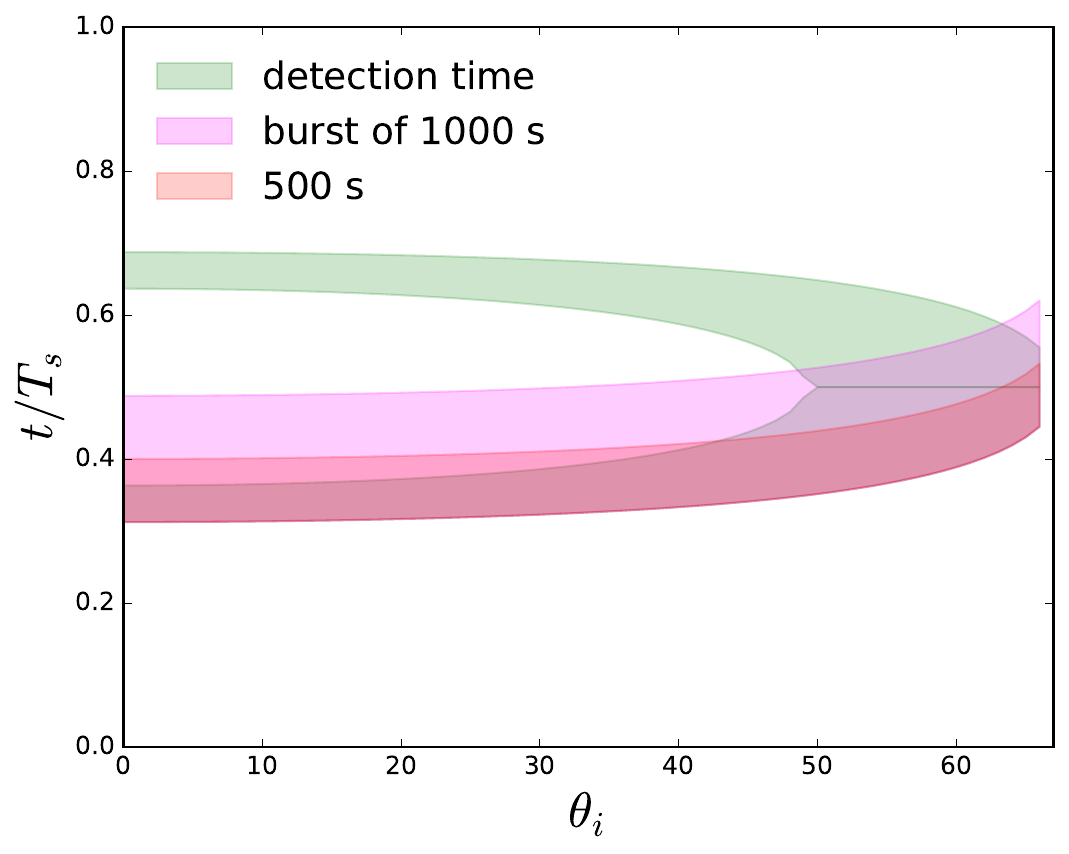}	
	\caption{The green band shows the fraction of the time during which the source is observable during astronomical night relative to the orbital period for a given $\theta_i$ (see text). The pink band shows the burst time of $10^3~{\rm s}$ relative to the orbital period of $T_s=5,700$~s. The red band shows the relative time of $500$~s to $T_s$.}	
	\label{fig:times}
\end{figure}

Even if POEMMA is not pointing at the burst, with an alert, POEMMA can slew $90^{\circ}$ in $500$~s. For most locations, a $500$~s delay will not change the sensitivity to $10^3$~s bursts if the source alignment with the Earth is optimal, since the burst duration is longer than the amount of time the source is visible to POEMMA. This last feature, and the result that POEMMA is potentially more sensitive to well-positioned neutrino sources with short bursts than to long bursts is demonstrated in Fig.~\ref{fig:times}. For this example, we consider sources with an RA of $0^{\circ}$ and for which a line from the Earth to the source is at an angle of $\theta_i$ relative to POEMMA's orbital plane. 
All other source locations can be mapped to this configuration if we are free to choose $t_0$ in Eq.~(\ref{eq:avga}). The green shaded band in Fig.~\ref{fig:times} shows the fraction of an orbit when a source is behind the Earth with neutrino trajectory elevation angles in the range $\beta_{\rm tr}=1^{\circ}-35^{\circ}$. The source first sets below the horizon, then rises above the limb of the Earth as viewed from the POEMMA satellites. Considering the example of a source within POEMMA's orbital plane ($\theta_i=0^{\circ}$), the green shaded band indicates two time intervals for which Earth-emerging neutrinos will have elevation angles in the range $\beta_{\rm tr}=1^{\circ}-35^{\circ}$. The region between the green bands represents the time when the neutrino fluence is strongly attenuated by the Earth. Before the first green interval and after the second interval, the source is not behind the Earth. For $\theta_i\simeq 50^{\circ}$, the source dips below the horizon and $\beta_{\rm tr} \leq 35^{\circ}$ for one extended interval. Given the inclination of POEMMA's orbital plane of $28.5^{\circ}$, when $\theta_i>68.5^{\circ}$, the source is never below the Earth's horizon for POEMMA. In Figs.~\ref{fig:sensitivity-sunmoon} and \ref{fig:sensitivity-burst}, the dashed lines bracket the sensitivities (including the effect of the Sun and Moon for long bursts) for $\theta_i\leq 50^{\circ}$ (the dark purple region), and the dotted lines extend to $50^{\circ}<\theta_i<68.5^{\circ}$ with the light purple region.

For long bursts, $\langle A(E_{\nu})\rangle$ is determined with $T_s$, the full range of the $y$-axis in Fig.~\ref{fig:times}. For short bursts, the fraction of the $y$-axis equivalent to $10^3~{\rm  s}$ is shown with the pink band. The time average of the effective area is the probability-weighted green band with normalization of $10^3~{\rm s}$. If the burst begins at $t=0$ for $\theta_i=0^\circ$, a $10^3~{\rm s}$ burst will not be observed at all. On the other hand, if the burst begins within $\sim 500-700$~s of the viewing window (either green band) the sensitivity is the optimal value. This is true for most of the angles $\theta_i$. The dark pink band shows a window of $500$~s. If the source is optimally placed, a $500$~s delay from slewing the instrument to the position of the source will not change the sensitivity.

\section{Neutrino Estimates from Flaring Astrophysical Sources and Neutrino Horizons}
\label{sec:3}

In this section, we use the time-averaged effective area calculated in Section~\ref{sec:2} to estimate the numbers of neutrino events that would be detectable by POEMMA for several models of astrophysical transients. As the nearby matter distribution is fairly anisotropic, Sec.~\ref{sec:3a} discusses our methodology for determining the galaxy-luminosity weighted effective area that we use to calculate the number of neutrino events expected for a given source model as discussed in Sec.~\ref{sec:3b}. In Sec.~\ref{sec:3b}, we also determine POEMMA's sky coverage in terms of detecting neutrinos according to the two astrophysical models pictured in Figs.~\ref{fig:sensitivity-sunmoon}~and~\ref{fig:sensitivity-burst} and featuring two scenarios for neutrino ToO observations. To provide some context for bench-marking POEMMA's capability for ToO observations relative to currently operating and other proposed future neutrino observatories, we perform similar sky coverage calculations for IceCube and \grandacro\ and compare with our findings for POEMMA. In Sec.~\ref{sec:3c}, we define the neutrino horizon, the maximum distance at which POEMMA will be able to detect a neutrino for a given source class, used to calculate the cosmological event rate for determining the occurrence of transient events, modeled as a Poisson process. In Sec.~\ref{sec:3d}, we provide descriptions for the most promising modeled source classes as determined by the Poisson probability of detecting at least one ToO during the proposed mission lifetime for POEMMA of $3$--$5$ years. We discuss additional transient neutrino source models in Appendix~\ref{app:c}. We summarize our findings for a selection of models for candidate astrophysical neutrino sources in Table~\ref{table:events-bytype}.

\subsection{Effective Area Averaged Over the Sky}\label{sec:3a}

\begin{figure*}[htb]
\centering
    \includegraphics[trim = 28mm 20mm 25mm 20mm, clip, width=0.63\columnwidth]{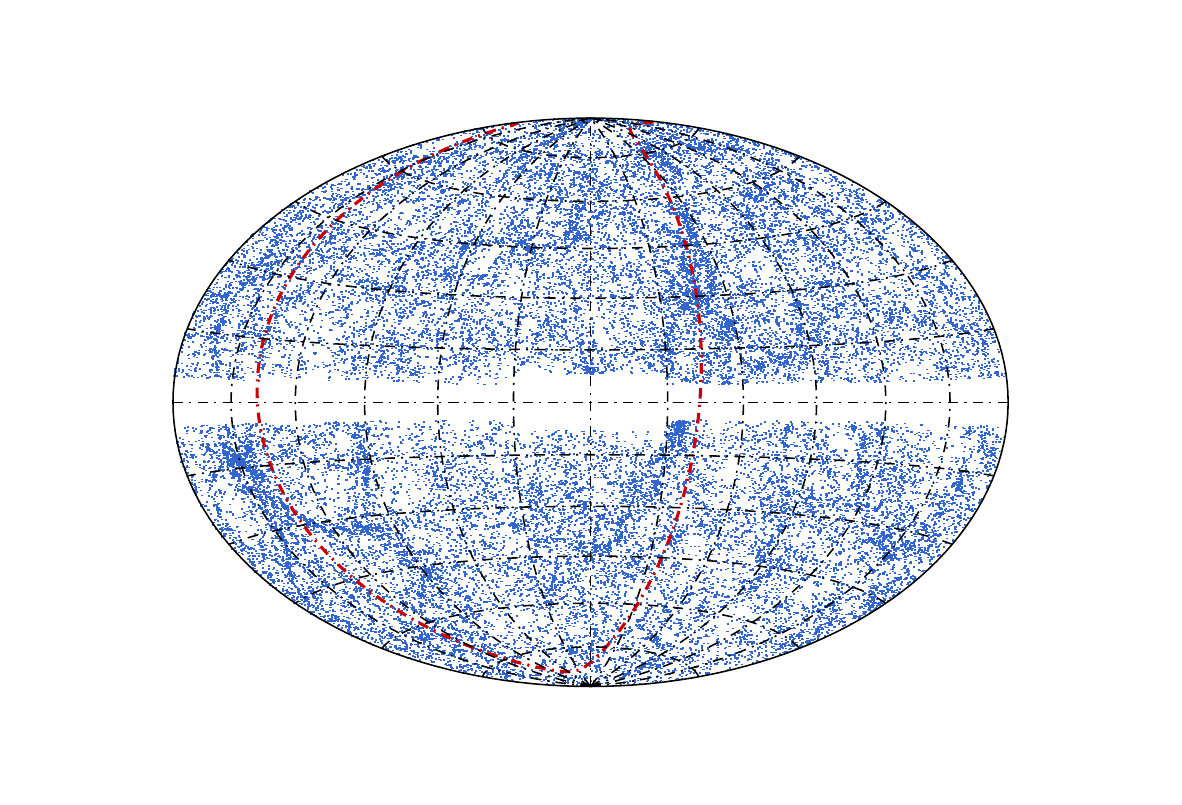}\includegraphics[trim = 30mm 43mm 18mm 32mm, clip, width=0.74\columnwidth]{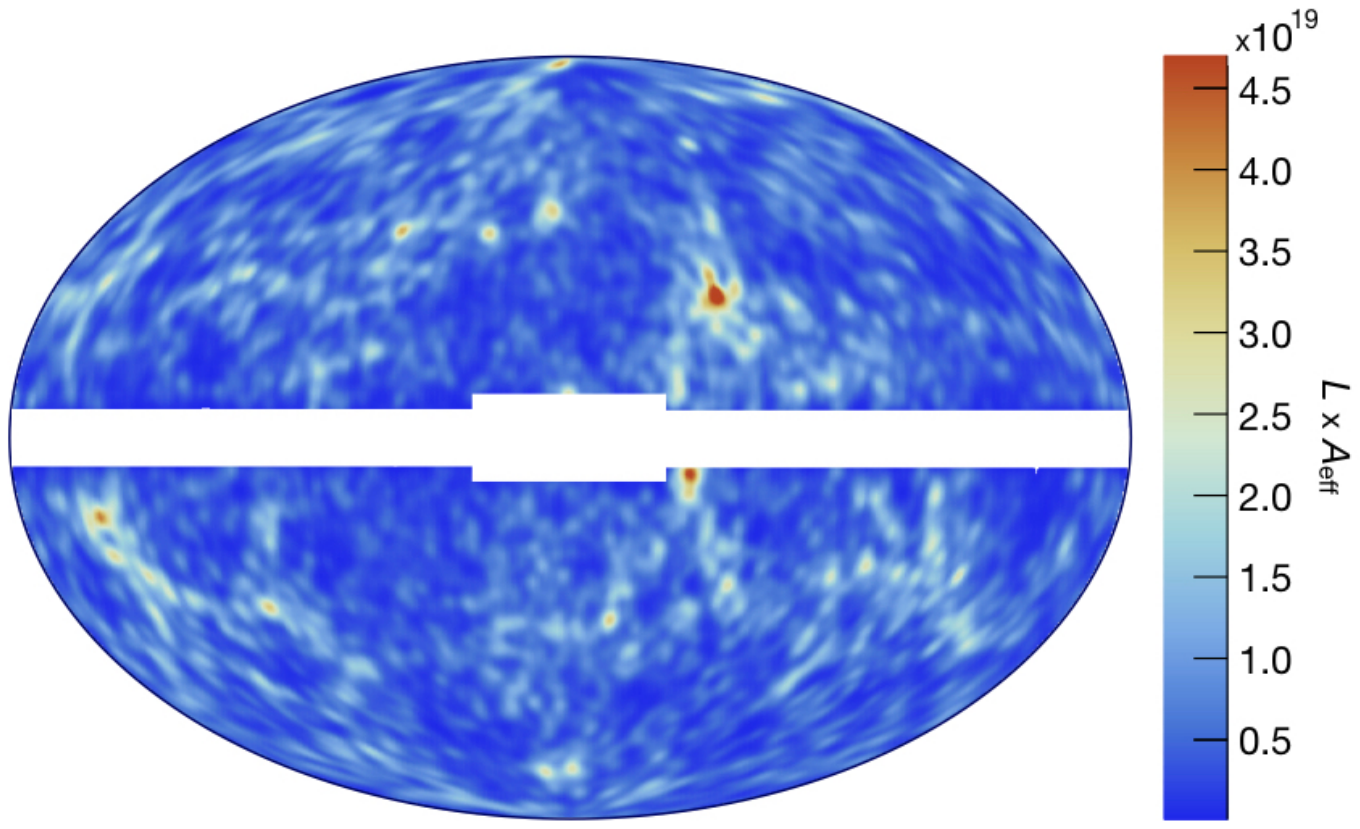}\includegraphics[trim = 30mm 43mm 20mm 32mm, clip, width=0.73\columnwidth]{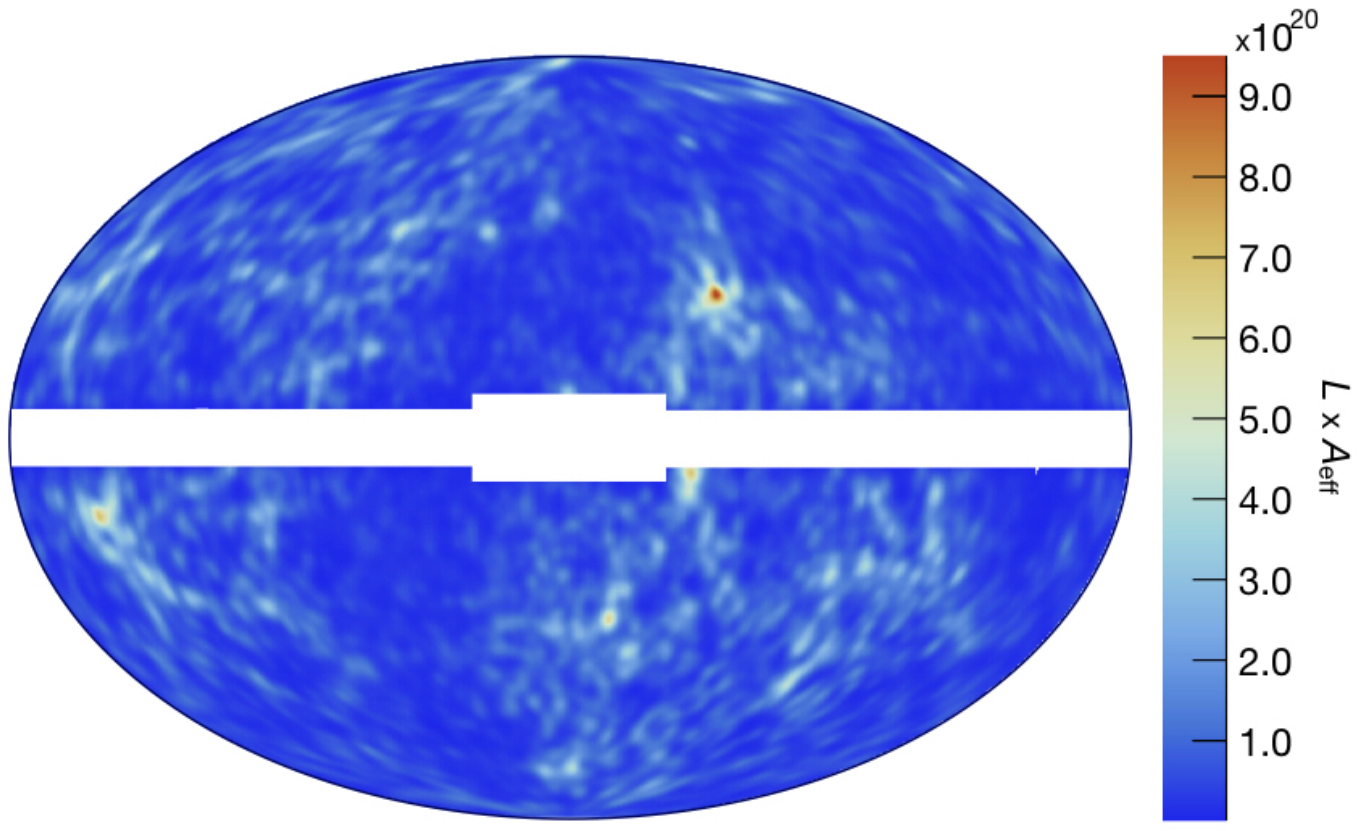}
    \caption{\textit{Left}: Sky plot of galaxies in the 2MRS catalog~\cite{2012ApJS..199...26H} in galactic coordinates. Overdensities seen in the plot are due to nearby clusters of galaxies. For reference, the supergalactic plane is plotted as the red dot-dashed line. \textit{Middle}: Sky plot of the smoothed 2MRS catalog galaxy luminosity weighted effective area in units of $L_0\cdot$cm$^2$ for $E_{\nu_{\tau}} = 10^{9}$ GeV for long bursts. \textit{Right}: As at left for short bursts.}
    \label{fig:2mrsskyplots}
\end{figure*}

As evidenced in Fig.~\ref{fig:nusensitivityskyplot}, the effective area of POEMMA varies considerably over the sky due to the orbital characteristics of the satellites and the influence of the Sun and the Moon (see Sec.~\ref{sec:2}). 
To calculate the expected numbers of neutrinos from models of astrophysical neutrino sources, we compute the average effective area over the sky as a function of redshift:
\begin{equation}
{\cal A} \left(E_\nu,z\right) = \frac{\int \left<A\left(E_\nu,\theta,\phi\right)\right>_{T_{0}}p\left(\theta,\phi,z\right)\, d\Omega}{\int p\left(\theta,\phi,z\right) \, d\Omega}\,,
    \label{eq:Aeffofz}
\end{equation}
where $p\left(\theta,\phi,z\right)$ is the weighting function expressing the probability of finding a source at a given redshift, $z$, and sky location, $\left(\theta,\phi\right)$, where $\theta = \frac{\pi}{2} - b$ and $\phi = l$ are expressed in galactic longitude and latitude, $\left(l,b\right)$ and $d\Omega = \sin\theta\, d\theta \, d\phi$. 

The weighting function is determined by the distribution of matter in the universe, which while being statistically isotropic out to high redshifts, is relatively anisotropic out to the distances within which POEMMA is most likely to detect neutrinos. As such, we model the weighting function using the 2MASS Redshift Survey (2MRS) of galaxies in the nearby universe (see Fig. \ref{fig:2mrsskyplots}) \cite{2012ApJS..199...26H}. The 2MRS catalog includes a sample of nearly $45,000$ galaxies selected from the original 2 Micron All-Sky Survey (2MASS) \cite{2006AJ....131.1163S}. The resulting 2MRS redshift catalog consists of galaxies with apparent magnitudes $K_s \leq 11.75$ mag in the near infrared and galactic latitudes $|b| \geqslant 5^{\circ}$ ($|b| \geqslant 8^{\circ}$ near the Galactic bulge). Galaxy redshifts are provided as measured radial velocities in the solar system barycenter reference frame. In order to compute cosmological redshifts for each galaxy, radial velocities are corrected to the cosmic microwave background (CMB) reference frame through
\begin{multline}
    V_{\rm corr} = V_{\rm uncorr} + V_{\rm apex}\sin \left(b\right)\sin\left(b_{\rm apex}\right)\\
     + V_{\rm apex}\cos\left(b\right)\cos\left(b_{\rm apex}\right)\cos\left(l-l_{\rm apex}\right),
\end{multline}
where $l_{\rm apex} = 264.14^{\circ}$, $b_{\rm apex} = +48.26^{\circ}$, and $V_{\rm apex} = 371.0$ km~s$^{-1}$, which accounts for the motion of the Galaxy with respect to the CMB \cite{Fixsen:1996nj}. For those 2MRS galaxies with positive corrected radial velocities, redshifts are then determined using
\begin{equation}
    V_{\rm rad} = V_{\rm corr} = c \int^{z}_{0} \frac{dz'}{E\left(z'\right)},
\end{equation}
where $E\left(z'\right) = \sqrt{\Omega_M\left(1+z'\right)^3 + \Omega_k\left(1+z'\right)^2 + \Omega_{\Lambda}}$ with $\left(\Omega_M,\Omega_k,\Omega_{\Lambda}\right)$ being cosmological parameters related to the matter density of the universe, the curvature of the universe, and the dark energy density, respectively (\textit{c.f.}, Refs. \cite{Davis:2014jwa,Hogg:1999ad,1993ppc..book.....P}).\footnote{For this paper, we take $\Omega_M = 0.3153$, $\Omega_{\Lambda} = 0.6847$, $\Omega_k = 1 - \left(\Omega_M + \Omega_{\Lambda}\right) = 0$, and $H_0 = 67.36$ km s$^{-1}$ Mpc$^{-1}$ \cite{Aghanim:2018eyx}. We have verified that if we adopt the value of $H_0$ derived from from the maser-cepheid-supernovae distance ladder~\cite{Riess:2011yx}  our results are not significantly altered.} For those 2MRS galaxies with negative corrected radial velocities (only $25$ galaxies out of the full sample), rather than using redshifts, we instead determine their distances by following a procedure similar to that discussed in Ref.~\cite{2017MNRAS.470.2982L}. Most of the 2MRS galaxies have been associated with known nearby galaxies, and distances are provided in the Extragalactic Distance Database (EDD)~\cite{2009arXiv0902.3668T}. For the four 2MRS galaxies that remain unassociated, we used the distances of their nearest neighbors from the list of $25$ 2MRS galaxies with negative corrected radial velocities. 

With redshifts or distances associated with every galaxy in the 2MRS catalog, we construct maps of the weighting function in bins of redshift. In so doing, we consider two options for assigning weights to the galaxies in the catalog: (1) assigning the same weight to every galaxy; (2) weighting each galaxy according to its luminosity. Galaxy luminosities, $L$, are computed from their absolute magnitudes, $M$ by
\begin{equation}
    \frac{L}{L_{0}} = 10^{-0.4M},
\end{equation}
where $L_{0}$ is the zero-point luminosity in the $K_s$ bandpass (taken to be the luminosity of Vega in the $K_s$ band). The absolute magnitude is computed from $K_s$ apparent magnitudes using
\begin{equation}
M = m + \Delta m - A_{K}\left(l,b\right) - k\left(z\right) - e\left(z\right) - DM\left(z\right),
\end{equation}
where $m$ is the apparent magnitude in the $K_s$ bandpass, $\Delta m = 0.017$ is the zero-point offset required to calibrate the 2MASS with the standard Vega system \cite{Cohen:2003ga}, $A_{K}\left(l,b\right)$ is the correction for extinction due to dust in the Milky Way (already included in 2MRS apparent magnitudes), $k\left(z\right)$ is the k-correction due to cosmological redshifting of the spectrum, $e\left(z\right)$ corrects for evolution in galaxy spectra arising from stellar populations aging over the redshift distribution of the survey \cite{Bernardi:2003rg}, 
\begin{equation}
DM\left(z\right) = 5\log_{10}\left(\frac{d_{L}}{10\ {\rm pc}}\right)
\end{equation}
is the distance modulus, and 
\begin{equation}
d_L = \frac{c}{H_0}(1+z) \int^{z}_{0}\frac{dz'}{E\left(z'\right)}
\label{dL}
\end{equation}
is the luminosity distance. For the k- and evolution-corrections, we adopt the values given in Ref. \cite{Bell:2003cj}:
\begin{align}
    k\left(z\right) = -2.1z\\
    e\left(z\right) = 0.8z\,.
\end{align}
Many studies of redshift surveys such as the 2MRS make use of isophotal apparent magnitudes\footnote{\textit{I.e.}, from fluxes integrated within the isophotal radius, the distance from the center along the semi-major axis beyond which the surface brightness falls below a given value.}, which would require an aperture correction that would convert these observed aperture magnitudes to some proper diameter (\textit{c.f.}, Ref. \cite{2017MNRAS.470.2982L}). For our study, we use the extrapolated total apparent magnitudes provided in the 2MRS catalog; hence, the aperture correction is not needed~\cite{2017MNRAS.470.2982L,1999coph.book.....P}.

In addition to enabling the calculation of galaxy luminosities, the calculated absolute magnitudes also enabled the construction of volume-limited samples in every redshift bin. In each bin, we calculated the limiting absolute magnitude for which a galaxy at the highest redshift in the bin would have an observed apparent magnitude at the survey limit (\textit{i.e.}, $K_s = 11.75$ mag). We then included only those galaxies with calculated absolute magnitudes that were less than the limiting absolute magnitude for that bin. This corrects for the possible bias in favor of fainter galaxies that could only be detected at the lower redshifts in the bin.

Finally, the weighting function maps are created by smoothing our constructed 2MRS samples with a Gaussian with $\sigma = {\theta_{\rm Ch}^{\rm app}}/{\sqrt{2\ln{2}}}$, where $\theta_{\rm Ch}^{\rm app}\sim 1.5^{\circ}$ is an approximation of the effective Cherenkov angle. The effective area averaged over the constructed weighting functions is then calculated for each redshift bin according to Eq.~(\ref{eq:Aeffofz}). Sample maps for the entire 2MRS catalog are provided in Fig. \ref{fig:2mrsskyplots}.

\subsection{Expected Numbers of Neutrino Events from Modeled Astrophysical Neutrino Fluences}\label{sec:3b}

With the average effective area computed as a function of energy and redshift, the expected number of neutrino events from an astrophysical source at redshift $z$ is given by
\begin{equation}
    N_{\rm ev} = \int_{\Delta E_\nu}  \phi_{\nu_\tau}(E_\nu) \  {\cal A}\left(E_\nu,z\right) \ dE_\nu\,,
\label{eq:numevents}
\end{equation}
where $\phi_{\nu_\tau} (E_\nu)$ is the single-flavor ($N_\nu = 1$) neutrino fluence in units of energy per unit area. The observed energy-squared scaled tau-neutrino fluence is given by
\begin{equation}
    E^2_\nu \ \phi_{\nu_\tau} (E_\nu) = \frac{\left(1+z\right)}{4\pi d^2_L} \ \frac{Q}{3} \ E_{\rm src}^2  \ \Delta t_{\rm src} \,,
\label{eq:phiobs}
\end{equation}
where $Q$ is the all-flavor neutrino source emission rate as measured by a fundamental observer at the source redshift in units of neutrinos per energy interval per time interval, $\Delta t_{\rm src}$ is the event duration at the source redshift, $E_{\rm src}$ is the emission energy, and we assume that the relevant quantities for calculating the fluences are \textit{isotropic equivalent} quantities and that neutrino oscillations will yield equal flavor ratios on Earth (for derivation of Eq.~(\ref{eq:phiobs}), see Appendix~\ref{app:b}). For any astrophysical model that provides an observed fluence for a source at a given redshift or luminosity distance, the observed fluence can be computed for any redshift using Eq.~(\ref{eq:phiobs}) by calculating the intrinsic neutrino source emission rate and then rescaling to the new redshift.  The expected number of neutrino events predicted by the astrophysical model is then given by Eq. (\ref{eq:numevents}). 

\begin{figure*}[htb]
\centering
    \includegraphics[trim = 27mm 43mm 19mm 40mm, clip, width=0.69\columnwidth]{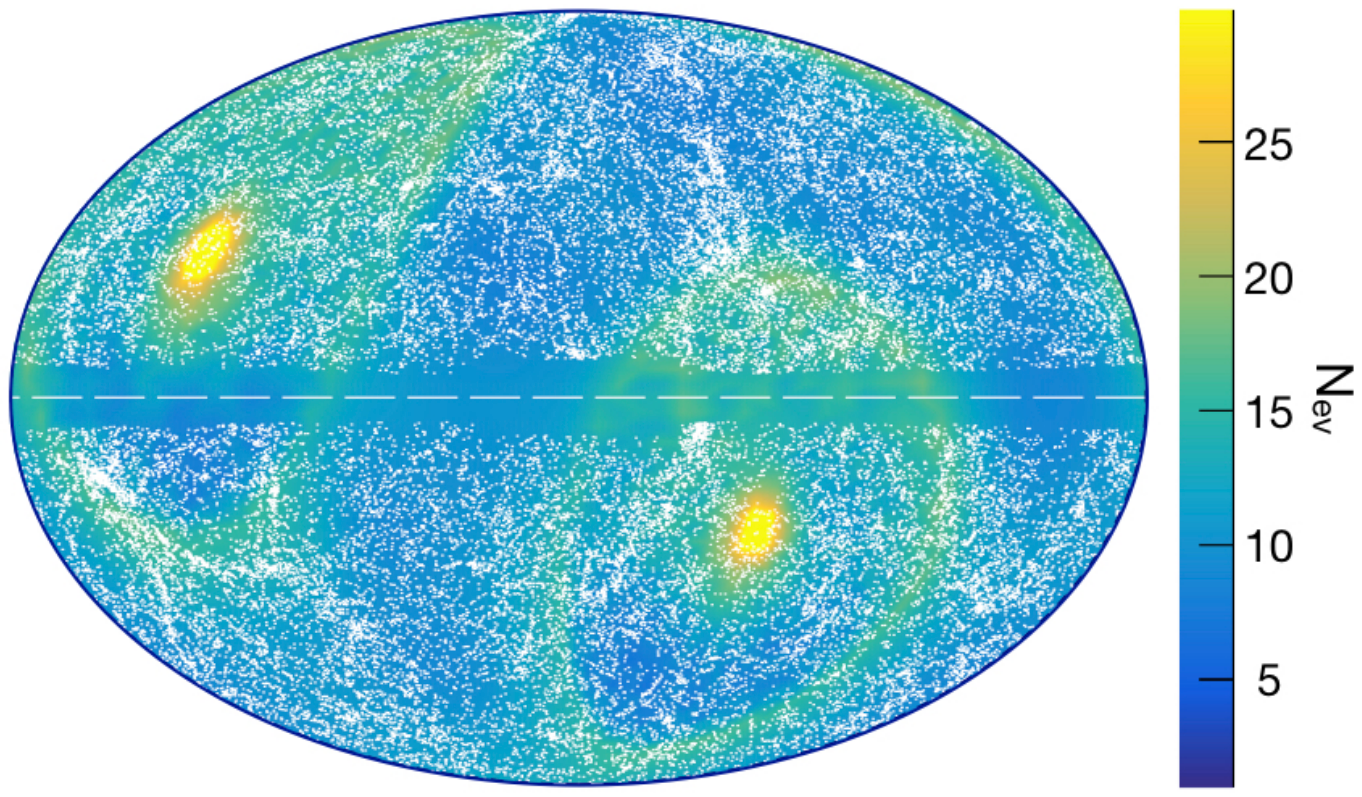}\includegraphics[trim = 27mm 43mm 19mm 40mm, clip, width=0.69\columnwidth]{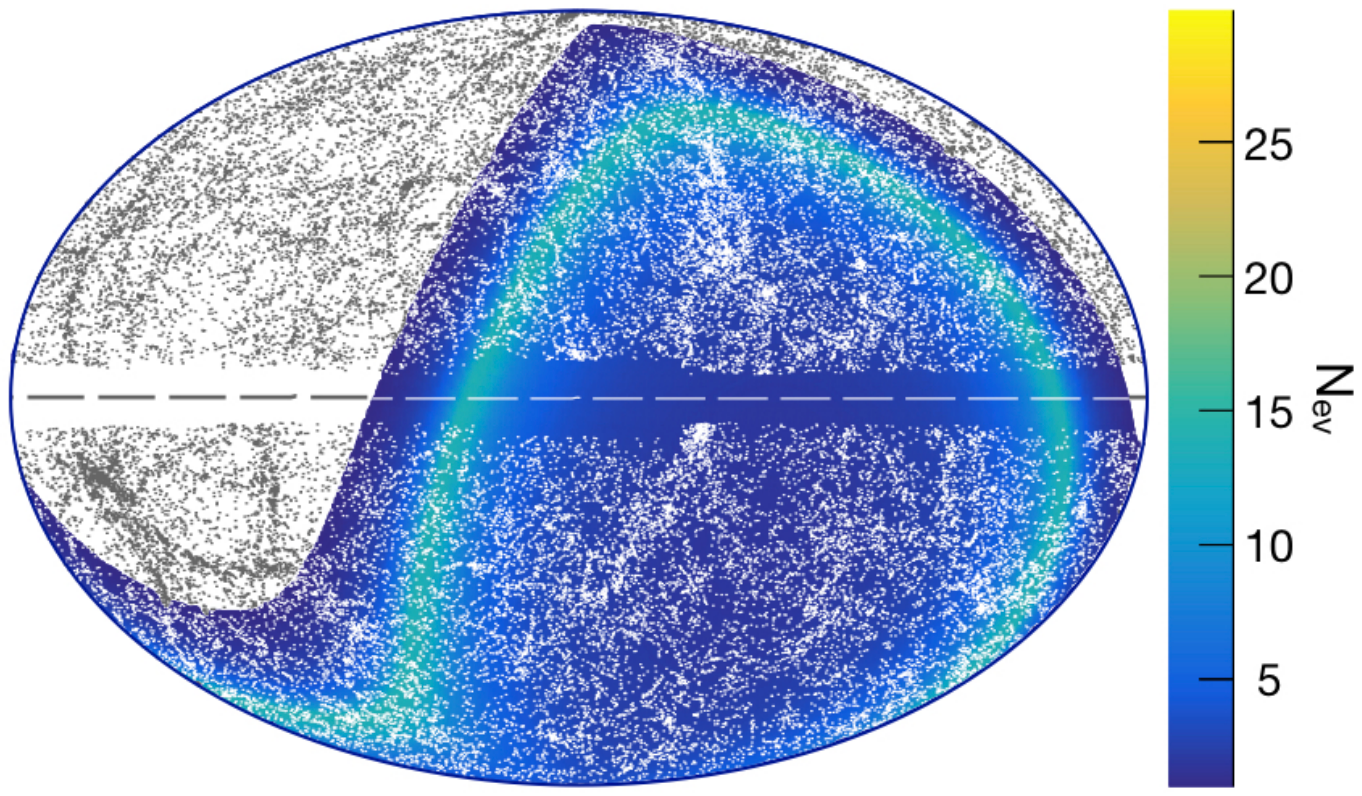}\includegraphics[trim = 27mm 43mm 19mm 40mm, clip, width=0.69\columnwidth]{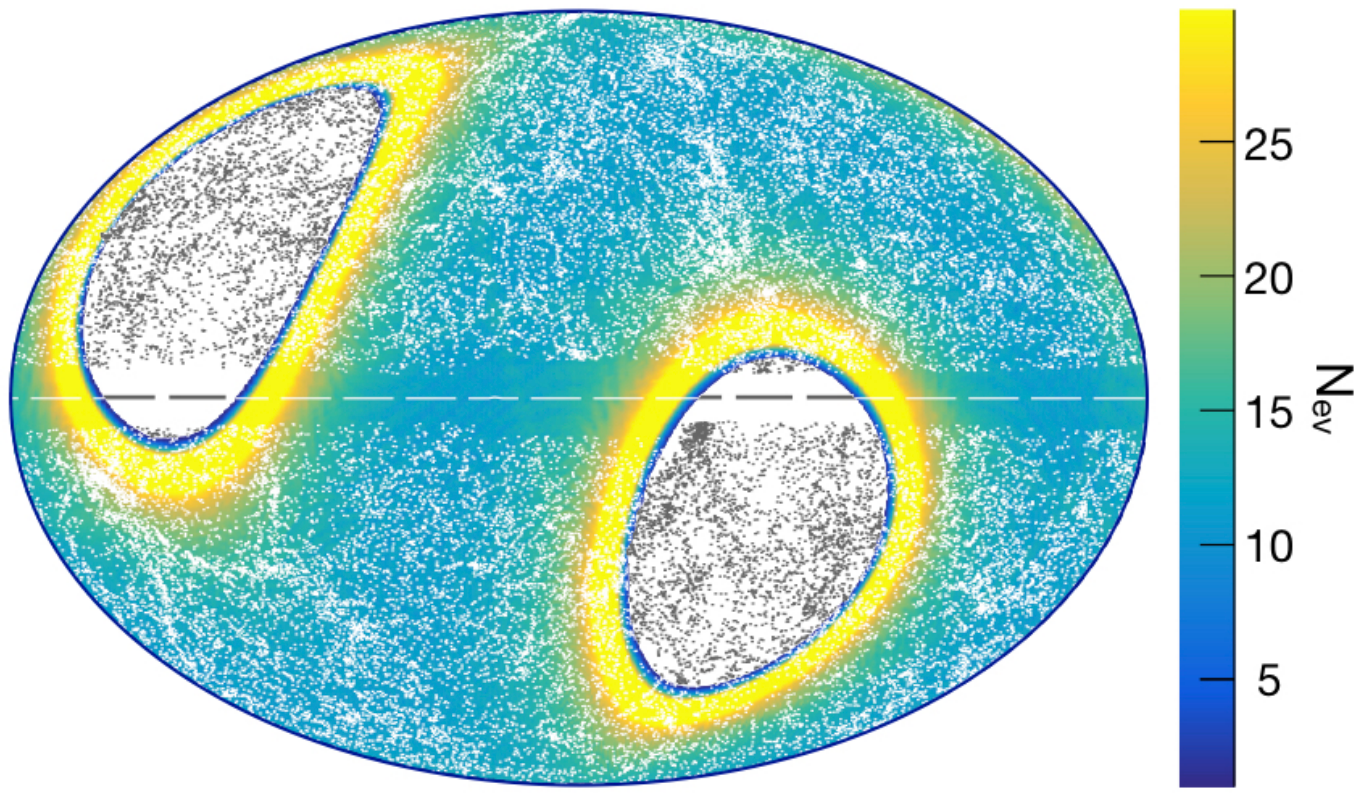}
    \caption{\textit{Left}: Sky plot of the expected number of neutrino events as a function of galactic coordinates for POEMMA in the long-burst scenario of BNS merger, as in the Fang \& Metzger model~\cite{Fang:2017tla}, and placing the source at $5$~Mpc. Point sources are galaxies from the 2MRS catalog~\cite{2012ApJS..199...26H}. \textit{Middle}: Same as at left for IceCube for muon neutrinos. \textit{Right}: Same as at left for \grandacro. Areas with grey point sources are regions for which the experiment is expected to detect less than one neutrino.}
    \label{fig:nevskyplotslong}
\end{figure*}
\begin{figure*}[htb]
\centering
    \includegraphics[trim = 27mm 43mm 19mm 40mm, clip, width=0.69\columnwidth]{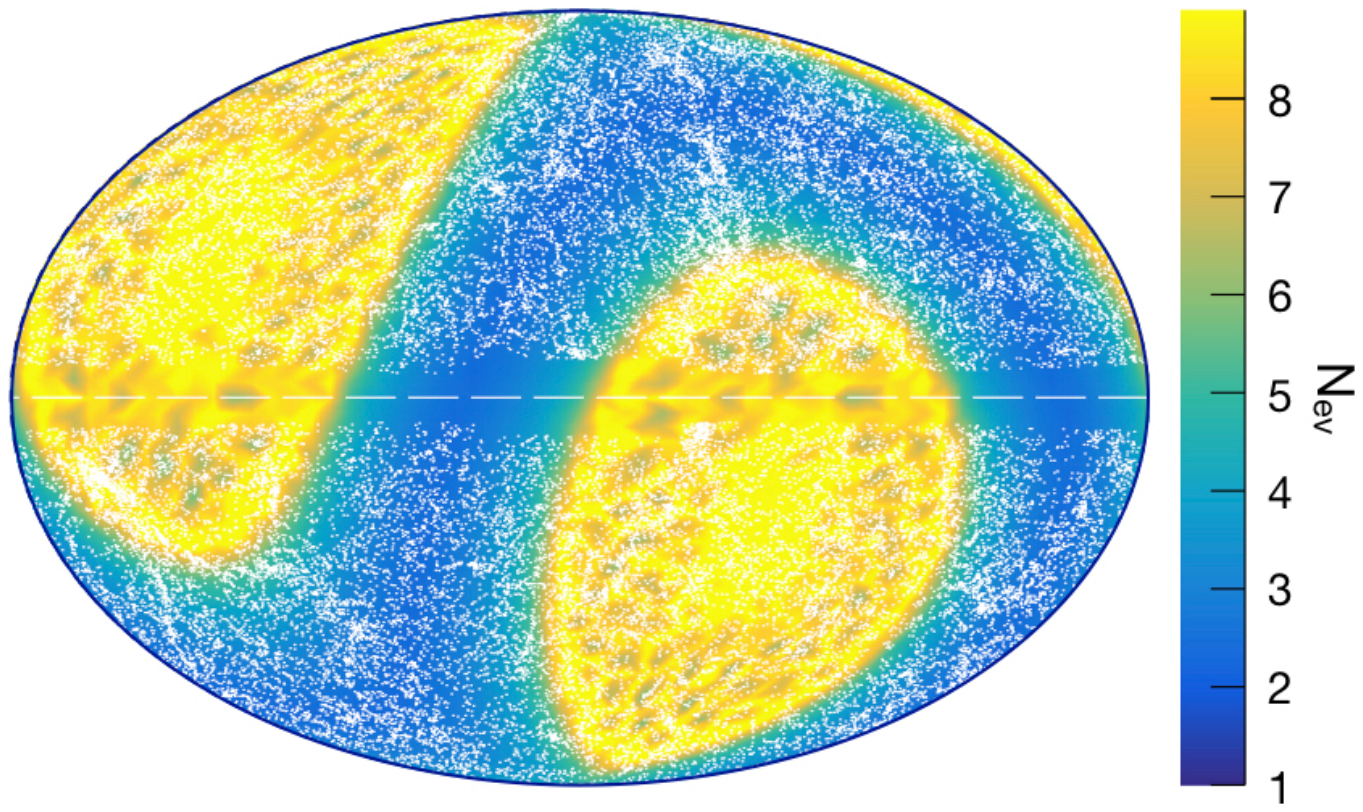}\includegraphics[trim = 27mm 43mm 19mm 40mm, clip, width=0.69\columnwidth]{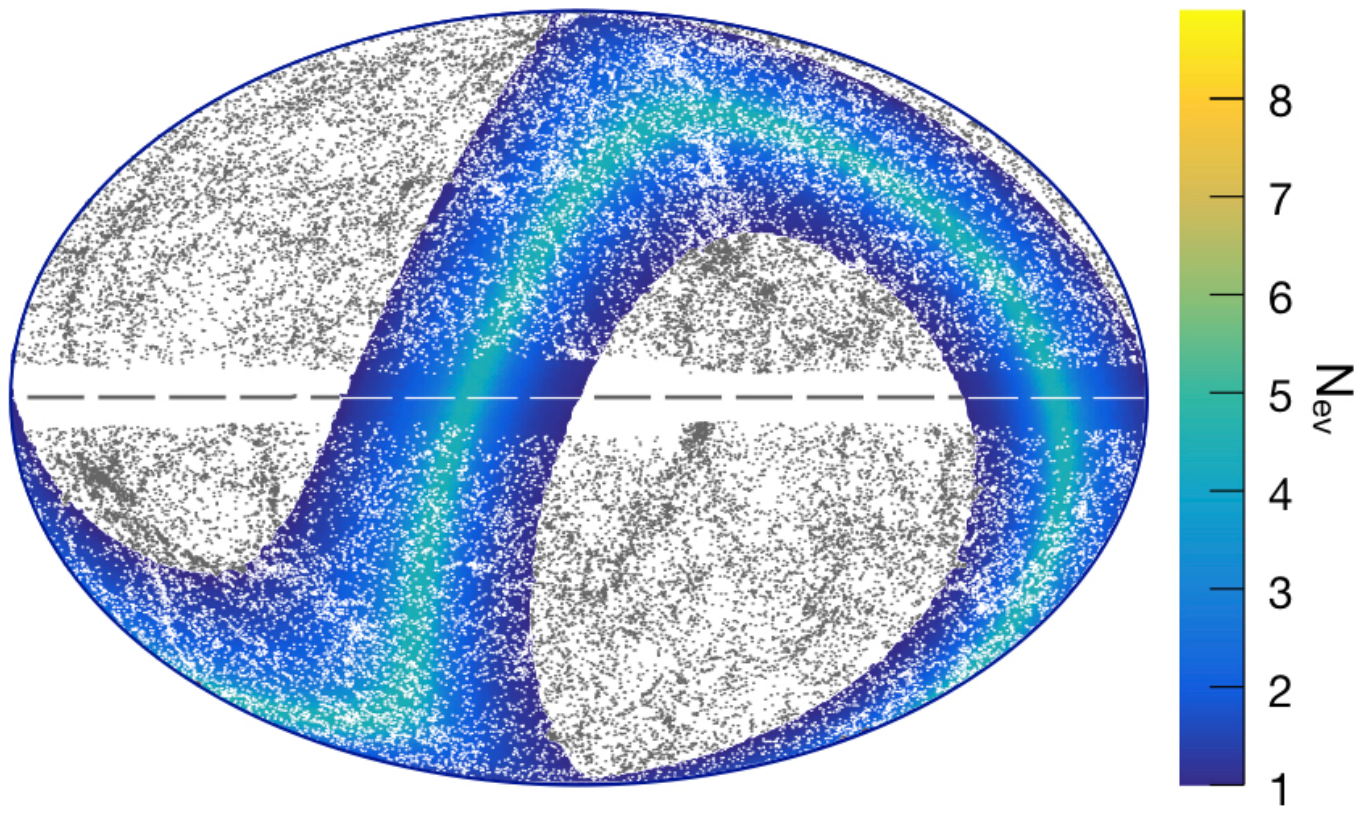}\includegraphics[trim = 27mm 43mm 19mm 40mm, clip, width=0.69\columnwidth]{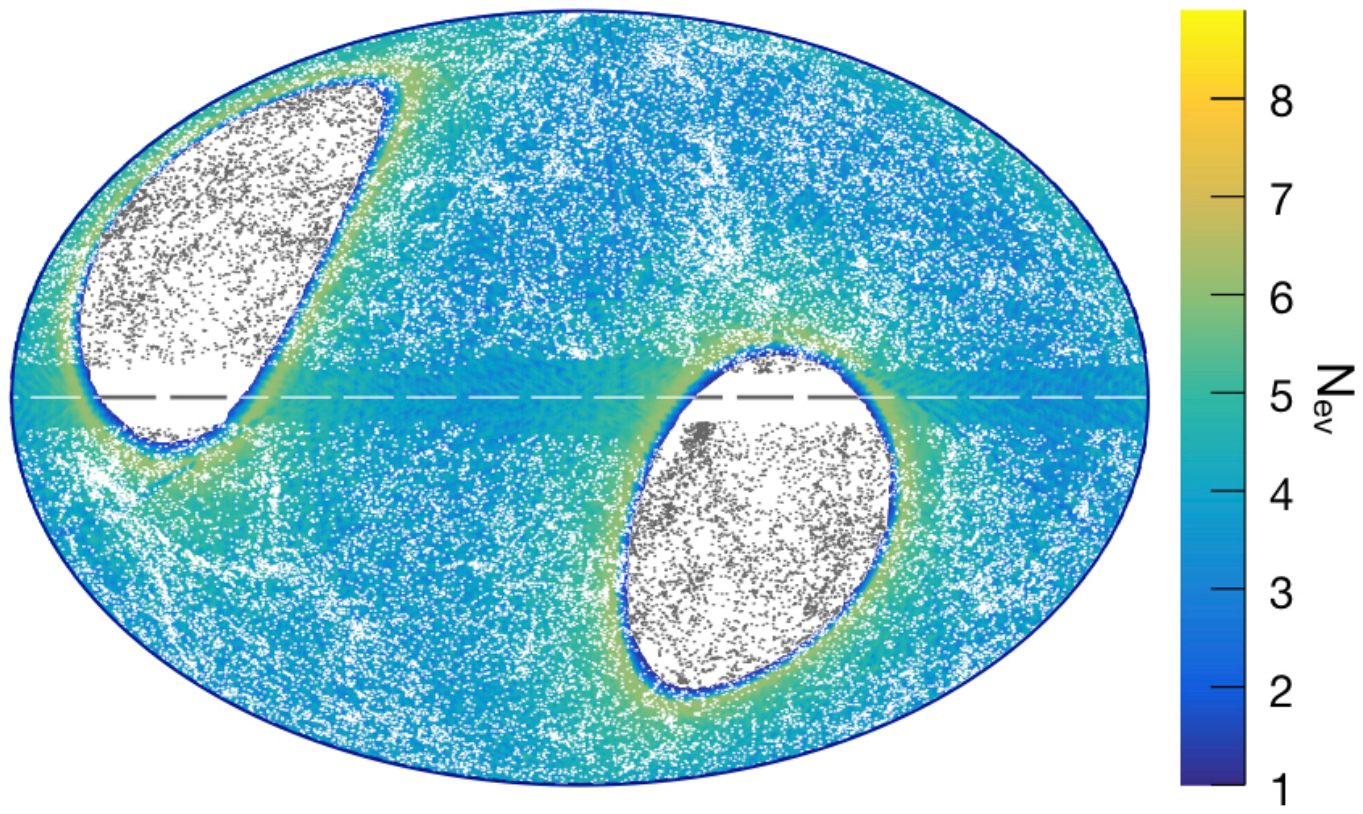}
    \caption{\textit{Left}: Sky plot of the expected number of neutrino events as a function of galactic coordinates for POEMMA in the ``best-case'' short-burst scenario of an sGRB with moderate EE, as in the KMMK model~\cite{Kimura:2017kan}, and placing the source at $40$~Mpc. Point sources are galaxies from the 2MRS catalog~\cite{2012ApJS..199...26H}. \textit{Middle}: Same as at left for IceCube for muon neutrinos. \textit{Right}: Same as at left for \grandacro. Areas with grey point sources are regions for which the experiment is expected to detect less than one neutrino.}
    \label{fig:nevskyplotsshort}
\end{figure*}

\begin{table*}[htb]
\caption{Percentage of the sky for which various neutrino experiments will be able to detect $1.0$ or $6.0$ neutrinos for one long ToO scenario (BNS merger) and one ``best-case'' short burst scenario (sGRB with moderate EE emission).}
\centering
\begin{tabular}{|C{1.25in}|C{0.5in}|C{0.5in}|C{0.5in}|C{0.5in}|C{0.5in}|C{0.5in}|}
\hline
\hline
 \multirow{2}{1.25in}{\centering Model} & \multicolumn{2}{c|}{POEMMA} & \multicolumn{2}{c|}{IceCube*} & 
 \multicolumn{2}{c|}{\grandacro*}\\
\cline{2-7}
& $1.0\ \nu_{\tau}$ & $6.0\ \nu_{\tau}$ & $1.0\ \nu_{\mu}$ & $6.0\ \nu_{\mu}$ & $1.0\ \nu_{\tau}$ & $6.0\ \nu_{\tau}$\\
\hline
Fang \& Metzger~\cite{Fang:2017tla} BNS merger at $5$~Mpc & $100$\% & $100$\% & $70$\% & $18$\% & $82$\% & $81$\% \\
\hline
KMMK~\cite{Kimura:2017kan} sGRB Mod. EE at $40$~Mpc & $100$\% & $49$\% & $50$\% & $0$\% & $81$\% & $2$\% \\
\hline
\hline
\multicolumn{7}{l}{(*) Sky coverage for short bursts is not reflective of instantaneous FoV (see text).}\\
\end{tabular}{}
\label{table:Nevskyno}
\end{table*}

Though Eq.~(\ref{eq:numevents}) is expressed in terms of the average effective area as a function of energy and redshift, we can also determine the expected number of neutrino events as a function of celestial position by replacing ${\cal A}\left(E_\nu,z\right)$ with $\left<A\left(E_\nu,\theta,\phi\right)\right>_{T_{0}}$, the time-averaged effective area as a function of celestial position from Eq.~(\ref{eq:avga}). In Figs.~\ref{fig:nevskyplotslong} and \ref{fig:nevskyplotsshort}, we plot the expected numbers of neutrino events as functions of galactic coordinates for POEMMA for a long burst scenario (BNS merger according to the Fang \& Metzger model in Ref.~[\citenum{Fang:2017tla}] and Fig.~\ref{fig:sensitivity-sunmoon}; for further details on the model see Sec.~\ref{sec:3d}) and a short burst scenario (sGRB with moderate levels of extended emission according to the KMMK model in Ref.~[\citenum{Kimura:2017kan}] and Fig.~\ref{fig:sensitivity-burst}; for further details on the model see Appendix~\ref{app:c}), respectively. For comparison, we provide analogous sky plots for IceCube and \grandacro\ in their respective energy ranges ($10$~TeV--$1$~EeV for IceCube and $10^8$--$3 \times 10^{11}$~GeV for \grandacro) in Figs.~\ref{fig:nevskyplotslong} and \ref{fig:nevskyplotsshort}. As the location on the sky of a given source as viewed by the instrument varies as a function of time, we compute time-averaged effective areas as a function of galactic coordinates for IceCube and \grandacro\footnote{The \grandacro\ effective area as a function of elevation angle was provided through private communication with Olivier Martineau-Huynh.} in Figs.~\ref{fig:nevskyplotslong} and \ref{fig:nevskyplotsshort}.

For all three experiments, we calculate the percentage of the sky in which the expected number of neutrinos meets or exceeds the thresholds corresponding to two scenarios for neutrino ToO observations: (i) multi-messenger follow-up observations in which the experiment detects one neutrino coincident both spatially and in time with an electromagnetic transient event (\textit{e.g.},
as with \icevent~coincident with blazar \txsblazar~\cite{IceCube:2018dnn}; \iceventtde~coincident with tidal disruption event AT2019dsg~\cite{Stein:2020xhk}) and/or a gravitational wave event, and (ii) neutrino-only observations in which the experiment detects a significant number of neutrino events in the absence of coincident multi-messenger observations via electromagnetic or gravitational messengers. In the second scenario, we set the threshold to be the number of events for which the lower limit of the $5\sigma$ unified confidence interval (calculated using the methodology provided by Feldman \& Cousins;~\cite{Feldman:1997qc}) exceeds the expected number of background events for POEMMA (see Appendix~\ref{app:adoubleprime}), thereby ruling out a background-only model. As the expected number of background events increases with the length of the observation, we base these calculations on observations of long-duration events and include a trials factor of $\sim 100$ observations. Based on these considerations, we set the threshold in the second scenario to six events. We note that separate event thresholds should be set for IceCube and \grandacro; however, as we are not as familiar with the backgrounds for these experiments, we take their backgrounds to be zero and assume the same threshold of six events. Table~\ref{table:Nevskyno} provides the calculated sky percentages for the three experiments.

For long bursts, we averaged the effective area over the operation lifetime for IceCube\footnote{For years beyond 2012, we assumed that the effective area was the same as that provided for 2012.} and over a $24$-hour period for \grandacro; as such, the holes in the IceCube and \grandacro\ skyplots (areas with grey point sources) are regions for which the experiment has limited or no effective area and/or exposure for the range of energies in which it can detect neutrinos from the source model. For instance, the hole in the northern celestial hemisphere for IceCube arises due to a suppression in the effective area at high zenith angles due to attenuation by the Earth for events above $\sim 10$ PeV. \grandacro\ will be sensitive to tau neutrinos with zenith angles between $85^{\circ}$ and $95^{\circ}$ ($360^{\circ}$ in azimuth); hence, the holes in the \grandacro\ skyplot in Fig.~\ref{fig:nevskyplotslong} are those regions of the sky which never enter its FoV, while the slices with enhanced numbers of neutrino events are those regions of the sky which spend the most time in the FoV, and this is where \grandacro\ can expect to see the most neutrinos. For the scenario of a BNS merger at $5$~Mpc, Fig.~\ref{fig:nevskyplotslong} shows that POEMMA will be sensitive to neutrinos from all over the sky, while IceCube and \grandacro\ will be sensitive to $\sim 70$\% and $\sim 82$\% of the sky, respectively. For the higher threshold of $\sim 6$ neutrinos, POEMMA will be able to achieve this level in $\sim 100$\% of the sky, giving it a distinct advantage over IceCube ($\sim 18$\%) and slightly better sky coverage than even \grandacro\ ($\sim 81$\%). On the other hand, while POEMMA will see more neutrinos than IceCube for most regions of the sky, the regions in which IceCube and \grandacro\ will detect the most neutrinos (roughly $10$\% for both IceCube and \grandacro) are larger than that for POEMMA ($\lesssim 1$\%), and \grandacro\ can expect to see more neutrinos in their best region ($\sim 60$ events for \grandacro\ compared with $\sim 36$ for POEMMA and $\sim 14$ for IceCube). However, we note that while the POEMMA plot accounts for the decrease in observing time due to the Sun and the Moon, no background was assumed for either IceCube or \grandacro; as such, the estimates for IceCube and \grandacro\ are somewhat optimistic, particularly in comparison with POEMMA.

For short bursts, given that neither IceCube nor \grandacro\ will be able to slew to a given target as POEMMA will, the observational scenario for these experiments is not completely analogous to that considered in this paper for POEMMA. For the purposes of comparison, we constructed their sky plots in Fig.~\ref{fig:nevskyplotsshort} by assuming that the burst starts at a time for which the effective area at a given set of sky coordinates is at a maximum. We then average the effective area over the assumed timescale for short bursts ($\sim 10^3$~s). In this manner, we compare these ``best-case'' scenarios for IceCube and \grandacro\ to our best-case scenario for POEMMA for short bursts. However, both IceCube and \grandacro\ will be limited in their capability to follow up short bursts due to their inability to slew. This is less of a disadvantage for IceCube than for \grandacro\ since IceCube is sensitive to muon neutrinos in a greater range of zenith angles than \grandacro\ is sensitive to tau neutrinos. The band of zenith angles for \grandacro\ results in an instantaneous FoV of $\sim 4.4$\% of the sky, so the probability of this ``best-case'' scenario occurring is relatively low. On the other hand, while POEMMA's instantaneous FoV ($\sim 30^{\circ} \times 9^{\circ}$) is smaller than that of \grandacro\ ($\sim 360^{\circ} \times 10^{\circ}$), POEMMA's orbital speed (one orbit in $95$~min.) and quick re-pointing capability ($\sim 90^{\circ}$ in $500$~s) will allow it to access regions of the sky outside of its instantaneous FoV faster than \grandacro, which is restricted to the rotation speed of the Earth. With this combination of capabilities, POEMMA will be able access to $\sim 21$\% of the sky in $500$~s ($\sim 37$\% in $10^3$~s)~\cite{Guepin:2018yuf}, a key advantage over \grandacro\ in terms of sky coverage on such short timescales.

As in Fig.~\ref{fig:nevskyplotslong}, holes in the IceCube and \grandacro\ sky plots in Fig.~\ref{fig:nevskyplotsshort} appear where the experiment has limited or no effective area and/or exposure for the range of energies in which it can detect neutrinos from the source model. In this scenario, a hole in the southern celestial sphere for IceCube appears because the range of energies in which it can detect neutrinos for the KMMK model is smaller than that for the Fang \& Metzger model at the distances considered (\textit{c.f.}, Figs.~\ref{fig:sensitivity-sunmoon} and \ref{fig:sensitivity-burst}). Even considering the ``best-case'' scenarios for IceCube and \grandacro, POEMMA has a distinct advantage in detecting these types of short burst events. Not only will POEMMA be sensitive to neutrinos from the entire sky (compared with $\sim 50$\% for IceCube and $\sim 81$\% for \grandacro), POEMMA can expect to see more neutrinos (maximum number of $\sim 10$ events versus $\sim 5$ for IceCube and $\sim 6$ for \grandacro). For the higher threshold of $\sim 6$ neutrinos, POEMMA will be able to achieve this level in $\sim 49$\% of the sky, compared with $\sim 0$\% for IceCube and $\sim 2$\% for \grandacro.

\subsection{Probability of ToOs for Modeled Astrophysical Neutrino Sources}\label{sec:3c}

\begin{figure}[htb]
\centering
	\includegraphics[trim = 50mm 20mm 50mm 20mm, clip, width=0.95\columnwidth]{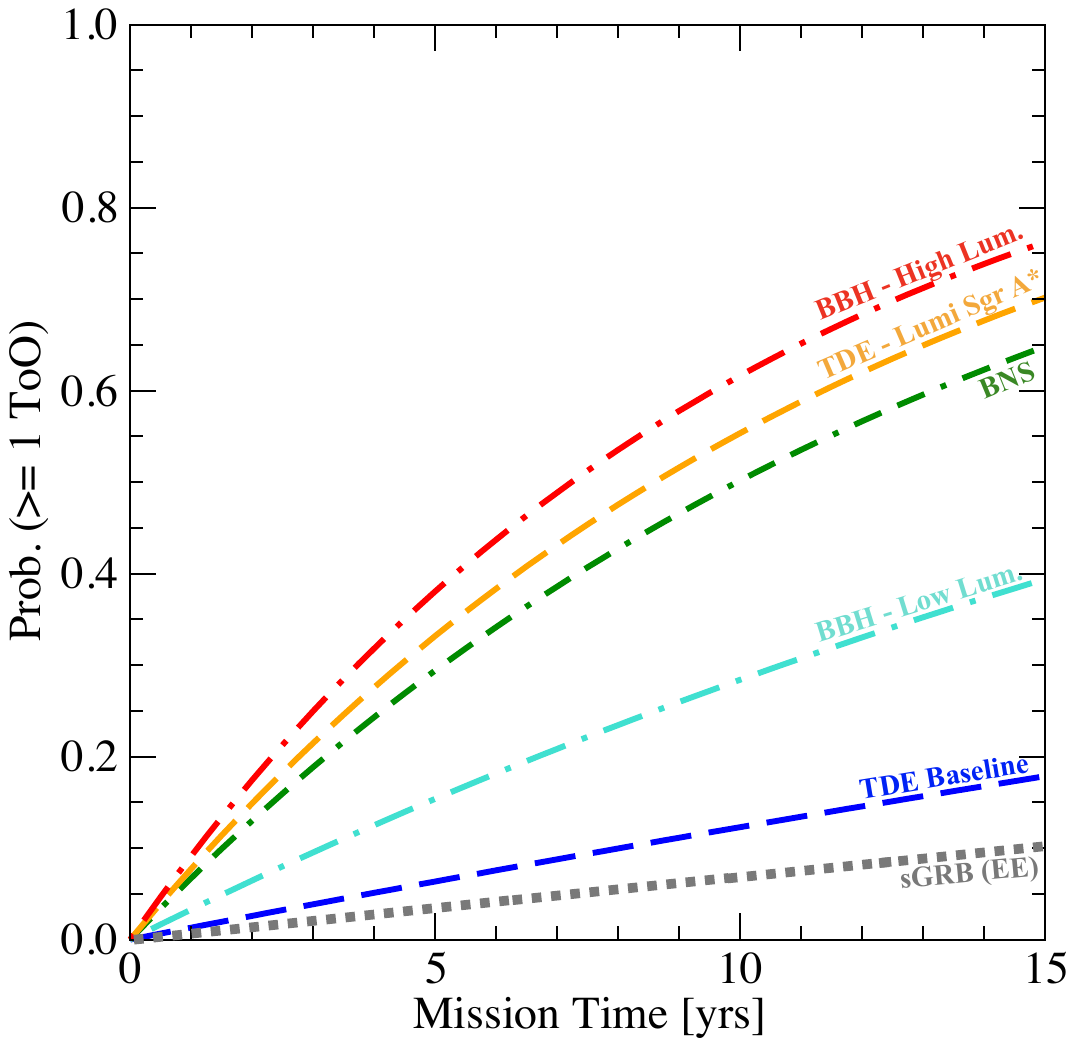}
	\caption{The Poisson probability of POEMMA observing at least one ToO versus observation time for several modeled source classes. Featured source models are TDEs from Lunardini and Winter~\cite{Lunardini:2016xwi}, BNS mergers from Fang and Metzger~\cite{Fang:2017tla}, BBH mergers from Kotera and Silk~\cite{Kotera:2016dmp}, and sGRBs with moderate EE from KMMK~\cite{Kimura:2017kan}.}
	\label{fig:poisson}
\end{figure}

In order to determine the modeled source classes that are most likely to result in ToOs for POEMMA, we model the occurrence of transient events as a Poisson process. The probability of POEMMA observing at least one ToO for a given source model as a function of time, $t$, is then given by:
\begin{equation}
    P\left(\geqslant 1 \mbox{ ToO}\right) = 1 - P\left(0\right) = 1 - \mathrm{e}^{-rt}\,,
\end{equation}
where $r$ is the expected rate of ToOs for the source model as determined from the cosmological volume in which neutrinos would be detectable by POEMMA and from cosmological event rates for the source class taken from the literature (see model descriptions provided in Sec.~\ref{sec:3d}). The cosmological volume is determined from the neutrino horizon, $z_{\rm hor}$, which we calculate from Eq. (\ref{eq:numevents}) by determining the redshift at which $N_{\rm ev}$ is set equal $1.0$. In Fig.~\ref{fig:poisson}, we plot the probability that POEMMA will observe at least one ToO versus observation time for several of the source models considered in this paper.

In Table~\ref{table:events-bytype}, we provide the calculated number of neutrino events for several models of astrophysical transient source classes assuming a source at the 
\begin{table*}[p]
\centering
\caption{Average expected numbers of neutrino events above $E_{\nu}>10^7$ GeV detectable by POEMMA for several models of transient source classes assuming source locations at the Galactic Center (GC) and at $3$ Mpc. The horizon distance for detecting $1.0$ neutrino per ToO event is also provided. Source classes with observed durations $> 10^3$~s are classified as long bursts. Those with observed durations $\lesssim 10^3$ s are classified as short bursts. Models in boldface type are those models for which POEMMA has $\gtrsim 10$\% chance of observing a ToO during the proposed mission lifetime of $3$~--~$5$~years. Models in italics are the same but for a mission lifetime of $10$~years.}
 \begin{tabular}{|C{0.95in}|C{0.8in}|C{0.8in}|C{1.3in}|C{2.0in}|} 
 \hline
 \hline
  \multicolumn{5}{c}{Long Bursts} \\
 \hline
 \hline
 Source Class & No. of $\nu$'s at GC & No. of $\nu$'s at $3$ Mpc & Largest Distance for $1.0\ \nu$ per event & Model Reference \\ [0.5ex] 
 \hline
TDEs & $1.4 \times 10^5$ & $0.9$ & $3$ Mpc & Dai and Fang\  \cite{Dai:2016gtz}
 average \\ 
 \hline
 TDEs & $6.8 \times 10^5$ & $4.7$ & $7$ Mpc & Dai and Fang \cite{Dai:2016gtz} 
  bright \\
 \hline
 \textbf{TDEs} & $\boldsymbol{2.7 \times 10^8}$ &  $\boldsymbol{1.7 \times 10^3}$ & $\boldsymbol{128}$ \textbf{Mpc} & \textbf{Lunardini and Winter}~\cite{Lunardini:2016xwi}
 $\boldsymbol{M_{\rm SMBH} = 5 \times 10^6 M_{\odot}}$\ \ \ \ \ \ \ \ \textbf{Lumi Scaling Model} \\
 \hline
 \textit{TDEs} & $\mathit{7.7 \times 10^7}$ & $\mathit{489}$ & $\mathit{69}$ \textit{Mpc} & \textit{Lunardini and Winter}~\cite{Lunardini:2016xwi}\ \ \ \ \ \ \ \  \textit{Base Scenario} \\
 \hline
Blazar Flares & NA* & NA* & $47$ Mpc & 
RFGBW
\cite{Rodrigues:2017fmu}
-- FSRQ proton-dominated advective escape model \\
 \hline
 lGRB Reverse Shock (ISM) & $1.2 \times 10^{5}$ & $0.8$ & $3$ Mpc & Murase \cite{Murase:2007yt} \\
 \hline
 lGRB Reverse Shock (wind) & $2.5 \times 10^{7}$ & $174$ & $41$ Mpc & Murase \cite{Murase:2007yt} \\
 \hline
 \textbf{BBH merger} & $\boldsymbol{2.8 \times 10^{7}}$ & $\boldsymbol{195}$ & $\boldsymbol{43}$ \textbf{Mpc} & \textbf{Kotera and Silk}~\cite{Kotera:2016dmp}~\textbf{(rescaled)}\  
 \textbf{Low Fluence} \\
 \hline
 \textbf{BBH merger} & $\boldsymbol{2.9 \times 10^{8}}$ & $\boldsymbol{2.0 \times 10^3}$ & $\boldsymbol{137}$ \textbf{Mpc} & \textbf{Kotera and Silk}~\cite{Kotera:2016dmp}~\textbf{(rescaled)}\  
 \textbf{High Fluence} \\
 \hline
 \textbf{BNS merger} & $\boldsymbol{4.3 \times 10^{6}}$ & $\boldsymbol{30}$ & $\boldsymbol{16}$ \textbf{Mpc} & \textbf{Fang and Metzger} \cite{Fang:2017tla} \\
 \hline
 BWD merger & $25$ & $0$ & $38$ kpc &  
 XMMD \cite{Xiao:2016man} \\
 \hline
 Newly-born Crab-like pulsars (p) & $190$ & $0$ & $109$ kpc & Fang \cite{Fang:2014qva} \\
 \hline
 Newly-born magnetars (p) & $2.5 \times 10^4$ & $0.2$ & $1$ Mpc & Fang \cite{Fang:2014qva} \\
 \hline
 Newly-born magnetars (Fe) & $5.0 \times 10^{4}$ & $0.3$ & $2$ Mpc & Fang \cite{Fang:2014qva} \\
 \hline
 \hline
 \multicolumn{5}{c}{Short Bursts}\\
 \hline
 \hline
  Source Class & No. of $\nu$'s at GC & No. of $\nu$'s at $3$ Mpc & Largest Distance for $1.0\ \nu$ per event & Model Reference \\ [0.5ex] 
 \hline
 sGRB Extended Emission (moderate) & $1.1 \times 10^{8}$ & $800$ & $90$ Mpc & KMMK \cite{Kimura:2017kan} \\
 \hline
 \hline
 \multicolumn{5}{l}{(*) Not applicable due to a lack of known blazars within $100$~Mpc.}\\
\end{tabular}
\label{table:events-bytype}
\end{table*}
Galactic Center (GC) and at $3$ Mpc (roughly the distance to the nearest starburst galaxy, NGC253). To provide a sense of the maximum distance at which a given source class is detectable by POEMMA, we include its neutrino horizon expressed as a luminosity distance as determined from a model taken from the literature. The results for long bursts include the average impacts of the Sun and the Moon and hence, provide a reasonable estimate of POEMMA's capability in detecting such sources. For short bursts, we do not account for the Sun and Moon due to strong variations in their effects over the course of POEMMA's orbital period. Furthermore, for these scenarios, the source was placed at the optimal sky position for POEMMA observations. As such, the results for short bursts should be regarded as reflecting the best possible scenarios for POEMMA observations. The models in boldface type are those for which POEMMA has at least a $10$\% chance of seeing a ToO within the proposed mission lifetime of $3$ -- $5$ years and hence, are the most promising source classes for POEMMA. Other source classes listed in Table~\ref{table:events-bytype} would be detectable by POEMMA if located reasonably close by, but would likely require mission lifetimes of $10$ years (source classes in italics) or more for a reasonable chance of detecting one ToO. Based on the results from this study and studies of ToOs with other neutrino observatories provided in the literature, we expect these latter sources to be challenging to observe by any currently operating or planned neutrino observatory.

\subsection{Most Promising Candidate Neutrino Source Classes for POEMMA}\label{sec:3d}

In the remainder of this section, we provide brief discussions of the most promising astrophysical candidate neutrino source classes in terms of their expected ToO rates for POEMMA (boldface and italicized models in Table~\ref{table:events-bytype}; for a discussion of the additional source classes, see Appendix~\ref{app:c}). We should note that our list of sources and corresponding models is not intended to be an exhaustive list or present a complete characterization of the sources in question. Several of the source classes have been proposed as possible neutrino emitters going back several decades. Furthermore, the relevant parameter spaces for the characteristics of these sources can be quite large and uncertain, particularly in the presumed regime of neutrino production. Rather, our intent with this list is to provide a rough idea of POEMMA's capability in detecting neutrinos from commonly-invoked source candidates and identify the most promising source classes for POEMMA. For each of the most promising source candidates, we discuss their contributions to the diffuse astrophysical neutrino flux in light of IceCube measurements below $5$~PeV~\cite{Stettner:2019tok} and constraints at higher energies~\cite{Aartsen:2018vtx}.

\sideheader{Jetted Tidal Disruption Events} During a tidal disruption event (TDE), a massive black hole rips apart an orbiting star, accreting its material and producing a flare of radiation that can last for months or even years~[\citenum{1975Natur.254..295H,1988Natur.333..523R}; for detailed reviews, see \textit{e.g.}, \citenum{2015JHEAp...7..148K,2017ApJ...838..149A}]. As demonstrated by Swift J1644+57, some TDEs result in powerful, relativistic jets~\cite{2011Sci...333..203B,2011Natur.476..421B,2011Natur.476..425Z}. With the abundance of baryons from the disrupted stellar material, jetted TDEs are natural candidates for proton and nuclei accelerators, possibly capable of reaching ultra-high energies~\cite{Farrar:2008ex,2014arXiv1411.0704F,2017MNRAS.466.2922P,Guepin:2017abw} and producing very-high and ultra-high energy neutrinos~\cite{Dai:2016gtz,Lunardini:2016xwi,2011PhRvD..84h1301W,2016PhRvD..93h3005W,Guepin:2017abw}. In order to evaluate the capability of POEMMA for detecting neutrinos from jetted TDEs, we use models from Lunardini and Winter in Ref.~\cite{Lunardini:2016xwi}, which explored the relationship between key jet characteristics and the mass of the SMBH. Alternative models of TDE neutrino production are available in the literature~\cite[\textit{c.f.},][]{2011PhRvD..84h1301W,2016PhRvD..93h3005W,Senno:2016bso,Biehl:2017hnb,Guepin:2017abw} can exhibit differences related to modeling parameters such as the jet luminosity, the baryon loading, and the comoving event rate.

For the purposes of this study, we consider two models from Ref.~\cite{Lunardini:2016xwi}: the Base Case model in which no dependence on SMBH mass is included, and a Lumi Scaling model in which the jet bulk Lorentz factor, variability timescale, and X-ray luminosity scale with SMBH mass. We note that neither model violates IceCube measurements of the diffuse astrophysical flux~\cite{Aartsen:2017mau} and if correct, both models would predict significant contributions to the astrophysical flux from jetted TDEs, particularly at energies $\gtrsim 10^6$~GeV~\cite{Lunardini:2016xwi}. For the Lumi Scaling model, we took $M_{\rm SMBH} = 5 \times 10^6 M_{\odot}$, as motivated by estimates of the mass of Sgr A*~\cite[see \textit{e.g.},][]{2016ApJ...830...17B}, and the neutrino fluence was determined by interpolating between the $10^6 M_{\odot}$ and the $10^7 M_{\odot}$ models. For a TDE at the galactic center, these models predict that POEMMA will detect $\sim 8 \times 10^7$ and $\sim 3 \times 10^8$ neutrinos for the Base and Lumi Scaling Scenarios, respectively. In addition to the neutrino fluence, Lunardini and Winter~\cite{Lunardini:2016xwi} also modeled the cosmological rate of TDEs, finding the local rate of jetted TDEs to be $\mathcal{R} \simeq 0.35$--$10$~Gpc$^{-3}$~yr$^{-1}$ depending on assumptions for the minimum SMBH mass. For both models, these rates imply diffuse neutrino fluxes that are consistent with current IceCube measurements~\cite{Stettner:2019tok}. For the Lumi Scaling model, the neutrino horizon for POEMMA is $\sim 130$~Mpc with a corresponding Poisson probability of detecting at least one such event of $\gtrsim 21$ -- $33\%$ over the proposed mission lifetime of $3$ -- $5$ years or up to $\sim 55\%$ for an extended mission lifetime of $10$ years. For the Base model, the neutrino horizon is closer ($\sim 70$~Mpc), resulting in a Poisson detection probability of $\sim 10\%$ over the course of an extended mission lifetime of $10$ years.

\sideheader{Binary Neutron Star Mergers} 
Strong magnetic fields and rapid rotation in pulsars combine to induce electric fields that naturally accelerate particles~\cite[see \textit{e.g.},][]{Gunn:1969ej,1992MNRAS.257..493B,Blasi:2000xm,2003ApJ...589..871A,2012ApJ...750..118F}, with ultra-high energies possibly being achievable in newly-born magnetars (pulsars with magnetic field strengths $\gtrsim 10^{14}$~G; for detailed review, see~[\citenum{Kaspi:2017fwg}]) with spin periods $\sim$ milliseconds~\cite[see \textit{e.g.},][]{Blasi:2000xm,2003ApJ...589..871A,2012ApJ...750..118F,Fang:2017tla}. Accelerated UHECRs produce neutrinos through interactions with the surrounding ambient medium and radiation fields. In Ref.~\cite{Fang:2017tla}, Fang and Metzger modeled the time-dependent neutrino production in the magnetosphere of a rapidly spinning magnetar resulting from a BNS merger. Their model predicts that PeV--EeV neutrinos could be detectable for days and even months following the merger. Alternatively, the BNS merger could result in a spinning black hole which could accrete marginally bound merger debris, resulting in unbound winds or wide-angle jets that accelerate particles to ultra-high energies~\cite{Decoene:2019eux}. In this paper, we only explore the scenario in which the BNS merger remnant is a rapidly spinning magnetar.

Following the announcement of the observation of a BNS merger~\cite{TheLIGOScientific:2017qsa,Monitor:2017mdv} by Advanced LIGO~\cite{Aasi:2014mqd} and Advanced Virgo~\cite{TheVirgo:2014hva}, the ANTARES, IceCube, and Pierre Auger Observatories conducted a search for high-energy neutrinos positionally coincident with the merger arriving within $\pm 500$~s of the merger time and within a $14$-day period following the merger~\cite{ANTARES:2017bia}. No neutrinos were found, though at a distance of $\sim 40$ Mpc, the neutrino fluences predicted by Fang and Metzger would have been undetectable with these neutrino experiments. As shown in Fig.~\ref{fig:sensitivity-sunmoon}, POEMMA will have an advantage in searching for neutrinos from BNS merger events due to its capability to rapidly re-point for follow-up and to revisit a source location every orbit and also due to the fact that POEMMA is most sensitive at the energies at which the neutrino fluences are expected to peak ($\sim$~hundreds PeV). Using the Fang and Metzger model, we predict that POEMMA will be able to detect $\sim$~tens of neutrinos up to distances $\sim$~few Mpc, with a neutrino horizon of $\sim 16$ Mpc. Taking the upper limit of the LIGO-Virgo event rate for BNS mergers ($\mathcal{R} \sim 110$--$3840$~Gpc$^{-3}$~yr$^{-1}$; [\citenum{LIGOScientific:2018mvr}]), the Poisson probability of POEMMA detecting at least one such event is $\gtrsim 20$ -- $30\%$ over the proposed mission lifetime of $3$ -- $5$ years or up to $\sim 50\%$ for an extended mission lifetime of $10$ years. 

We note that the BNS merger rates reported by LIGO-Virgo are higher than that used in the Fang and Metzger analysis and the combined neutrino fluence from the cosmological population of BNS mergers may overproduce the IceCube upper limit on the diffuse neutrino flux above $5$ PeV~\cite{Aartsen:2018vtx} depending on source evolution and maximum redshift. As the calculated neutrino horizon for BNS mergers is very local, the use of the local BNS rate as measured by LIGO-Virgo is appropriate, but it is worth noting that with only two confirmed detections, the BNS merger rate is unconstrained, particularly beyond the LIGO-Virgo BNS horizon ($\sim 130$~Mpc).\footnote{\url{https://emfollow.docs.ligo.org/userguide/capabilities.html}} Alternatively, it is also worth considering the possibility that a large fraction of BNS mergers may not result in a long-lived or stable magnetar that would produce neutrinos. Such a scenario would reduce the diffuse neutrino flux from BNS mergers, but it would also reduce the predicted ToO rates for POEMMA.

\sideheader{Binary Black Hole Mergers} Analogous to BNS mergers, binary black hole (BBH) systems are also potential reservoirs of power; for instance, the rotational energy of a spinning black hole in a magnetized disk can be extracted to power jets~\cite{Blandford:1977ds}. However, unlike in the case of BNS mergers, black holes in BBH systems lack a companion that can be tidally disrupted and reorganized into an accretion disk~\cite{Perna:2019pzr}. As such, BBH mergers are generally expected to release energy solely in the form of gravitational waves. On the other hand, reported candidate electromagnetic counterparts to LIGO-Virgo BBH events~\cite{Connaughton:2016umz,Graham:2020gwr} have spurred interest in BBH merger scenarios that would give rise to multi-messenger counterparts, including the possibility of pre-existing material still being present at the time of the merger~\cite[see \textit{e.g.},][]{Loeb:2016fzn,Perna:2016jqh,Murase:2016etc,Woosley:2016nnw,Janiuk:2016qpe,Bartos:2016dgn,deMink:2017msu,Khan:2018ejm,Martin:2018iov,Graham:2020gwr} or the possibility of charged black holes~\cite[see \textit{e.g.},][]{Liebling:2016orx,Zhang:2016rli,Liu:2016olx,Fraschetti:2016bpm}. In Ref.~\cite{Kotera:2016dmp}, Kotera and Silk take the further step of suggesting that if BBH mergers can form accretion disks and associated jets or magnetohydrodynamic outflows, they could possibly accelerate CRs to ultra-high energies, which would produce neutrinos via interactions with the surrounding environment. While such a scenario would make BBH mergers promising candidate sources of neutrinos, it is as yet unclear whether enough material is present at the time of the BBH merger in order to provide an environment for accelerating particles or even to emit electromagnetic radiation, and no definitive detections of electromagnetic counterparts to BBH mergers have been reported to date~\cite{Anchordoqui:2016dcp}. As such, we acknowledge that the models that predict neutrino emission from BBH mergers are highly speculative.

For the purposes of predicting the capability of POEMMA for detecting neutrinos from BBH mergers, we use the neutrino flux suggested by Kotera and Silk~\cite{Kotera:2016dmp}. In deriving the neutrino flux, they estimated the Poynting flux that can be generated by stellar BHs and, in calculating the maximum neutrino flux, they assumed the Poynting flux can be entirely tapped into UHECRs. The Kotera and Silk neutrino flux includes a parameter, $f_{\nu}$, for the optical depth to neutrino production. For our calculations, we set $f_{\nu}$ equal to $1/3$ in order to not violate IceCube upper limits on the diffuse neutrino flux above $5$~PeV~\cite{Aartsen:2018vtx}. The Kotera and Silk model requires that each individual source supply a fixed amount of energy in the form of CRs in order to reproduce the observed CR flux above $10^{19}$~eV, resulting in a predicted neutrino fluence for each individual source that depends on the BBH merger rate. For the purposes of our calculations, we consider two scenarios -- a High Fluence scenario based on the lower limit of the LIGO-Virgo BBH merger rate ($9.7$~Gpc$^{-3}$~yr$^{-1}$;~[\citenum{LIGOScientific:2018mvr}]) and a Low Fluence scenario based on the upper limit of the LIGO-Virgo BBH merger rate ($101$~Gpc$^{-3}$~yr$^{-1}$;~[\citenum{LIGOScientific:2018mvr}]). For these scenarios, we predict that POEMMA will detect $\sim$ hundreds or $\sim$ thousands of neutrinos for events occurring within $\sim$ few Mpc in the Low Fluence and High Fluence cases, respectively. For the neutrino horizon, we expect POEMMA to be able to detect neutrinos out to $\sim 40$ Mpc in the Low Fluence scenario and out to $\sim 120$ Mpc in the High Fluence scenario. Based on these horizons and on the LIGO-Virgo BBH merger rate, the Poisson detection probability for POEMMA detection of such an event is $\sim 7$ -- $20\%$ over the proposed mission lifetime of $3$ -- $5$ years and $\sim 20$ -- $34\%$ over an extended mission lifetime of $10$ years.

\section{Conclusions}
\label{sec:4}

In this paper, we have explored several scenarios for neutrino ToO observations with POEMMA, calculating its sensitivity and evaluating prospects for detecting neutrinos from several candidate transient astrophysical source classes. While at any particular time only transient sources below the limb of the Earth as viewed from the satellites are relevant to tau-neutrino induced upward-going air shower signals, POEMMA and other space-based instruments will have full-sky coverage over the orbital period of the satellites and the precession period of the orbital plane. The slewing capability of POEMMA in time frames of on the order of $500$~s will permit rapid response to short-duration transient events over a large region of the sky ($\sim 21$\%).

As compared with the standard limb-viewing configuration for diffuse neutrino flux measurements (POEMMA-limb mode, which is limited to $7^{\circ}$ below the horizon; \citenum{PhysRevD.100.063010}), POEMMA's ToO observation modes provide access to a broader range in \taon elevation angles before neutrino flux attenuation in the Earth obscures a neutrino source. Our results here are based on elevation angles $\beta_{\rm tr} \leq 35^{\circ}$, equivalent to viewing from the satellites to an angle of $\sim 20^{\circ}$ below the limb. The capability for tracking the source means that the best case sensitivities for POEMMA are as much as two orders of magnitude better than those of Auger as reported in Ref.~\cite{ANTARES:2017bia} with all-sky coverage. Based on the calculations performed here, we predict that POEMMA will have reasonable chance to observe TDEs, BBH mergers, and BNS mergers within a $3$ -- $5$-year observation period. Long bursts within luminosity distances specified in Table \ref{table:events-bytype} will be observable by POEMMA, regardless of location. For short duration bursts, the sensitivity will be better than for long bursts if the source is well placed relative to the Earth and POEMMA. However, short bursts may not be observable if the source does not dip below the Earth's horizon, or if the burst occurs when the Sun and/or Moon interfere with observing.

For long-duration events, POEMMA will have the option of maneuvering its satellites closer together (ToO-stereo mode) in order to lower its energy threshold. In most cases, ToO observations will be multi-messenger follow-up observations with POEMMA responding to alerts issued by electromagnetic or gravitational-wave detectors. In these cases, the decision to maneuver the satellites closer together will hinge in large part on the source class, the distance, and expectations for the duration of the event. A BNS merger event such as \bnsevent/\bnseventgrb\ occurring within one or two sigma of the predicted horizon distance of $16$~Mpc would be a good example of a priority target that might warrant satellite maneuvers. As slewing the telescopes without changing their separation requires minuscule amounts of propellant, there is no limit to the number of ToOs POEMMA can follow up in ToO-dual mode. For sky localizations with large error circles (as in gravitational-wave events with fewer than three detectors), POEMMA's large field-of-view ($\sim 30^{\circ} \times 9^{\circ}$) will enable relatively efficient tiling. However, tiling very large error circles will reduce the observation time for each individual tile, so source localizations to within a factor of a few times POEMMA's field-of-view would be another broad requirement for follow-up.

For the purposes of this study, we have assumed that the neutrino burst will be closely coincident in time and space with the event and/or other neutral messengers, such as gamma rays or gravitational waves. Murase and Shoemaker~\cite{Murase:2019xqi} recently explored possible time delays and angular signatures in the neutrino signal resulting from beyond SM interactions between high-energy neutrinos and the cosmic neutrino background and/or dark matter particles. In POEMMA's energy range (beginning at $\sim 10$~PeV or $\sim 30$~PeV in ToO-stereo and ToO-dual modes, respectively) and at the neutrino horizon distances calculated in this paper, we expect the effects from these types of interactions to be minuscule; however, we note that any time delay in the neutrino burst would be helpful to POEMMA by providing more time for re-pointing and re-positioning the satellites for the ToO observation.

In any ToO scenario, whether neutrino detectors following up electromagnetic and/or gravitational-wave alerts or vice versa, multi-messenger observations of transient astrophysical phenomena will not be possible without a high-quality alert system incorporating all three messengers. We note that there is already an elaborate multi-messenger network consisting of all-sky/wide-field instruments sensitive to electromagnetic radiation (\textit{e.g.}, \textit{Swift}, \textit{Fermi}, INTEGRAL, etc.), neutrinos (\textit{i.e.}, IceCube and ANTARES), and gravitational waves (\textit{i.e.}, LIGO, Virgo, KAGRA). These instruments provide timely notifications of transient astrophysical events via the Gamma-ray Coordinates Network/Transient Astronomy Network (GCN/TAN)\footnote{\url{https://gcn.gsfc.nasa.gov/about.html}} and/or Astronomer's Telegram (ATel)\footnote{\url{http://www.astronomerstelegram.org/}} in order to enable such rapid responses. Alerts from LIGO and Virgo are also made available via the Gravitational-Wave Candidate Event Database.\footnote{\url{https://gracedb.ligo.org/}} 

In the coming decade and beyond, the contemporary multi-messenger network will only flourish as maintaining and further developing a well-coordinated network is a top priority for the high-energy astrophysics community. Several wide-field electromagnetic missions (\textit{e.g.}, Transient Astrophysics Observatory, Transient Astrophysics Probe, All-sky Medium Energy Gamma-ray Observatory, BurstCube, etc.) and ground-based and space-based gravitational-wave detectors (\textit{e.g.}, Einstein Telescope, Cosmic Explorer, Laser Interferometer Space Antenna) have been proposed for operations over a time frame that will overlap with POEMMA. We envision POEMMA playing an essential, complementary role, particularly at ultra-high energies, in the next-generation multi-messenger network.

The source models described here, with associated numbers of events, follow from standard model (SM) processes. The ANITA Collaboration has reported two unusual events, which qualitatively look like air showers initiated by energetic ($\sim 500~{\rm PeV}$) particles that emerge from the ice along trajectories with large elevation angles~\cite{Gorham:2016zah,Gorham:2018ydl}. However, at these high energies, neutrinos are expected to interact inside the Earth with a high probability. For the angles inferred from ANITA observations, the ice would be well screened from up-going neutrinos by the underlying layers of Earth, challenging SM explanations~\cite{Romero-Wolf:2018zxt,Fox:2018syq,Aartsen:2020vir}. Several beyond SM physics models have been proposed to explain ANITA events~\cite{Cherry:2018rxj,Anchordoqui:2018ucj,Huang:2018als,Collins:2018jpg,Chauhan:2018lnq,Anchordoqui:2018ssd,Heurtier:2019git,Hooper:2019ytr,Cline:2019snp,Esteban:2019hcm,Heurtier:2019rkz}, but systematic effects in the data analysis may play a larger role than originally anticipated~\cite{deVries:2019gzs,Shoemaker:2019xlt,Gorham:2020zne}. POEMMA will have detection capabilities for such events. For example, a $600$~PeV EAS will yield a signal of more than $10^4$ photons/m$^2$ for $35^{\circ}$ Earth-emergence angle, implying a PE signal that is a factor of $500$ times greater than the $10$~PE threshold.  Relative to ANITA, POEMMA will have a factor of $\sim 10$ increase in acceptance solid angle since these EASs are so bright. POEMMA, in tracking neutrino sources, will also be sensitive to non-standard model particles that generate up-going EASs.

Our results herein provide a first assessment of the prospects for detecting neutrinos with POEMMA for commonly-invoked candidate astrophysical neutrino source classes given their current modeled neutrino fluences. As the multi-messenger network evolves and expands with the addition of next-generation detectors across the electromagnetic, gravitational-wave, and neutrino sectors, we envision that our methodology will provide a framework for evaluating the prospects of future experiments detecting neutrinos from candidate transient astrophysical sources, as well as for developing a more detailed survey strategy for space-based neutrino detectors such as POEMMA.

\vskip 0.20in

\noindent
{\bf Acknowledgements}
\vskip 0.20in
We would like to thank Roopesh Ojha and Elizabeth Hays for helpful discussions about AGNs and ToOs. We would also like to thank Francis Halzen and Justin Vandenbroucke for helpful discussions of IceCube's effective area and sensitivity. We would similarly like to thank Olivier Martineau-Huynh for helpful discussions of the \grandacro's effective area and Foteini Oikonomou for a careful reading of the manuscript and helpful comments. 
\begin{figure}[htb]
\centering
	\includegraphics[width=0.95\columnwidth, trim=5.5cm 10.25cm 4.9cm 10cm, clip]{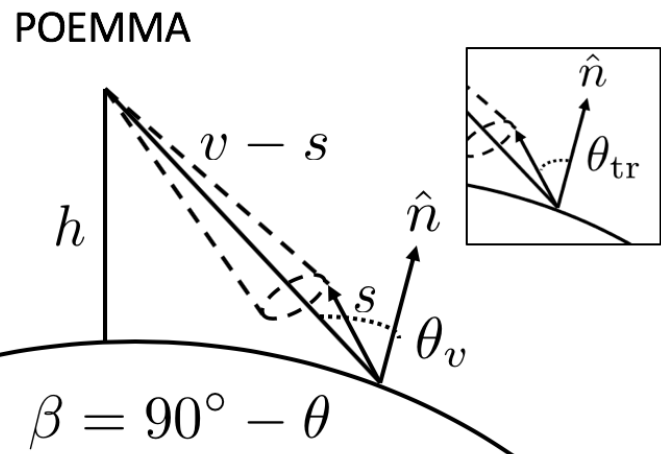}
	\caption{The effective area (dashed disk on the figure) for a \taon air shower that begins a path length $s$ from the point of emergence on the Earth. The local zenith angle of the line of sight, of distance $v$, is $\theta_v$. The inset shows the emergence angle of the \taon $\theta_{\rm tr}$.}
	\label{fig:geometry}
\end{figure}
We would also like to thank Kyle Rankin (New Mexico State University) for performing analytic and GMAT flight dynamics calculations used to quantify the satellite separation maneuvers, Simon Mackovjak for assistance with the NSB evaluation, and Austin Cummings for his simulation work, analysis, and discussion regarding the above-the-limb UHECR background assessment. We would also like to thank 	
Kenji Shinozaki for discussions of the impact of atmospheric refraction on Cherenkov signals above and below the Earth's limb. We would also like to thank our colleagues of the Pierre Auger and POEMMA collaborations for valuable discussions. This work is supported in part by US Department of Energy grant DE-SC-0010113, NASA grant 17-APRA17-0066, NASA awards NNX17AJ82G and 80NSSC18K0464, and the U.S. National Science Foundation (NSF Grant PHY-1620661). CG is supported by the Neil Gehrels Prize Postdoctoral Fellowship.

\vskip 0.25in

\appendix
\section{POEMMA detection for $\beta_{\rm tr}<35^\circ$}
\label{app:a}

\begin{figure}[htb]
\centering
	\includegraphics[width=0.99\columnwidth, trim = 3.0mm 2.3mm 2.5mm 0mm, clip]{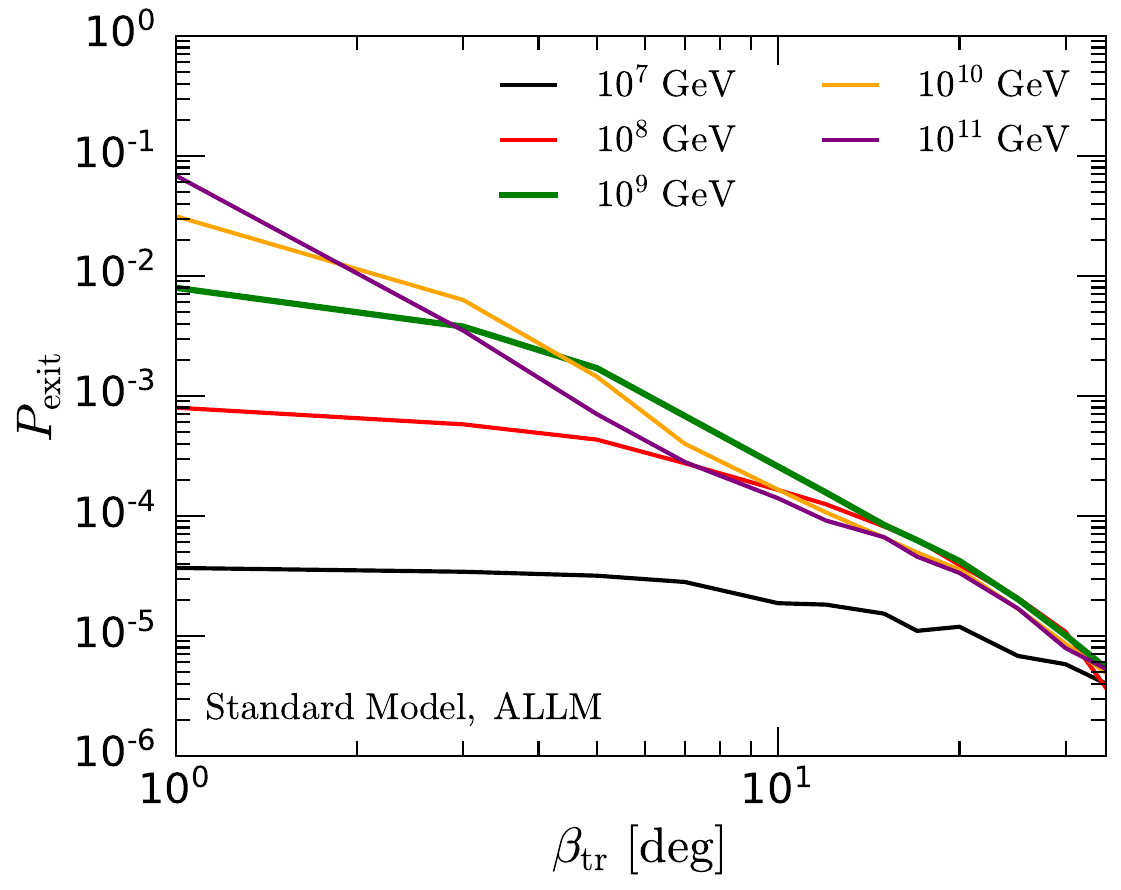}
	\caption{The exit probability for a $\nu_\tau$ of a given energy to emerge as a $\tau$-lepton as a function of elevation angle $\beta_{\rm tr}=1^\circ - 35^\circ$.}
	\label{fig:pexit}
\end{figure}

Many of the details required for the evaluation of the POEMMA effective area follow from the discussion of the sensitivity to the diffuse flux in Ref.~\cite{PhysRevD.100.063010}. Figure~\ref{fig:geometry} shows the configuration of POEMMA at altitude $h=525$~km and a \taon emerging at a local zenith angle $\theta_{\rm tr}$. In practice, we consider angles $\theta_{\rm tr}$ close ($\lsim \theta_{\rm Ch}^{\rm eff}\sim 1.5^\circ$) to the local zenith angle $\theta_v$ of the line of sight as required for detection of the showers. The difference in angles $\theta_{\rm tr}$ and $\theta_v$ in Fig.~\ref{fig:geometry} is exaggerated for clarity.

\begin{figure}[htb]
\centering
	\includegraphics[width=0.95\columnwidth]{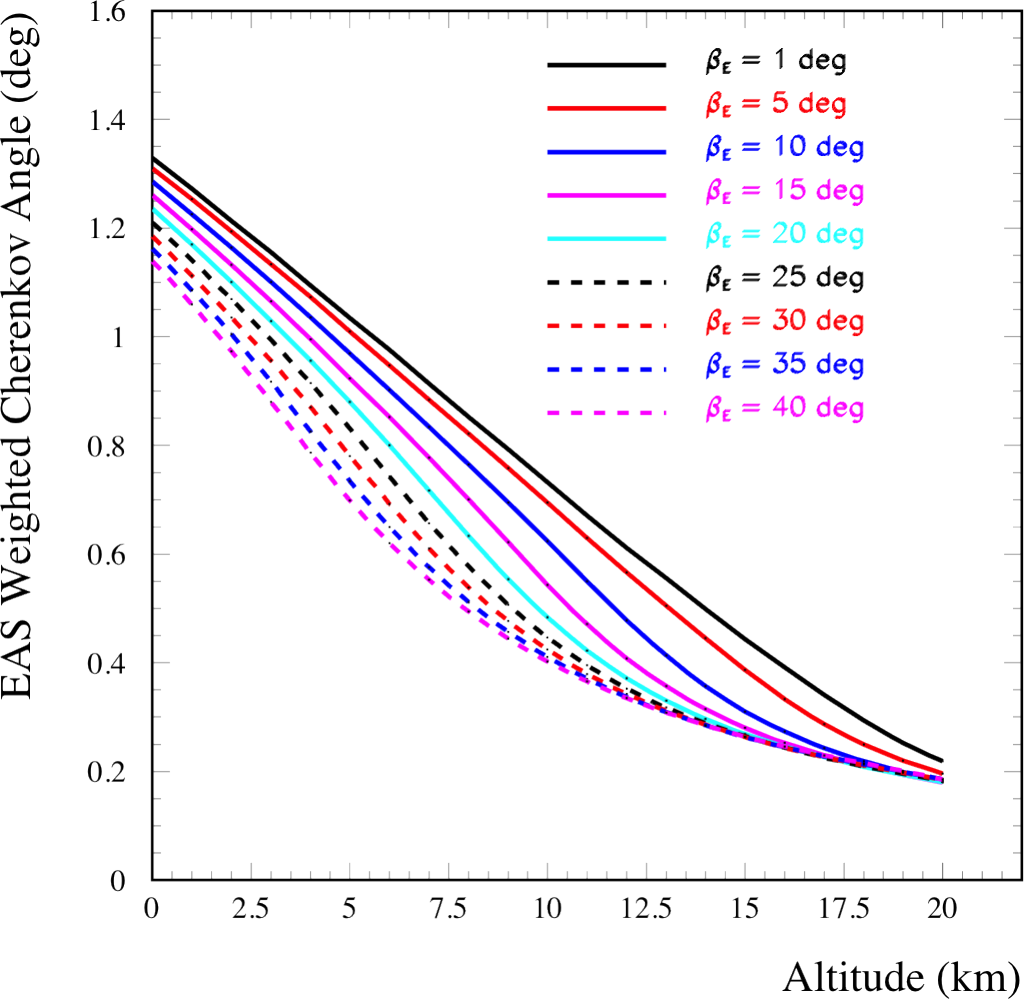}
	\caption{The effective Cherenkov angle of the air shower as a function of altitude of the \taon decay and elevation angle $\beta_{\rm tr}$ for an upward-moving $100$~PeV EAS.}
	\label{fig:Cang}
\end{figure}

For \taon air showers, it is common to use the local elevation angle to describe the trajectory rather than the local zenith angle. The elevation angles, labeled with $\beta$, are defined by angles relative to the local tangent plane, e.g., $\beta_{\rm tr}= 90^\circ -\theta_{\rm tr}$.

The \taon decay at a distance $s$ is viewable for decays within a cone of opening angle $\theta_{\rm Ch}^{\rm eff}$. 
The effective area  for the \taon air shower that begins $s$ from the point of emergence on the Earth is shown
by the dashed disk on the figure. The area of
the disk is expressed in Eq.~(\ref{eq:ACh}).

\begin{figure}[htb]
\centering
	\includegraphics[width=0.97\columnwidth]{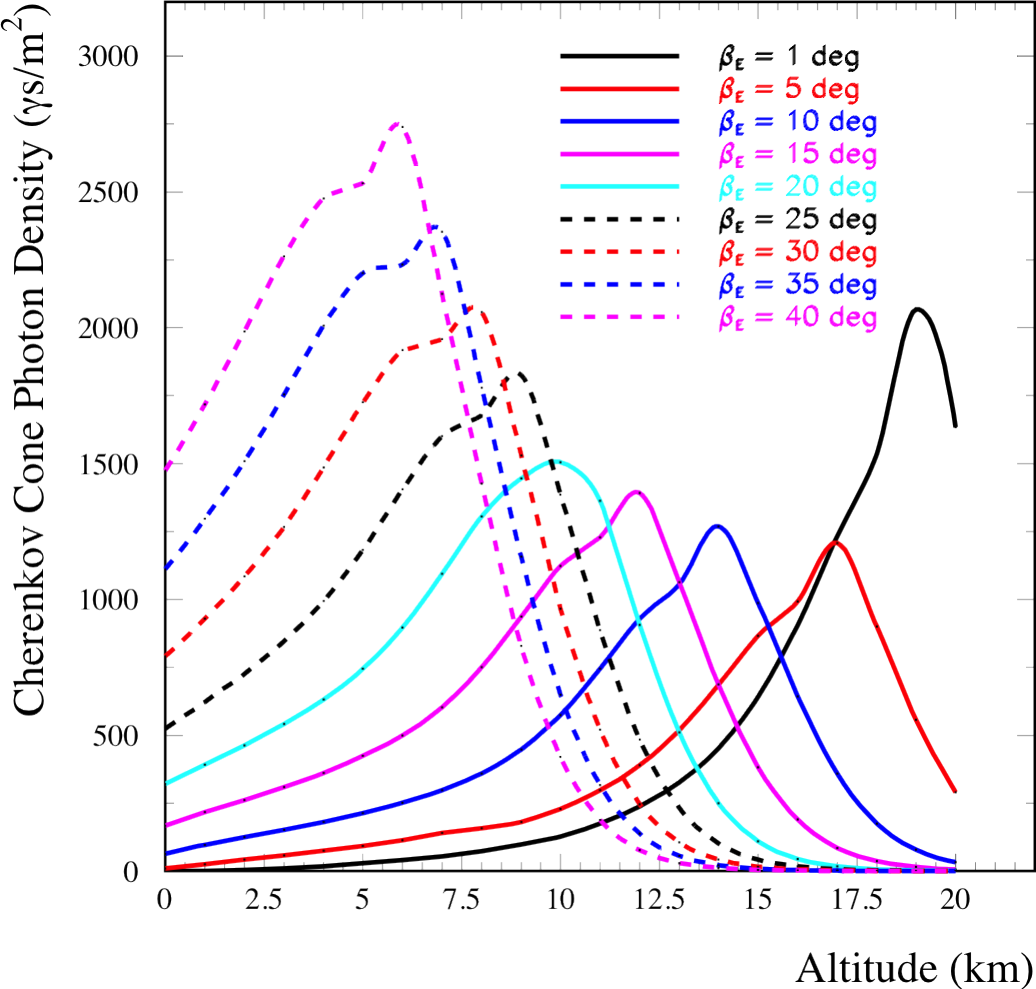}
	\caption{The photon density at POEMMA a function of altitude of the \taon decay and elevation angle $\beta_{\rm tr}$ for $100$~PeV upward-moving EAS.}
	\label{fig:photd}
\end{figure}

For the ToO neutrino sources, the slewing capabilities of POEMMA allow for a larger range of viewing below the limb, or alternatively, a larger range of elevation angles $\beta_{\rm tr}$. We show the \taon exit probability for angles up to $\beta_{\rm tr}=35^\circ$ in Fig.~\ref{fig:pexit}. Neutrino attenuation becomes increasingly important for larger $\beta_{\rm tr}$ and higher neutrino energies. Tau neutrino regeneration is included here, namely, multiple iterations of $\nu_\tau\to \tau$ production for weak scattering with nucleons, and $\tau\to \nu_\tau$ regeneration through decays.

Figures~\ref{fig:Cang} and \ref{fig:photd} are EAS parameter inputs to the detection probability calculated by a neutrino sensitivity Monte Carlo. They are derived from modeling of the upward EAS development, Cherenkov signal generation, and atmospheric attenuation of the Cherenkov signal (see Ref.~\cite{PhysRevD.100.063010}). The EAS development is modeled using shower-universality~\cite{1982JPhG....8.1475H,1982JPhG....8.1461H} and provides an average EAS profile for a given energy and $\beta_{\rm tr}$, with the assumption that $50$\% of the energy of the \taon goes into the EAS.  The Cherenkov angle is calculated from the modeling as a function of altitude and $\beta_{\rm tr}$, which is sampled in the POEMMA neutrino sensitivity Monte Carlo. The Cherenkov angle variations shown in Fig.~\ref{fig:Cang} are mainly due to the fact that the atmosphere density decreases as function of altitude, \textit{e.g.}, the index of refraction of air decreases as altitude increases, with an additional effect because EAS development at larger $\beta_{\rm tr}$ spans larger ranges of altitudes. The Cherenkov photon yield, shown in Fig.~\ref{fig:photd} for $100$~PeV EASs is more complicated. This is best illustrated by examining the variation in photon yield for EASs starting at sea level as a function of $\beta_{\rm tr}$.  At the lowest altitudes, the Cherenkov light attenuation is dominated by aerosol scattering due to the aerosol distribution having a scale height of $\sim 1$~km. As $\beta_{\rm tr}$ increases, a larger fraction of the EAS development occurs at higher altitudes where the aerosol contribution becomes smaller, thus leading to a larger Cherenkov photon density at $525$~km. This effectively leads to a lower energy threshold for tau-induced EAS detection for larger $\beta_{\rm tr}$. Note that the EAS Cherenkov (and fluorescence) light below $\sim 300$~nm is effectively eliminated by ozone attenuation when viewed from space. In regards to the altitude variation, for given $E$ and $\beta_{\rm tr}$ there is an altitude where the atmosphere becomes too rarefied to support EAS development. This leads to the turn over of the photon densities at higher altitudes shown in Fig.~\ref{fig:photd}.  Note that the neutrino sensitivity Monte Carlo effectively uses the results shown in Figs.~\ref{fig:Cang} and \ref{fig:photd} to generate the EAS signals for a specific \taon decay by interpolating the Cherenkov angle and photon density results to obtain those for a given \taon EAS geometry, with linearly scaling as a function of shower energy for the photon yield.

\section{POEMMA in ToO-stereo and ToO-dual modes}
\label{app:aprime}

The ability to reorient its neutrino detectors in a relatively short time makes POEMMA effective in its detection of transient neutrino sources. POEMMA's observing strategy employs a dual detection system: cosmic-ray detection mode for detecting fluorescence signals from cosmic ray interactions with stereo viewing at a satellite separation of $300$~km, and neutrino detection mode with a $25$~km separation when pointing to the Earth's limb so that both telescopes view the same Cherenkov light pool. Short neutrino bursts may occur when POEMMA is in cosmic-ray mode. In this appendix, we briefly describe considerations in changing the satellite separation to allow both telescopes to view the same Cherenkov light pool, and considerations in setting the PE threshold for short-duration neutrino bursts when the detectors, $300$~km apart, cannot view the same light pool.  These conditions, which we denote ToO-stereo when the two POEMMA satellites observe an event in the same Cherenkov light pool and ToO-dual when the satellites have a larger separation and measure the Cherenkov signals from a ToO source separately, have different energy thresholds because of the effects of the night-sky air glow background in the $300$ -- $900$ nm wavelength band. We conclude the appendix with a discussion of additional potential backgrounds for POEMMA ToO observations.

Once an external transient astrophysical event alert is received, the POEMMA satellites are designed to quickly slew, $90^{\circ}$ in $500$~s, to re-orient the POEMMA telescopes into to view near the limb of the Earth and optimize the orientation for the detection of tau neutrinos. The nature of the satellite orbits and the spacecraft avionics allow slewing maneuvers to occur with a negligible amount of propulsion, thus the number of slewing operations available over the mission is not limited by consumables such as propellant. The actual mode, e.g., ToO-stereo or ToO-dual, depends on the initial separation of the POEMMA spacecraft.  In the case that the satellite separation is $\lsim 50$ km, the slewing will put POEMMA into ToO-stereo mode. In the case POEMMA is in UHECR-stereo mode, with a satellite separation $\gsim 100$ km, the slews will put POEMMA into ToO-dual mode. The POEMMA spacecraft carry extra propulsion to perform satellite separation maneuvers during the mission. Flight dynamic studies have quantified the number of these available for the entire mission as a function of the separation distance and time scale of the maneuver.  Assuming a 300 km initial separation moving to a 25 km separation, and then back to the original 300 km studies show that the re-positioning can occur $\sim$40 times for the mission, assuming the time scale is $\sim1$ day to perform both separation changes.  
\begin{figure}[htb]
\centering
	\includegraphics[width=0.95\columnwidth, trim = 2.75mm 0mm 2.5mm 0mm, clip]{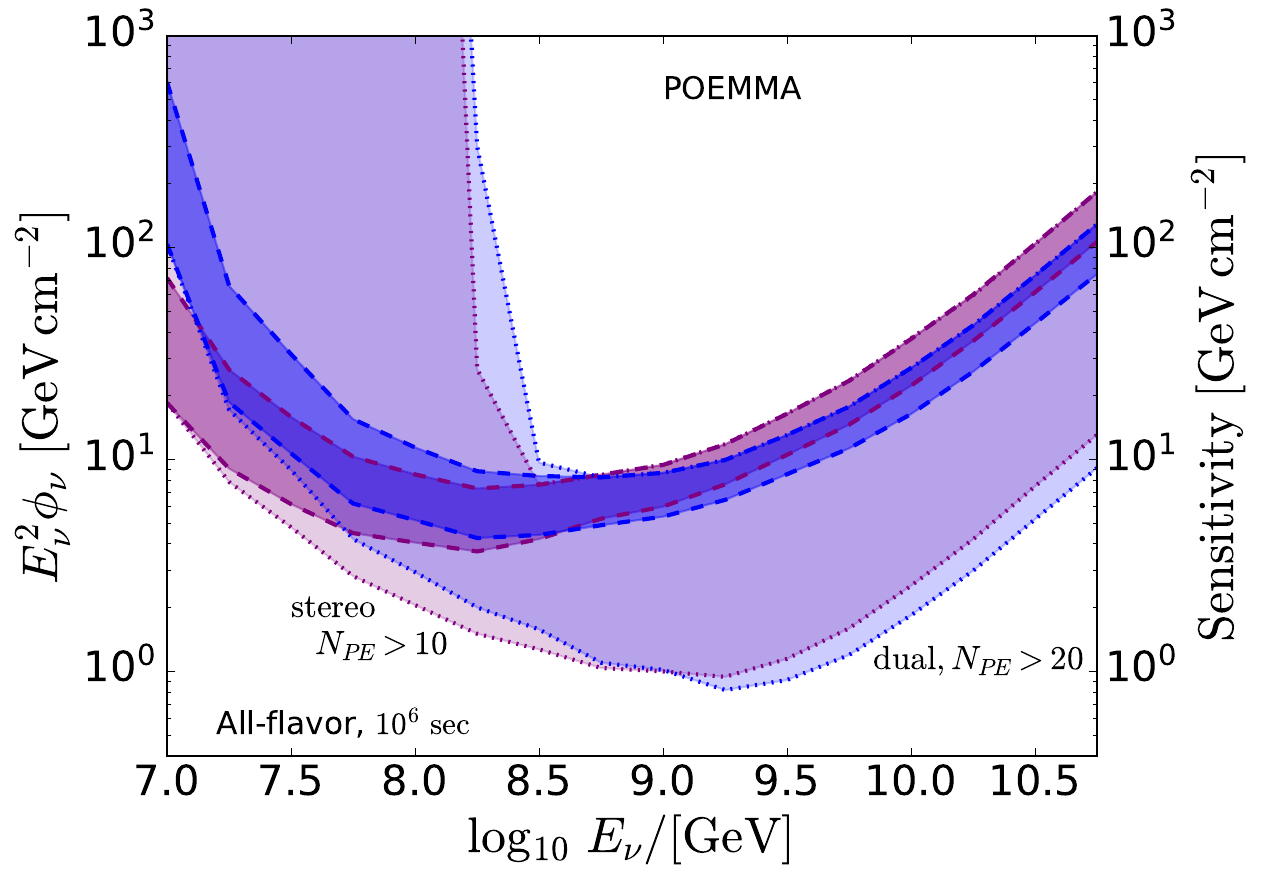}
	\includegraphics[width=0.99\columnwidth, trim = 2.75mm 0mm 2.5mm 0mm, clip]{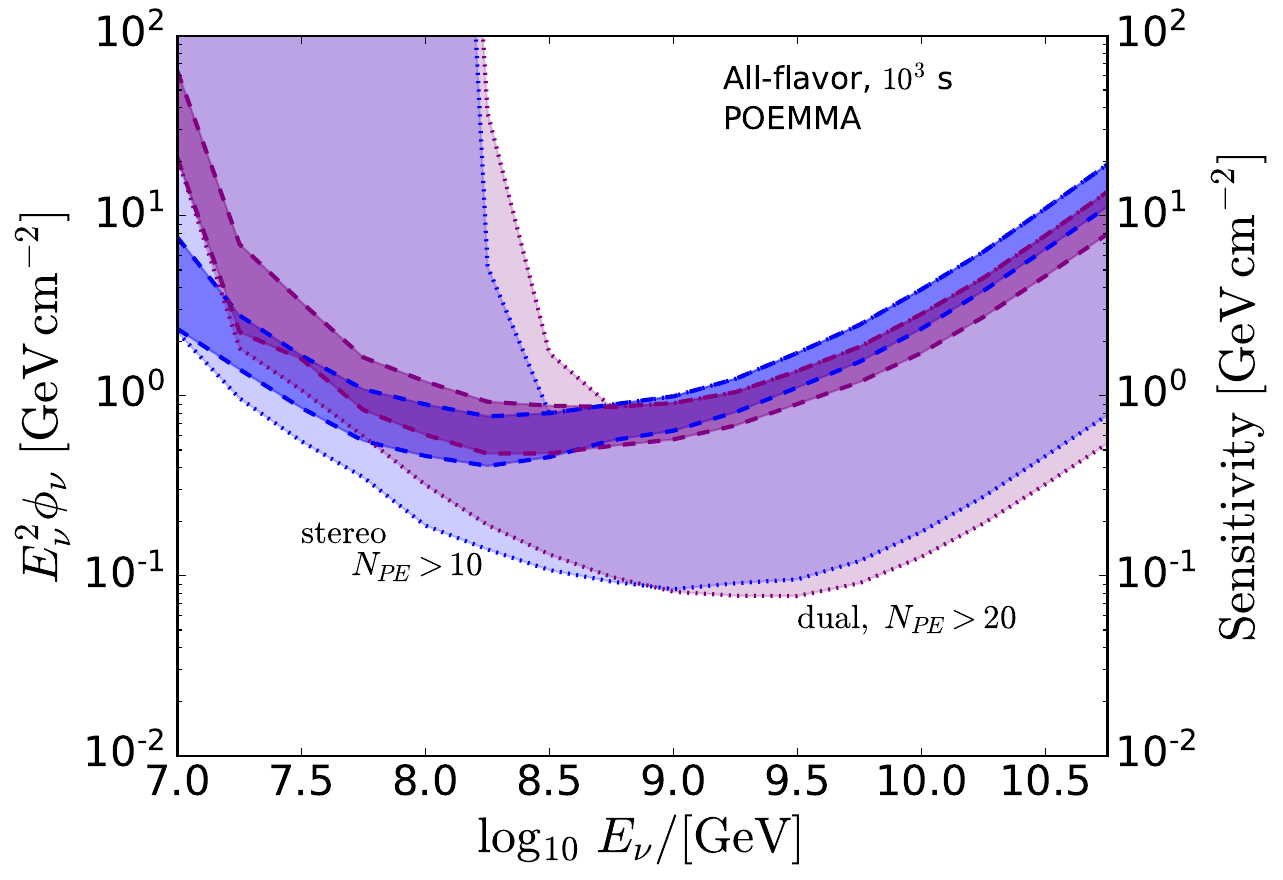}
	\caption{The POEMMA all-flavor $90$\% unified confidence level sensitivity per decade in energy with the default (purple) and alternate (blue) satellite configuration. \textit{Upper}: Sensitivity in the ToO-dual (blue) and ToO-stereo (purple) configurations for $10^{6}$~s burst, accounting for the effects of the Sun and the Moon ($f_t=0.3$; see Sec.~\ref{sec:2}). \textit{Lower}: Sensitivity in the ToO-dual (purple) and ToO-stereo (blue) configurations for $10^{3}$~s burst, assuming observations during astronomical night ($f_t=1$).}
		\label{fig:DualStereo}	
\end{figure}
If the duration for the initial maneuver to reduce satellite separation is reduced to $\sim 7$~hours, then $\sim 12$ maneuvers can be performed over the mission lifetime, assuming $1$~day to bring the satellites back to the $300$~km separation after the ToO observation. The altitude variation for the spacecraft performing the separation change is $500$ -- $550$ km, which has minimal effect on EAS signal (both Cherenkov and fluorescence) detection thresholds during the maneuver.

The performance for short- and long-duration ToO observations is determined in part by the flight dynamics performance. There is a benefit to bringing the two POEMMA spacecraft to a separation of $\sim 25$~km in order to put both instruments into the Cherenkov light pool. The nearly simultaneous measurement of the Cherenkov signal with both telescopes within a time spread of $\sim 20$~ns allows for a lower energy threshold for POEMMA by using coincidence timing to reduce the effects of the air glow background in the $300 - 900$~nm Cherenkov signal band.  Calculations using POEMMA's response to the Cherenkov signals, assuming $2.5$~m$^2$ effective telescope area, $20$\% PE 
conversion efficiency, pixel FoV of 0.084$^{\circ}$, assuming $20$-ns timing coincidence, and the average night-sky air glow background rate in the $300 - 900$~nm band have determined that a PE threshold of $10$ PEs yields a false positive rate of $\sim$ a fraction of an event per year~\cite{PhysRevD.100.063010}. For long bursts, characterized by time scales of $\sim 10^6$~s, we 
assume the satellites are in ToO-stereo mode and set
$N_{\rm PE}^{\rm min}=10$.

For short bursts, characterized by times scales of $\sim 10^3$~s, a lower PE threshold enabled by coincidence timing may not be achievable if the satellites are not already in ToO-stereo mode or POEMMA-limb viewing mode (satellites pointed towards the limb and $\sim 2^{\circ}$ above for diffuse neutrino and UHECR measurements and separated by $\sim 25$~km). In ToO-dual mode, even with a separation of $300$~km, POEMMA will still be able to detect neutrino signals, albeit at a higher PE threshold. We find that for the assumptions listed above, a PE threshold of $N_{\rm PE}^{\rm min}=20$ for POEMMA in ToO-dual mode will maintain a similarly low false positive rate.

To demonstrate the impact of the different PE thresholds on POEMMA's sensitivity, we plot the all-flavor neutrino sensitivity at the $90$\% unified confidence level in both the ToO-dual and ToO-stereo configurations for long and short bursts in Fig.~\ref{fig:DualStereo}. The purple shaded regions show our default values (ToO-stereo mode for long bursts and ToO-dual mode for short bursts), and the blue shaded regions show the PE threshold for the alternative configuration. At low energies, the lower PE threshold in ToO-stereo mode improves the sensitivity. At higher energies, the higher PE threshold of the ToO-dual configuration is somewhat mitigated by the doubled light-pool area. While we use the $N_{\rm PE}^{\rm min}=20$ threshold case for our short burst analyses, we note that if a short burst occurs when the POEMMA satellites are already in the ToO-stereo configuration, the sensitivity in the case of $N_{\rm PE}^{\rm min}=10$ would be applicable. The difference in PE thresholds corresponds to approximately an order of magnitude improvement in sensitivity at $10$~PeV.

\section{POEMMA's angular resolution and additional backgrounds for ToO observations}
\label{app:adoubleprime}

The angular resolution when observing the Cherenkov signal from an EAS is defined by the instantaneous field of view (iFoV), \textit{e.g.}, pixel angular span, of the optics of the POEMMA Cherenkov Camera (PCC). The iFoV of the PCC is $0.084^\circ$, which corresponds to a particular area on the ground monitored by the PCC for emergent EASs. When POEMMA is viewing near the Earth limb in ToO neutrino observation mode, the distance to the ground is $\sim 2000$ km, which yields the linear distance scale of $4$ km on the ground that is monitored for a given iFoV. As determined by simulation studies of the optical Cherenkov signal measurable by POEMMA for upward-moving \taon-generated EASs~\cite[\textit{c.f.},][]{PhysRevD.100.063010}, the viewed size of transverse component of the visible portion of the EAS is $< 1$ km. This implies that the \taon EAS Cherenkov signal will be confined to a single pixel in the PCC, even considering the point-spread-function (PSF) of the optics (see Fig. 3 in Ref.~\cite{PhysRevD.101.023012}).  Thus the direction to the observed Cherenkov EAS signal is known to iFoV ($0.084^\circ$) and with an RMS error of $0.084^\circ/\sqrt(12) \approx 0.024^\circ$.
The error on reconstructing the direction of an EAS trajectory also depends on the maximum viewing angle away from the trajectory, $\theta_v$ in Fig. \ref{fig:geometry}, where the Cherenkov signal is measurable.  This depends on the EAS development and the location in the atmosphere the EAS, which determines the Cherenkov angle. 
\begin{figure}[htb]
\centering
	\includegraphics[trim = 5mm 0mm 10mm -1mm, clip, width=0.95\columnwidth]{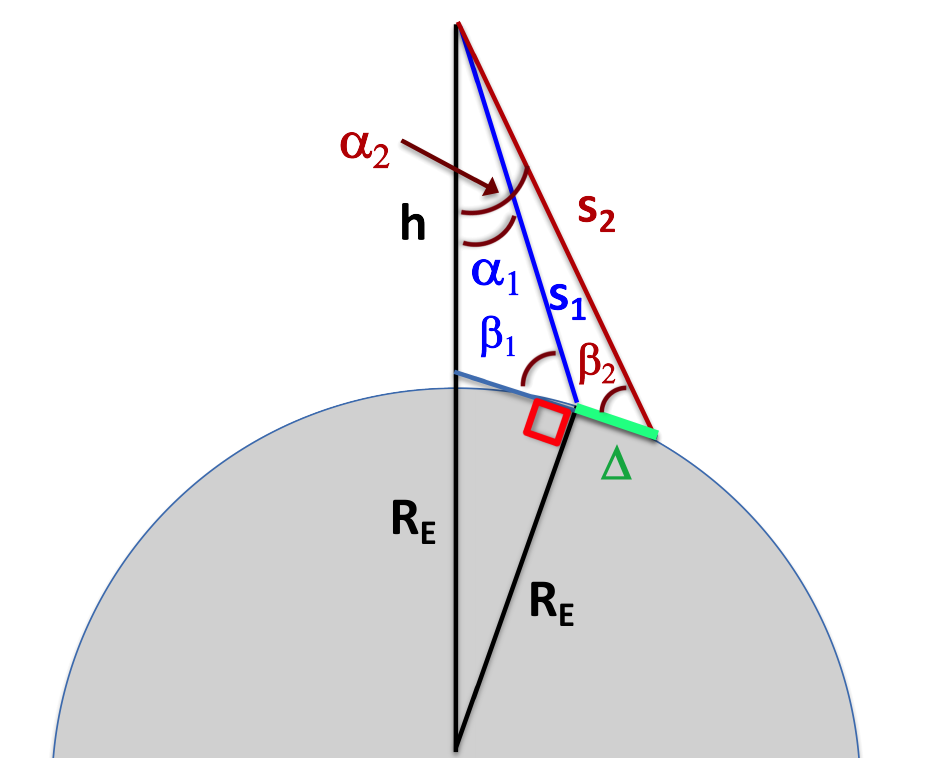}
	\caption{The geometry for a space-based detector at altitude $h$ detecting Cherenkov radiation from a downward-going UHECR EAS reflected by the ground. The span of EAS Cherenkov spot on the ground is $\Delta$, which has angular extent $\Delta\alpha = \alpha_2 - \alpha_1$ as viewed from the detector. $s_1$ and $s_2$ are the path lengths from the detector to the near side and far side of the Cherenkov spot, respectively. $\beta_1$ and $\beta_2$ are the elevation angles for $s_1$ and $s_2$, respectively.}
	\label{fig:reflgeo}
\end{figure}
For upward-moving $\tau$-lepton-induced EASs, simulations of optical Cherenkov signals measurable by POEMMA have shown that the maximum viewing angle from the EAS trajectory is determined the highest energy (brightest) events with $> 99\%$ satisfying $\theta_{\rm Ch}^{\rm eff} \le 3.0^\circ$. It should be noted that the maximum angle viewed away from the trajectory of the EAS is $\sim 1.2^\circ$ for $E_\nu \lsim 300$ PeV. Above 1 PeV, the direction of a \taon generated in a neutrino interaction is virtually colinear to that of the incident neutrino. Thus the error on determining the direction to the cosmic neutrino source is $\lsim 3.0^\circ$.

Aside from the night-sky air glow background, the other potential sources of background for POEMMA during ToO observations are due to the cosmic diffuse neutrino flux and background signals from the UHECR flux. For the diffuse neutrino flux, we can estimate the expected number of background events using the IceCube differential $90\%$ confidence upper limit for energies $\gtrsim 5$~PeV~\cite{Aartsen:2018vtx}. Based on this differential limit, and taking the assumed timescale for a long observation ($10^6$~s) and the effective Cherenkov angle for the highest energy events ($\lsim 3.0^{\circ}$), we expect $2.0 \times 10^{-4}$ background events during such a ToO observation. 

Several factors result in the above background estimates being quite \textit{conservative}. First, the limitations in IceCube's sensitivity above $10$~PeV result in an upper limit that becomes less constraining with energy, resulting in a larger assumed background flux at higher energies. If instead, we extrapolate the IceCube best-fit diffuse astrophysical muon-neutrino spectrum (through-going muon neutrinos from the 9.5-yr Northern-hemisphere data, assuming equal numbers of tau neutrinos; \citenum{Stettner:2019tok}), this corresponds to $4.0 \times 10^{-5}$ background events per long ToO observation for the Cherenkov angle of $3.0^{\circ}$. Second, the assumed Cherenkov angle of $3.0^{\circ}$ is only valid at the highest energies of the energy range relevant for POEMMA; at lower energies, the Cherenkov angle will be smaller, $\lsim 1.5^{\circ}$.

\begin{figure}[htb]
\centering
	\includegraphics[trim = 43mm 13mm 43mm 15mm, clip, width=0.95\columnwidth]{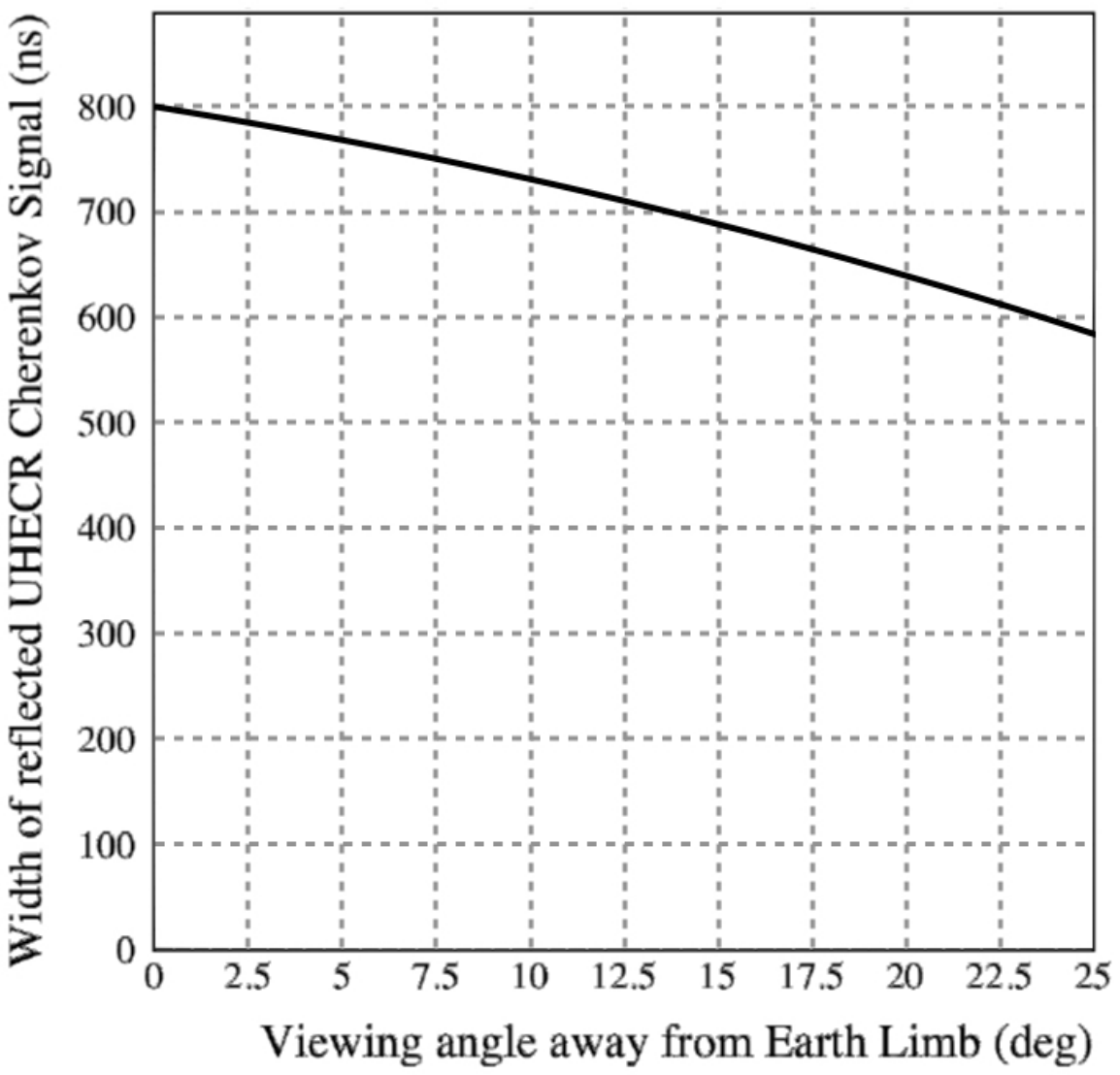}
	\caption{The width of the pulse from a ground-reflected UHECR Cherenkov signal over the range of angles (as measured from the Earth's limb, \textit{i.e.,} $\alpha_H - \alpha$, where $\alpha_H = \arcsin(R_E/\left(R_E + h\right)
	)\simeq 67.5^{\circ}$ is the nadir angle of the Earth's limb for a detector at altitude $h = 525$~km) at which POEMMA will view ToO astrophysical neutrino sources.}
	\label{fig:tspread}
\end{figure}

We expect two possible contributions from UHECRs to the background for ToO observations: (i) UHECR Cherenkov signals reflected off of the ground, and (ii) Cherenkov signals generated by above-the-limb UHECRs during ToO observations close to the Earth's limb. First we discuss the reflected Cherenkov signals from downward-going UHECR EASs. As detailed in the pivotal works of Patterson and Hillas \cite{1983JPhG....9.1433P,1983JPhG....9..323P} the Cherenkov lateral distribution (CLD) generated by a downward-moving UHECR EAS is a filled disk with diameter $\Delta \approx 250$ meters and with power law tails. While the amount of Cherenkov light collected within the disk is proportional to the energy of the UHECR, the value of the disk diameter is relatively insensitive to the UHECR energy, nuclear composition, altitude (at least to $\sim$5.2 km on Mount Chacaltaya \cite{2014NIMPA.763..320T}, and rather insensitive to the UHECR incidence angles. This finite and nearly constant width of the UHECR reflected Cherenkov pulse sets a minimum time scale of $\gsim 600$ ns for the observation of the signal regardless on the nature of the reflection (see Fig.~\ref{fig:tspread}), either Lambertian or specular, when the POEMMA Earth viewing constraints are considered.

As detailed in Ref.~\cite{PhysRevD.100.063010} for a space-based neutrino detector, the angle away from nadir, defined as $\alpha$, at which the detector points corresponds to the specific viewing angle on the ground; see Fig. \ref{fig:reflgeo})\footnote{As in Sec.~\ref{sec:2}, we take $\beta \simeq \beta_{\rm tr}$, where $\beta_{\rm tr}$ is the elevation of the particle or signal trajectory and must be within $\theta^{\rm eff}_{\rm Ch}$ in order to be detectable. Monte Carlo simulations have demonstrated that taking $\beta \simeq \beta_{\rm tr}$ is a good approximation to the more detailed evaluation in which $\beta \neq \beta_{\rm tr}$.} at which the detector will be able to detect Cherenkov signals:
\begin{equation}
    \cos \beta = \frac{R_E + h}{R_E}\sin \alpha \,,
\end{equation}
where $h$ is the altitude of the detector. 
For reflected UHECR Cherenkov signals to be observed by POEMMA, the Cherenkov signal will have to hit the ground at an angle that is within the range of viewing angles seen by the instrument. Since the angles are large ($48^\circ \lsim \alpha \le 67.5^\circ$ and bounded by the Earth limb), the duration of the Cherenkov pulse is extended in time due to the 250 m diameter of the CLD. Based on this geometry and assuming the CLD is generated instantaneously, the relation for the pulse duration is given by:
\begin{equation}
    \Delta t = \frac{s_2 - s_1}{c}\,
\end{equation}
where 
\begin{equation}
    s_1 = \left(R_E + h \right)\cos \alpha_1 - R_E \sin \beta_1\,,
\end{equation}
and
\begin{equation}
    s_2 = \sqrt{s^2_1 + \Delta^2 + 2 s_1 \Delta \cos \beta_1}\,.
\end{equation}
In Fig.~\ref{fig:tspread}, we plot the duration of the ground-reflected UHECR Cherenkov pulse as a function of POEMMA's viewing angle away from the limb of the Earth. The figure shows that at the viewing angles relevant for observations of transient astrophysical neutrino sources (for elevation angles up to $\beta = 35^{\circ}$, corresponding to viewing angles of up to $\sim 18^{\circ}$ away from the limb at $\alpha_{\rm Hor} =67.51^\circ$ as viewed by POEMMA at an altitude of $525$~km), the pulse widths for ground-reflected UHECR Cherenkov signals are $\gsim 600$ ns, which are much longer than the $\sim 20$~ns spread we expect from upward from \taon EASs.  Zenith angle effects will increase the time width by hundreds of ns. Detailed UHECR simulations show that this geometric argument is conservative and the time span of the reflected UHECR signal is $\gsim 1~\mu$sec (see Fig. 1 in Ref. \cite{2014AdSpR..53.1515B}.) In general, cloud height distributions are bi-modal with most probable values around 3 km and 15 km \cite{2014ClDy...43.1129Z}. Thus, the effects from scattering from low clouds are similar to that from the ground based on the measurements on Mount Chacaltaya \cite{2014NIMPA.763..320T}. Clouds above $\sim 10$ km altitude will not generate a reflected UHECR signal since the majority of the EAS develops at lower altitudes, due to the exponential nature of the atmosphere, with shower maximum $\sim 6$ km and is well below the most probable value for high clouds. As such, background events arising from UHECR Cherenkov signals reflected off of the ground or low clouds will be easily distinguishable from neutrino events.

In the case of reflections off of clouds, POEMMA's design\footnote{For more information on the design of POEMMA, see the NASA Astrophysics Probe study report~\cite{POEMMAConcept}} includes an atmospheric monitoring system consisting of two infrared cameras on each satellite that will allow real-time monitoring and analysis of cloud coverage. As such, this system will allow for rejection of background reflected signals via selection cuts for events that appear to originate from clouds. This would result in a slight reduction in exposure related to observing conditions, which has not been included in our calculations.

The second UHECR background to consider is the direct Cherenkov signals generated from UHECRs from above the Earth's limb that are observable during POEMMA observations near the Earth's limb. While the  modeling of the signals from these above-the-limb events is beyond the scope of this paper (and merits a paper on its own), we make a geometrical argument to provide a conservative evaluation of the impact to the ToO neutrino sensitivity. Initial stimulation studies have shown that the attenuation of the Cherenkov signal from these events constrains their visibility by POEMMA to a viewing angle $\sim 0.05^\circ$--$0.1^\circ$ above the limb. However, the atmosphere becomes too rarefied to generate an EAS around a viewing angle $\sim 1^\circ$ above the limb. Thus the acceptance for any observable above-the-limb UHECRs is constrained a narrow angular range. However, atmospheric refraction of the Cherenkov light will lead to the condition that the {\it above-the-limb} UHECR signal will appear as a {\it below-the-limb} signal mimicking that from a tau neutrino event. 
\begin{figure}[htb]
\centering
	\includegraphics[trim = 50mm 20mm 50mm 20mm, clip, width=0.95\columnwidth]{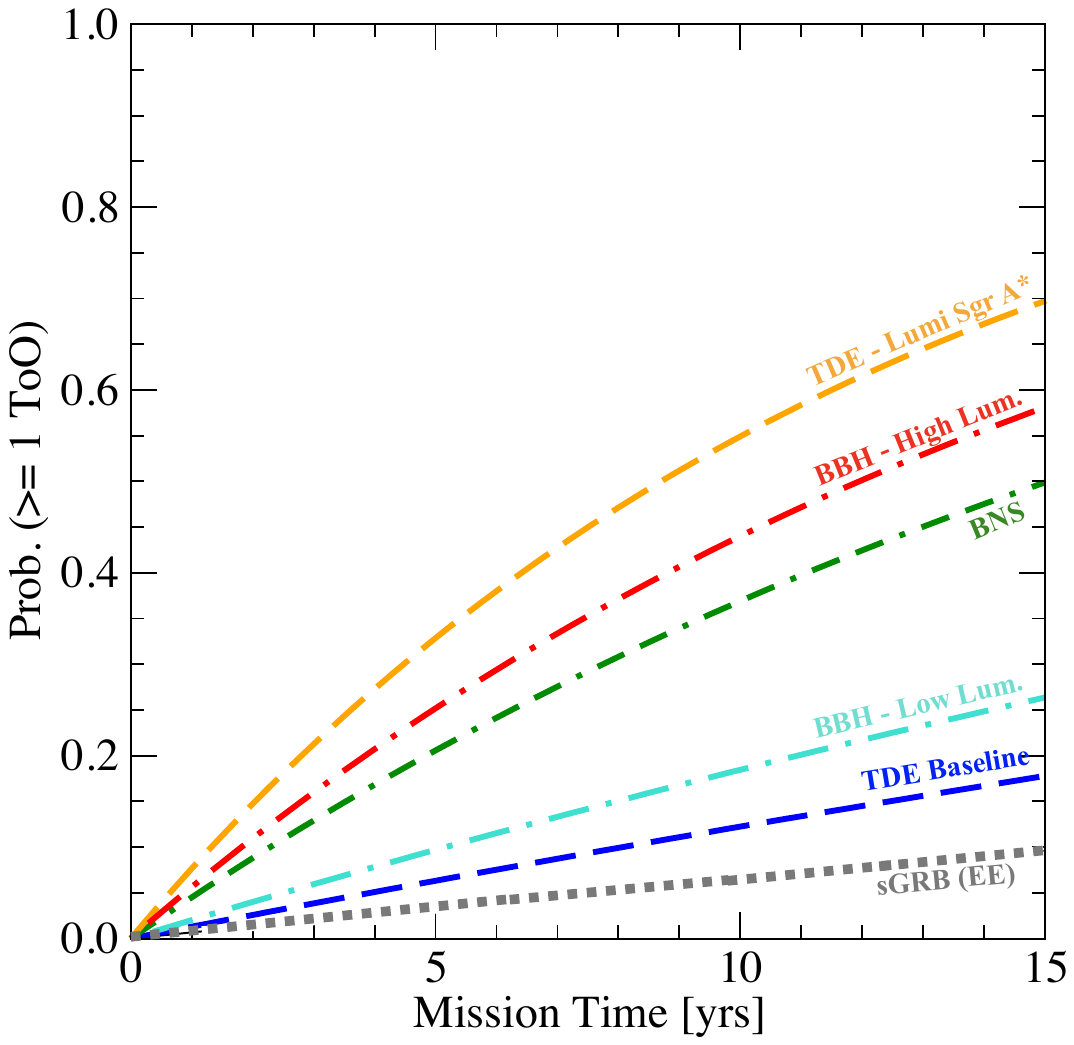}
	\caption{The Poisson probability of POEMMA observing at least one ToO versus observation time for the source classes featured in Fig.~\ref{fig:poisson} but assuming a cutoff in the modeled neutrino spectra at $10^9$~GeV.}
	\label{fig:poissonHEC}
\end{figure}
A Cherenkov signal even for the highest energy ($\gsim 1$~EeV) UHECRs is limited by atmospheric attenuation to begin to appear $0.05^\circ - 0.1^\circ$ above the Earth's limb. This range of angles above the limb is refracted by an amount $1^\circ - 0.75^\circ$, respectively ~\cite{1983ApOpt..22..721C}, such that they appear to originate $\lesssim 1^{\circ}$ below the limb. We can take a conservative approach and calculate the ToO neutrino sensitivities with the constraint that we only perform observations using an {\it observed} viewing angle $\geqslant 1^\circ$ (not taking into account atmospheric refraction) below the Earth-limb, or an Earth-emergence angle $\beta_{\rm tr}\gsim 7^\circ$. Atmospheric refraction has a similar impact on Cherenkov signals from \taon generated EASs, causing signals from EASs emerging $\lesssim 1^{\circ}$ from the limb to appear to originate from $\sim$ few tenths of a degree farther away from the limb. For these signals, the actual Earth-emergence angle for the \taon is $< 7^\circ$, reinforcing the conservative nature of this approach. 

In constraining the viewing angle during ToO observations as described above, we find that POEMMA's ToO sensitivities diminish to some extent for energies $\gtrsim 10^9$~GeV, while being preserved for energies below this scale. The resulting impact on POEMMA's capability to detect neutrinos depends on the predicted neutrino fluence for a given model; however, even for those models considered here that predict substantial amounts of neutrinos above $10^9$~GeV (e.g., BNS and BBH merger scenarios), the constraint on the viewing angle amounts to a modest reduction ($\sim 25$\%) in the number of neutrinos POEMMA would detect. For those models in which the neutrino spectrum falls off above $10^9$~GeV, the reduction amounts to $\lesssim$ few percent. To illustrate the impact on the prospects of POEMMA detecting a ToO, Fig.~\ref{fig:poissonHEC} plots the Poisson probability accounting for the decline in ToO sensitivity. The plot shows that POEMMA still has a $\gtrsim 10$\% chance of detecting a ToO during its $3$--$5$~year mission lifetime for all of the source classes previously identified as the most promising in Section~\ref{sec:3c}. We note that the upcoming flight of the Cherenkov telescope in the EUSO-SPB2 experiment \cite{2020NIMPA.95862164S} will provide key measurements of this and other backgrounds.

\section{Cosmological Fluences}\label{app:b}

\begin{figure*}[htb]
\centering
	 \includegraphics[trim = 27mm 43mm 20mm 40mm, clip, width=1.0\columnwidth]{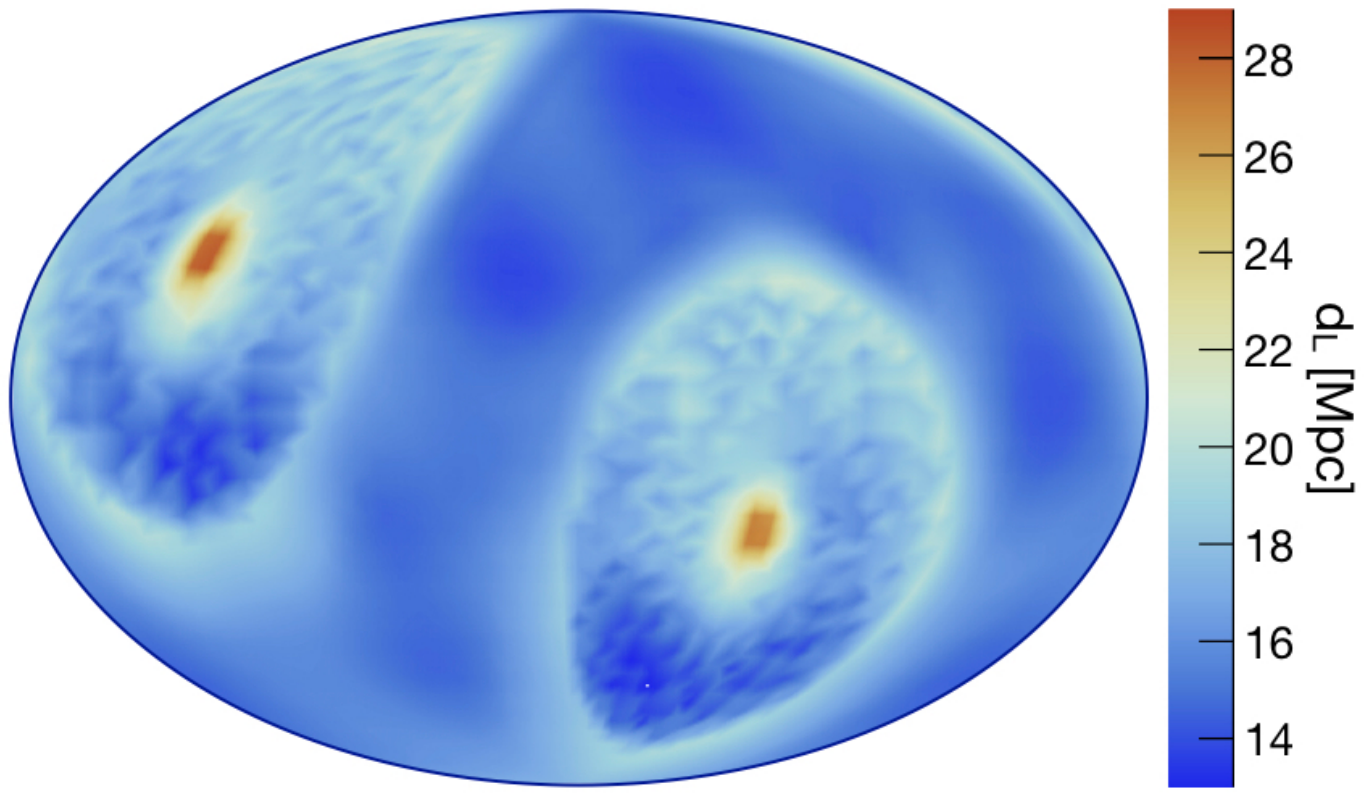} \includegraphics[trim = 27mm 43mm 15mm 40mm, clip, width=1.0\columnwidth]{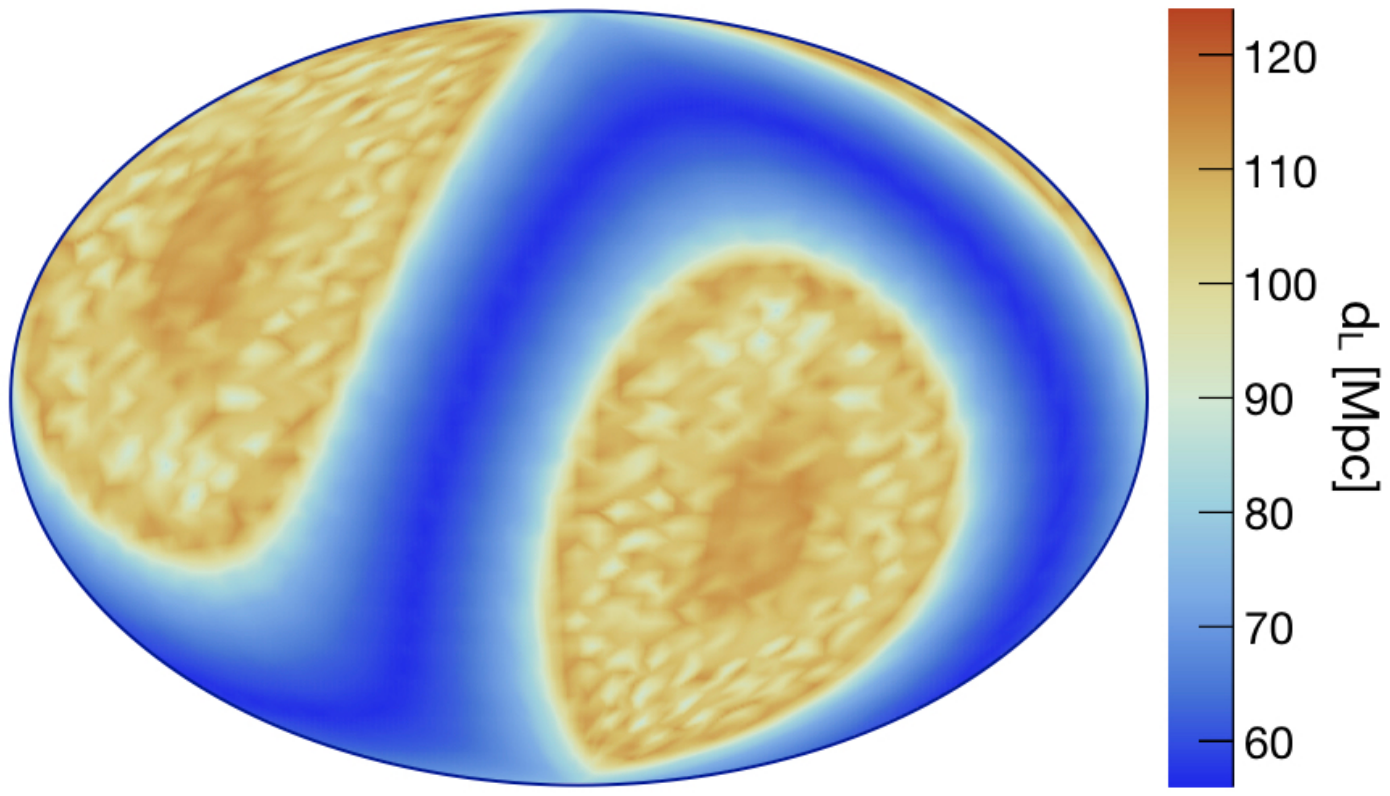}
	\caption{\textit{Left}: Sky plot of the neutrino horizon for the NS-NS merger model of Ref.~\cite{Fang:2017tla}. \textit{Right:} Same as at left for the sGRB EE neutrino model of Ref.~\cite{Kimura:2017kan}.}
	\label{fig:fmhorizon}
\end{figure*}

For $\Omega_k =0$, the comoving transverse distance $d_M$ is equivalent to the line-of-sight comoving distance
\begin{equation}
d_C = \frac{c}{H_0} \int_0^z \frac{dz'}{E(z')} \, ,
\end{equation}
i.e., $d_C = d_M$~\cite{Hogg:1999ad}. The luminosity distance $d_L$  is defined by the relationship between bolometric (i.e., integrated over all frequencies) energy-flux $S$ and bolometric luminosity $L$:
\begin{equation}
d_L = \sqrt{\frac{L}{4 \pi S}} \, .
\end{equation}
From Eq.~\ref{dL}, $d_L$ is related to $d_M$ by
\begin{equation}
d_L = (1+z) d_M \, .
\end{equation}

While sources often do not emit isotropically, we consider fluences based on \textit{isotropic equivalent} quantities. 
With this in mind, the total neutrino fluence at a line-of-sight distance $d_M$ can be written as
\begin{equation}
\phi_\nu (E_\nu) = \frac{d^2N_\nu}{dE_\nu \, dA_{\rm sph}} \,,
\end{equation}
where $A_{\rm sph}$ is the spherical area of radius $d_M$. 
The number of neutrinos crossing the area $A_{\rm sph}$ is then given by
\begin{equation}
    N_\nu =   4\pi d_M^2 \ \phi_\nu (E_\nu) \ \Delta E_\nu \, .
\end{equation}
On the other hand, the number of emitted neutrinos in a time interval $\Delta t_{\rm src}$ is found to be
\begin{equation}
N_{\rm src} = Q (E_{\rm src})  \, \Delta t_{\rm src} \, \Delta E_{\rm src} \, ,
\end{equation}
where $Q (E_{\rm src})$ is the (all-flavor) neutrino source emission rate and $E_{\rm src}$ indicates the emission energy.
Setting the number of neutrinos distributed over the sphere of area $A_{\rm sph}$ equal to the number of emitted neutrinos and re-arranging to isolate the fluence at the observation distance $d_M$, we obtain
\begin{equation}
\phi_\nu = \left(\frac{1}{4\pi d_M^2}\right) \ Q(E_{\rm src}) \ \Delta t_{\rm src} \ \frac{\Delta E_{\rm src}}{\Delta E_\nu}\,.
\end{equation}
Accounting for the redshift $z$, the energy scales as $E_{\rm src} = \left(1 + z\right)E_\nu$, 
and therefore the energy-squared scaled fluence at the observation point is
\begin{equation}
E^2_\nu \ \phi_\nu  = \frac{\left(1 + z\right)}{4\pi d^2_{L}} \ E^2_{\rm src} \ Q(E_{\rm src}) \  \Delta t_{\rm src}\,.
\label{Esquarephi}
\end{equation}
Finally, dividing Eq.~(\ref{Esquarephi}) by 3 to account for the fact that only 1/3 of the emitted neutrinos are of tau flavor we obtain the desired result displayed in Eq.~(\ref{eq:phiobs}). 
As such, for any model that provides an observed fluence and a source redshift or luminosity distance, one can determine $E^2_{\rm src} \, Q(E_{\rm src}) \,  \Delta t_{\rm src}$. We use Eq.~(\ref{eq:phiobs}) to calculate the observed single-flavor neutrino fluence at \textit{any} redshift $z$. The maximum redshift at which we can see the event, $z_{\rm hor}$, is the redshift at which $N_{\rm ev}$ in Eq.~(\ref{eq:numevents}) is equal to $1.0$. To provide a sense of how the variation in POEMMA's sensitivity with celestial position impacts the neutrino horizon, Fig.~\ref{fig:fmhorizon} provides skyplots of the neutrino horizons for one long-duration model and one short-duration model.
      
\section{Other Detectable Transient Source Classes}\label{app:c}

\sideheader{Blazar Flares} Active galactic nuclei (AGNs) are the most luminous persistent sources in the universe, powered by accretion of highly magnetized plasma onto SMBHs that can launch powerful relativistic jets. As they possess the characteristics necessary to accelerate particles to ultra-high energies (\textit{i.e.,} magnetic field strengths and spatial scales required to confine particles until they reach energies $\gtrsim 10^{18}$~eV; see \textit{e.g.}, [\citenum{1984ARA&A..22..425H,2011ARA&A..49..119K}]), AGN jets have long been proposed as candidate sources of the highest energy cosmic rays~\cite{Biermann:1987ep,1992PhRvL..69.2885P} with discussions of neutrino production having as long a history~\cite[see \textit{e.g.},][]{1978ApJ...221..990M,1979ApJ...232..106E,1989A&A...221..211M,1990ApJ...362...38B,1991PhRvL..66.2697S,1994APh.....2..375S,2001PhRvL..87v1102A,2003APh....18..593M,2010ApJ...721..630H,2013ApJ...768...54B,2013APh....43..155H,2014APh....54...61D,2014PhRvD..90b3007M,2015MNRAS.448.2412P,Romero:2016hjn,Rodrigues:2017fmu,Reimer:2018vvw,2018ApJ...864...84K,2019NatAs...3...88G,2019ApJ...876..109Z,2019MNRAS.483L..12C,Anchordoqui:2007tn,deBruijn:2020pky}. The recent IceCube detection of a high-energy neutrino ($E \gtrsim 300$~TeV) temporally and spatially coincident with a gamma-ray flare from blazar TXS~0506+056~\cite{IceCube:2018dnn} and the identification of a prior neutrino flare from the same source~\cite{IceCube:2018cha} provided the strongest evidence to date that AGNs produce neutrinos, as well as providing the first clues into the origins of the astrophysical neutrino flux and hints into the acceleration of hadrons to very-high energies and possibly beyond. 

For the purposes of this study, we consider a pure proton CR injection model with advective escape from Rodrigues, Fedynitch, Gao, Boncioli and Winter (RFGBW)~\cite{Rodrigues:2017fmu} for high-luminosity FSRQs. Based on the methodology presented in Sec.~\ref{sec:3}, we find that POEMMA's neutrino horizon for this model is $\sim 43$~Mpc. It is worth noting that the closest FSRQ in the Third Catalog of Hard \textit{Fermi}-LAT Sources (3FHL;~\citenum{TheFermi-LAT:2017pvy}) with a measured redshift is at a distance of $\sim 450$~Mpc. Expanding the search to include ``misaligned'' FSRQs (i.e., considering the whole parent population of Fanaroff-Riley Class II radio galaxies), the closest source in the First Catalog of FR II radio galaxies (FRIICAT;~\citenum{Capetti:2017fjb}) is at a distance of $\sim 200$~Mpc, though the sample size of the entire catalog is small ($122$ sources). As such, according to the RFGBW model and our analysis, we do not expect POEMMA to be able to detect neutrinos from FSRQ flares.

It is worth mentioning that we focus on FSRQs in this analysis because, as found by RFGBW, their photon field densities are high enough to result in efficient neutrino production, whereas less luminous blazars, such as BL Lacs, with lower photon field densities are typically not expected to efficiently produce neutrinos~\cite[see also,][]{PhysRevD.90.023007}. However, the first claimed astrophysical neutrino source, TXS 0506+056, has been classified as a BL Lac~\cite{Paiano:2018qeq}, leading some members of the high-energy community to revisit previously held assumptions regarding neutrino production in BL Lacs~\cite[\textit{c.f.},][]{Righi:2018xjr}. On the other hand, the classification of TXS 0506+056 as a BL Lac rather than an FSRQ has been called into question due to its multiwavelength properties and inferences about its Eddington ratio~\cite{Padovani:2019xcv}. Regardless, the closest BL Lac with measured redshift in the 3FHL catalog is at a distance of $\sim 130$~Mpc, though expanding the search to ``misaligned'' sources provides a handful of sources within $\sim 100$~Mpc, including well-known nearby radio galaxies such as Centaurus A ($\sim 4$~Mpc) and M87 ($\sim 20$~Mpc). If we assume lower neutrino fluences from BL Lacs, consistent with expectations prior to the TXS 0506+056 event, POEMMA's neutrino horizon for these sources should be quite a bit less than for FSRQs. Allowing for relativistic beaming in the case of ``misaligned'' sources would make detecting neutrino flares from even Centaurus A or M87 challenging. 

As a final consideration, it is worth pointing out that regardless of the classification for TXS 0506+056, its measured redshift is $z = 0.34$ corresponding to a luminosity distance of nearly $2$~Gpc. Based on the RFGBW FSRQ model and our analysis, we would expect IceCube's neutrino horizon to be $\sim 25$~Mpc; hence, IceCube's detection of a neutrino event associated with TXS 0506+056 is in tension with expectations of neutrino fluences for even FSRQs in this model. As such, if any kind of blazar produces neutrinos, the questions of the physics of neutrino production and which types of blazars produce them are very much open in light of the TXS 0506+056 event. Thus, in our view, the current landscape is far too uncertain to allow even a rough assessment of the prospects for POEMMA detecting neutrinos from a flaring blazar.

\noindent --- \textit{\textbf{Binary White Dwarf Mergers Mergers}} ---

\noindent In addition to BNS merger events and core-collapse supernovae, rapidly spinning magnetars can be produced by BWD mergers, making such mergers promising events for UHECR production~\cite{Piro:2016jaq}. Small amounts of surrounding material ($\sim 0.1 M_{\odot}$) allows UHECRs to escape the system more easily than in magnetars formed in core-collapse supernovae \cite{Piro:2016jaq}; on the other hand, the limited amount of surrounding material leads to lower neutrino fluxes \cite{Piro:2016jaq}. Alternatively, the magnetorotational instability that can develop in the debris disk surrounding the magnetar can lead to the formation of a hot, magnetized corona and high-velocity outflows \cite{Ji:2013sda,Beloborodov:2013kpa,2015ApJ...806L...1Z,Xiao:2016man}. Magnetic reconnection can accelerate cosmic rays that would interact with outflow material and radiation to produce high-energy neutrinos as modeled by Xiao et al. (XMMD) in Ref.~\cite{Xiao:2016man}. We adopt the XMMD model to determine the sensitivity of POEMMA to neutrinos from BWD mergers. The modeled neutrino fluences are very low -- for an event that occurs at the GC, we expect POEMMA to detect on the order of $20$ neutrinos, which is a substantially lower number than predicted by any of the other models. In fact, in order for POEMMA to detect neutrinos from these events, the source would have to be within the Galaxy. Based on an event rate provided in Ref.~\cite{Xiao:2016man} (see also Ref.~[\citenum{2012ApJ...749L..11B}]), which is comparable to the Type Ia supernova rate, we expect a ToO rate that would require POEMMA to operate for longer than a typical mission lifetime in order to detect one such event.

\sideheader{Non-jetted Tidal Disruption Events} In addition to launching relativistic jets, accretion processes in TDEs can also give rise to AGN-like winds \cite{2005ApJ...628..368O,2015ApJ...805...83M,2015MNRAS.454L...6M} and/or colliding tidal streams \cite{1999ApJ...519..647K,2016ApJ...830..125J} that could provide the conditions for accelerating protons and nuclei~\cite{Tamborra:2014xia,Dai:2016gtz} that would produce neutrinos. In these scenarios, neutrinos from non-jetted and/or misaligned jetted TDEs could be detectable \cite{Dai:2016gtz}. As such, we include estimates for the numbers of neutrino events and neutrino horizons for these scenarios in Table~\ref{table:events-bytype}.

In Ref.~\cite{Dai:2016gtz}, Dai and Fang modeled TDE neutrino fluences using parameters motivated by observations of nearby bright TDEs and allowing for the possibility of neutrino production outside of a relativistic jet. In modeling the neutrino fluence, Dai and Fang determined the total energy injected into cosmic rays over the duration of the TDE ($\mathcal{E}_{\rm CR}$). To that end, they adopted two approaches: one in which $\mathcal{E}_{\rm CR} \sim 10^{51}$~ergs and is presumed the same for every TDE, and one in which $\mathcal{E}_{\rm CR}$ is taken to be ten times the energy emitted in photons as determined from the observed X-ray or optical luminosity of nearby TDEs and a blackbody spectrum. It is worth noting that the value of $10^{51}$~ergs for the first approach is specifically the value required to produce the astrophysical neutrino flux measured by IceCube~\cite{Aartsen:2017mau} assuming a cosmological rate of $\mathcal{R} \sim 10^{-7}$~Mpc$^{-3}$~yr$^{-1}$,\footnote{This rate was calculated in Ref.~\cite{Dai:2016gtz} assuming an observed TDE rate of $\mathcal{R}_{\rm obs} \sim 10^{-5}$ per galaxy per year~\cite{Donley:2002mp}.} whereas values adopted in the second approach were calculated from observations and assuming a pion production efficiency of $f_{\pi} \sim 0.1$. For our calculations, we adopt the value of $\mathcal{E}_{\rm CR} \sim 10^{51}$~ergs for the first model (labelled ``average'' in Table~\ref{table:events-bytype}. In the second model (labelled ``bright'' in Table~\ref{table:events-bytype}), we adopt a similar approach to the second scenario presented by Dai and Fang, taking $\mathcal{E}_{\rm CR} \sim 10 \times E^{\rm obs}_{\rm rad} = 5 \times 10^{50}$~ergs (where the value for $E^{\rm obs}_{\rm rad}$ was adopted from values provided by Dai and Fang for nearby bright TDEs) but we take $f_{\pi} \sim 1$ since $f_{\pi}$ in non-jetted scenarios could be substantially different from $0.1$~\cite{Dai:2016gtz}. As such, our calculations for the second model are somewhat more optimistic than for the first model. Our calculated neutrino horizons ($z_{\rm hor} \sim 2.6$ and $5.9$~Mpc, respectively, for the ``average'' and ``bright'' scenarios) indicate that these events would have to be fairly nearby in order for POEMMA to detect neutrinos. Assuming the Dai and Fang cosmological rate of $\mathcal{R} \sim 10^{-7}$~Mpc$^{-3}$~yr$^{-1}$, the resulting ToO rate is rather low, requiring POEMMA to operate for longer than a typical mission lifetime in order to detect one such event. Higher rates suggested by some references in the literature~\cite[see \textit{e.g.},][]{Magorrian:1999vm} or by the upper limit of the Lunardini and Winter~\cite{Lunardini:2016xwi} rate (after correcting for the jet solid angle) would imply higher ToO rates, but still at the level of requiring a mission lifetimes that would be longer than typical.

\sideheader{Gamma-ray Bursts}
GRBs are associated with the deaths of massive stars and/or the birth of stellar-mass compact objects. The population of GRBs can be divided into two categories: long duration GRBs (lGRBs) with gamma-ray light curves lasting more than $2$ seconds, and short duration GRBs (sGRBs) with gamma-ray light curves that are shorter than $2$ seconds. lGRBs have been linked with core-collapse supernovae of massive stars ($\gtrsim 25 M_{\odot}$, whereas sGRBs are thought to arise from the merger of two neutron stars or the merger of a neutron star with a black hole. In either scenario, the phenomenology of GRBs can be described through the framework of the fireball model~\cite{1986ApJ...308L..43P,1986ApJ...308L..47G,1990ApJ...365L..55S,1993ApJ...405..278M}. In this model, the creation of a compact object releases a large quantity of gravitational energy in the form of an optically thick fireball of high-energy radiation and particles funneled into a relativistic jet. Similar to the source classes that have already been discussed in this paper, GRB jets could accelerate UHECRs and produce high-energy neutrinos. The pioneering works of Waxman in Ref.~\cite{Waxman:1995vg} and Waxman and Bahcall in Ref.~\cite{Waxman:1997ti} set the stage for extensive work in the literature on the topic of UHECR and neutrinos from GRBs [see \textit{e.g.}, \citenum{Guetta:2003wi,Murase:2006dr,Murase:2007yt,Hummer:2011ms,Baerwald:2013pu,Bustamante:2014oka,Kimura:2017kan,Vietri:1995hs,Dermer:2006bb,Wang:2007xj,Murase:2008mr,Globus:2014fka,Zhang:2017moz,Anchordoqui:2007tn}; for detailed review and more complete reference list see~\citenum{Meszaros:2014tta}].

In contrast to the process discussed earlier for producing neutrinos via BNS mergers, we now explore neutrino production in the sGRB that would occur during or immediately following the BNS merger. In Ref.~\cite{Kimura:2017kan}, KMMK modeled neutrino fluences from various phases of sGRBs, including the prompt phase and the extended emission phase accompanying $\sim 25$\% of sGRBs \cite{2011ApJS..195....2S}, for various assumptions for key GRB jet parameters. In Ref.~\cite{ANTARES:2017bia}, the ANTARES, IceCube, and Pierre Auger Collaborations compared their sensitivities to KMMK modeled fluences rescaled to a luminosity distance of $40$ Mpc. For sGRBs that are viewed on-axis, IceCube can constrain scenarios with more optimistic neutrinos fluences as long as the source is within $\sim 40$~Mpc. At the higher energies where Auger has sensitivity, the predicted neutrino fluences are substantially lower and would be undetectable for a source at $40$~Mpc in the case of neutrino emission from the extended emission phase.  

For our calculations for POEMMA, we consider the moderate extended emission model of KMMK. For sources located on the order of a few Mpc, we expect POEMMA to detect on the order of hundreds to thousands of neutrinos from the extended emission phase. For the neutrino horizon, we expect POEMMA to be able to detect neutrinos out to on the order of $120$~Mpc. Taking the local sGRB rate of $4$ -- $10$~Gpc$^{-3}$~yr$^{-1}$ \cite{Wanderman:2014eza} and multiplying by a factor of $0.25$ for the extended emission model (as only $25$\% of sGRBs have extended emission), we find that the resulting ToO rate would require a longer than typical mission lifetime in order for POEMMA detect one such event.

We also consider the possibility of detecting neutrinos from lGRBs. As in the case of sGRBs, neutrino production has been studied in all of the various phases of lGRBs. IceCube searches for neutrinos  coincident with GRBs have lead to stringent constraints on their contribution to the diffuse astrophysical neutrino flux and on the parameter space for GRB neutrino and UHECR production in single-zone fireball models~\cite{Aartsen:2017wea}; on the other hand, such searches are restricted to the prompt phase of the GRB, and hence, do not meaningfully address neutrino production in the GRB afterglow phase~\cite{Aartsen:2017wea}. As such, in determining the prospects of detecting neutrinos from lGRBs, we consider two models from Ref.~\cite{Murase:2007yt} of neutrino production in the lGRB early afterglow: one in which the circumburst environment is taken to be similar to the interstellar medium (ISM), and one in which the circumburst environment follows parameterized model in order to simulate an environment that would have included material that had been blown off of the massive progenitor star over the course of its lifetime (wind). Both models under consideration include target photons from the early afterglow and the overlapping prompt emission. The late prompt neutrino models that were also studied by Murase~\cite{Murase:2007yt} yield results that are similar to those for the wind model provided in Table~\ref{table:events-bytype}. As the wind model predicts higher neutrino fluences than the ISM model by roughly two orders of magnitudes, the results in the wind scenario are quite a bit more optimistic. An lGRB resembling the ISM model would have to be within $3$~Mpc in order to be detectable by POEMMA. On the other hand, for an lGRB resembling the wind model, we expect that POEMMA will be able to detect tens to hundreds of neutrinos for sources at distances on the order of a few Mpc. In this model, POEMMA will be able to detect neutrinos out to a distance of on the order of $40$~Mpc. Based on the local lGRB rate of $0.42$~Gpc$^{-3}$~yr$^{-1}$ \cite{Lien:2013qja}, we expect a longer than typical mission lifetime in order for POEMMA to detect one such event in either scenario.

\sideheader{Newly-born Pulsars and Magnetars from Core-Collapse Supernovae}
As noted earlier, newly born, rapidly spinning magnetars are promising candidate sources of UHECRs and neutrinos depending on the nature of the environment of the magnetar. The surrounding medium of a pulsar and a magnetar formed in a core-collapse supernova is likely to be distinct from that resulting from a BNS merger as the environment in the former is characteristic of stellar material from the exploding star whereas the environment of the latter would be characteristic of tidal debris from the merging neutron stars and the associated radiation~\cite{2014MNRAS.439.3916M}. In fact, CRs accelerated by core-collapse pulsars and magnetars will readily interact in the surrounding medium, preventing their escape as UHECRs; on the other hand, these interactions will produce high-energy neutrinos~\cite{Murase:2009pg,Fang:2013vla,Fang:2014qva}. In Ref.~\cite{Fang:2014qva}, Fang modeled neutrino production by newly-born core-collapse pulsars and magnetars under various assumptions for the magnetic field strength, spin period, and CR composition. In evaluating the sensitivity of POEMMA to detect neutrinos from these sources, we adopt three models from Ref.~\cite{Fang:2014qva}: a Crab-like pulsar model with pure proton composition, a magnetar model with pure proton composition, and a magnetar model with pure iron composition. In the Crab-like model, the lower magnetic fields and longer spin period limits the energy of the accelerated CRs, and very few of them are accelerated to ultra-high energies. As such, the neutrino fluence arising from Crab-like pulsars is expected to be very low; in fact, we find that such a source would have to be inside or very close to the Galaxy in order to be detectable by POEMMA. In contrast, the magnetar models result in higher neutrino fluences as more CRs are accelerated to ultra-high energies in these models. Our results for these two models are roughly similar, though the pure iron model results in slightly more neutrino events since the maximum energy for iron is $26$ times that of protons. For these models, we expect POEMMA to detect tens of thousands of neutrinos from a newly-born magnetar at the GC. The horizons for these models are on the order of $1$ -- $2$~Mpc, indicating that the magnetar would have to be fairly close to be detectable by POEMMA. In order to estimate the expected ToO rate, we use the local rate of superluminous supernovae expected to produce magnetars provided by Refs.~\cite{Quimby:2013jb,Villar:2018toe}, $\mathcal{R} \sim 21$~Gpc$^{-3}$~yr$^{-1}$. Based on this rate, we expect a ToO rate of $<< 1$ per $25$-year observation time with POEMMA. The rate for less luminous supernovae is many orders of magnitude higher: $\mathcal{R} \simeq \left(1.06 \pm 0.19\right) \times 10^{-4}$~Mpc$^{-3}$~yr$^{-1}$ \cite{Taylor:2014rlo}; however, the much smaller horizon for Crab-like pulsars implies a ToO rate that is comparable to those of the magnetar models considered here.

\bibliographystyle{utphys}
\bibliography{references}
\end{document}